\documentclass[onecolumn,preprint,preprintnumbers,nofootinbib,floats,12ptm,tightenlines]{revtex4}
\pdfoutput=1

\usepackage{amsmath,amssymb,color,mathrsfs,verbatim,epsfig, wasysym}
\usepackage[hyperfootnotes=false]{hyperref}
\usepackage{slashed}
\usepackage{multirow}
\usepackage{xspace}
\allowdisplaybreaks

\pdfoutput=1

\begin{document}

\baselineskip=18pt


\thispagestyle{empty}
\vspace{20pt}
\font\cmss=cmss10 \font\cmsss=cmss10 at 7pt


\hfill
\vspace{20pt}

\begin{center}
{\Large \textbf
{A new scalar resonance at $750$ GeV: \\[0.15cm] Towards a proof of concept in favor of strongly interacting theories
}}
\end{center}

\vspace{15pt}
\begin{center}
{\large Minho Son$^{\, a, b}$ and Alfredo Urbano$^{\, a}$}
\vspace{30pt}

\centerline{$^{a}$ {\small \it CERN, Theory division, CH-1211 Gen\`eve 23,
Switzerland.
}}
\vskip 3pt
\centerline{$^{b}$ {\small \it Department of Physics, Korea Advanced Institute of Science and Technology,
}}
\centerline{ {\small \it 335 Gwahak-ro, Yuseong-gu, Daejeon 305-701, Korea.
}}

\end{center}

\vspace{20pt}
\begin{center}
\textbf{Abstract}
\end{center}
\vspace{5pt} {\small
We interpret the recently observed excess in the diphoton invariant mass as a new spin-0 resonant particle. 
On theoretical grounds, an interesting question is whether this new scalar resonance belongs to a strongly coupled sector or a well-defined weakly coupled theory. 
A possible UV-completion that has been widely considered in literature 
is based on the existence of new vector-like fermions whose loop contributions---Yukawa-coupled to the new resonance---explain the observed signal rate. 
The large total width preliminarily suggested by data seems to  favor a large Yukawa coupling, at the border of a healthy perturbative definition. This potential problem can be fixed by introducing multiple vector-like fermions or large electric charges, bringing back the theory to a weakly coupled regime.
However, this solution risks to be only a low-energy mirage: 
Large multiplicity or electric charge can dangerously reintroduce the strong regime by modifying the renormalization group running of the dimensionless couplings. 
This issue is also tightly related to the (in)stability of the scalar potential. 
First, we study---in the theoretical setup described above---the parametric behavior of the diphoton signal rate, total width, and one-loop $\beta$ functions. 
Then, we numerically solve the renormalization group equations, taking into account the observed diphoton signal rate and total width, 
to investigate the fate of the weakly coupled theory. 
We find that---with the only exception of few fine-tuned directions---weakly coupled interpretations of the excess are brought back to a strongly coupled regime if the running is taken into account.
}

\vfill\eject
\noindent


\section{Introduction}
Both ATLAS and CMS announced an excess in the diphoton invariant mass distributions, using Run II data at $\sqrt{s} = 13$ TeV~\cite{ATLAS-CONF-2015-081,CMS-PAS-EXO-15-004}. ATLAS analyzed $3.2$ fb$^{-1}$ of data and reports the local significance of 3.9$\sigma$ for an excess peaked at 750 GeV whereas CMS, using $2.6$ fb$^{-1}$ of data, reports a local significance of 2.6$\sigma$ for an excess peaked at 760 GeV. The global significance is reduced to 2.3$\sigma$ and 1.2$\sigma$ for ATLAS and CMS respectively.

The observed excess is still compatible with a statistical fluctuation of the background,
and only future analysis will eventually reveal the truth about its origin.
In the meantime, it is possible to interpret the excess as the imprint of the diphoton decay of
a new spin-0 resonance (see~\cite{Harigaya:2015ezk,Mambrini:2015wyu,Angelescu:2015uiz,Knapen:2015dap,Franceschini:2015kwy,Buttazzo:2015txu,Pilaftsis:2015ycr,Gupta:2015zzs,Falkowski:2015swt,Petersson:2015mkr,Low:2015qep,Dutta:2015wqh,Kobakhidze:2015ldh,Cox:2015ckc,Ahmed:2015uqt,Cao:2015pto,Becirevic:2015fmu,No:2015bsn,McDermott:2015sck,Higaki:2015jag,Chao:2015ttq,Fichet:2015vvy,Demidov:2015zqn,Bian:2015kjt,Chakrabortty:2015hff,Bai:2015nbs,Csaki:2015vek,Kim:2015ron,Gabrielli:2015dhk,Curtin:2015jcv,Berthier:2015vbb,Kim:2015ksf,Bi:2015uqd,Huang:2015evq,Cao:2015twy,Heckman:2015kqk,Antipin:2015kgh,Ding:2015rxx,Barducci:2015gtd,Cho:2015nxy,Liao:2015tow,Feng:2015wil,Bardhan:2015hcr,Chang:2015sdy,Luo:2015yio,Chang:2015bzc,Han:2015cty,Chao:2015nsm,Bernon:2015abk,Carpenter:2015ucu,Megias:2015ory,Alves:2015jgx,Han:2015qqj,Liu:2015yec,Craig:2015lra,Cheung:2015cug,Das:2015enc,Davoudiasl:2015cuo,Allanach:2015ixl,Altmannshofer:2015xfo,Cvetic:2015vit,Patel:2015ulo,Gu:2015lxj,Chakraborty:2015gyj,Cao:2015xjz,Huang:2015rkj,Belyaev:2015hgo,Pelaggi:2015knk,Hernandez:2015ywg,Murphy:2015kag,deBlas:2015hlv,Dev:2015isx,Boucenna:2015pav,Chala:2015cev,Bauer:2015boy,Cline:2015msi,Dey:2015bur,Ellis:2015oso,Nakai:2015ptz,Molinaro:2015cwg,Backovic:2015fnp,DiChiara:2015vdm,Bellazzini:2015nxw} for similar or other possible interpretations). 
In this simple setup, the observed signal events are translated into a diphoton signal rate with central values
at  6 and 10 fb for the CMS and ATLAS analyses~\cite{ATLAS-CONF-2015-081,CMS-PAS-EXO-15-004}, respectively.

The postulated new scalar resonance is very likely  part of some unknown dynamics,  
related or not to the electroweak symmetry breaking. 
First and foremost, a crucial point is to understand whether this new dynamics is weakly or strongly coupled.
In either case, it will lead us to an exciting era beyond the Standard Model (SM). 
In the context of a weakly-coupled theory, a simple extension of the SM compatible with the excess considers the presence---in addition to the aforementioned scalar particle---of new vector-like fermions interacting with the scalar resonance via a Yukawa-like interaction. The new fermions mediate production of the new resonance via gluon fusion, and its subsequent diphoton decay.

The size of the new Yukawa coupling that successfully accounts for the signal rate in this framework is strongly correlated to the assumption on the total width. 
For instance, when assuming that the gluon PDF is mainly responsible for the production of the scalar resonance, 
the typical size of the total width from decay channels to gluons and photons is 
too small to explain the large total width, $\Gamma/M \sim$ 6\% in ATLAS~\cite{ATLAS-CONF-2015-081} (which corresponds to $\sim$ 45GeV).
 It is very unlikely that the above simple extension can produce a total width of order $O(1 {\rm GeV})$ 
 without invoking a large Yukawa coupling, large electric charge or large number of vector-like fermions. 
 
A couple of interesting questions naturally arise. The presence of a new scalar particle, interacting with new vector-like fermions with large Yukawa couplings and electric charges may introduce dangerous problems since the dimensionless parameters describing the new particles and their interactions are tightly connected by the Renormalization Group Equations (RGEs). By following the running from a low energy to a higher scale, the theory can develop several pathological behaviors, for instance violating perturbativity or generating unstable directions in the scalar potential.

The goal of this work is to survey the compatibility of simple models based on the presence 
of new vector-like fermions with {\it i)} the assumptions of a weakly coupled theory and  {\it ii)} the fit of the observed diphoton excess. To this end, we first study the parametric behavior of the diphoton signal rate, total width, and one-loop $\beta$ functions 
for all the relevant couplings. 
In full generality, we allow for a mixing between the scalar resonance and the SM Higgs. 
We numerically solve the RGEs, taking into account the observed diphoton signal rate and total width, 
to investigate the fate of the weakly coupled theory. 

In Section~\ref{sec:Diphoton} we discuss the general properties of the diphoton excess in the context of the new spin-0 resonance. In Section~\ref{sec:VectorLikeFrmions}, we study the parametric behavior of the diphoton signal rate and total width in a simple extension with new vector-like fermions. We show the parameter space compatible with the observed excess.
 In Section~\ref{sec:singlet}, we take the SM Higgs into account, and discuss the phenomenological implication.
  We briefly discuss the issue of the (in)stability of the scalar potential. In Section~\ref{sec:RGE} we provide one-loop $\beta$ functions, including that of the new Yukawa coupling, and matching conditions. 
  We discuss the parametric behavior qualitatively in terms of a large Yukawa coupling, a large electric charge, and a large number of vector-like fermions. We numerically solve the RGEs in several benchmark models, and discuss their features. Finally, we conclude in Section~\ref{sec:Summary}.

\section{Diphoton excess and new spin-0 resonance}\label{sec:Diphoton}
The cross section of diphoton production via $s$-channel exchange of a spin-0 resonance with mass $M$ and total width $\Gamma$, assuming narrow width, is 
\begin{equation}\label{eq:xsec:M}
 \sigma(pp\rightarrow S\rightarrow \gamma\gamma) = \frac{1}{M\Gamma s} \big [ C_{gg}\Gamma (S\rightarrow gg) + \sum_q C_{q\bar{q}}\Gamma(S\rightarrow q\bar{q}) \big ]\Gamma(S\rightarrow\gamma\gamma)~.
\end{equation}
In what follows, we use the short-hand notation $\Gamma(S\rightarrow\gamma\gamma) = \Gamma_{\gamma\gamma}$, $\Gamma(S\rightarrow gg) = \Gamma_{gg}$. If the main production process is due to gluon fusion (see \cite{Franceschini:2015kwy,Gupta:2015zzs} for related discussion), the cross section in Eq.~\ref{eq:xsec:M} reduces to
\begin{equation}\label{eq:xsec:Cgg}
 \sigma(pp\rightarrow S\rightarrow \gamma\gamma) \approx \frac{M}{\Gamma}\frac{1}{s} C_{gg}\frac{\Gamma_{gg}}{M}\frac{\Gamma_{\gamma\gamma}}{M}~.
\end{equation}
This assumption is favored by data, but it remains interesting to consider other production processes 
as well.\footnote{It will change the parametric dependence of the signal rates of the relevant channels and total width in terms on the involved parameters.} $C_{gg}$, $C_{q\bar{q}}$ in Eq.~\ref{eq:xsec:M}, \ref{eq:xsec:Cgg} are luminosity functions, and for gluon fusion we have
\begin{equation}
 C_{gg}=\frac{\pi^2}{8} \int_{M^2/s}^1 \frac{dx}{x}\ g(x) g\left (M^2/sx \right ) = 2137\ (174)\quad {\rm at} \quad \sqrt{s}=13\ (8)\,{\rm TeV}~,
\end{equation}
where $g(x)$ is the gluon parton distribution function and the values are estimated using MSTW2008NLO for $M = 750$ GeV. The observed signal rate, $\sim 8$ fb, implies
\begin{equation}
 \frac{\Gamma_{gg}}{M}\frac{\Gamma_{\gamma\gamma}}{M} \approx 1.6 \times 10^{-6}\frac{\Gamma}{M}~. 
\end{equation}
An additional piece of information that plays an important role in shaping any New Physics interpretation is the total width $\Gamma$. Recent ATLAS data from the run at $\sqrt{s} =$ 13 TeV~\cite{ATLAS-CONF-2015-081} indicates a total width of $\Gamma/M \sim 0.06$.

Production of the spin-0 resonance and its decay to diphoton can be studied in a model-independent way via the following effective Lagrangian,
\begin{equation}\label{eq:dim5SFF}
\frac{e^2}{16\pi^2}\frac{c_{s\gamma\gamma}}{M}S F_{\mu\nu}F^{\mu\nu}+\frac{g_s^2}{16\pi^2}\frac{c_{sgg}}{M}S G^a_{\mu\nu}G^{a\, \mu\nu}~.
\end{equation}
where loop suppression factors account for possible loop-induced origins of the effective operators. 
We assume that the scalar resonance is CP-even, and we expect that our finding also applies to the CP-odd case. Model-independent constraints on the effective couplings $c_{s\gamma\gamma}$ and $c_{sgg}$ in Eq.~\ref{eq:dim5SFF} appeared in the recent literature~\cite{Franceschini:2015kwy,Gupta:2015zzs,Falkowski:2015swt}. 
In the next Sections, we will rephrase these constraints in the context of a simple UV-complete model.

\section{On the role of vector-like fermions}\label{sec:VectorLikeFrmions}
A simple way to generate the dimension-5 operators in Eq.~{\ref{eq:dim5SFF}} is to introduce new colored vector-like fermions with electric charge. For instance, the new singlet $S$ may be coupled via a Yukawa-like interaction to a vector-like fermion $X$ described by the following Lagrangian
\begin{equation}
 \mathcal{L}_X  =  \overline{X}(i\slashed{D} - m_X)X  - y_X S \overline{X}X~.
\end{equation}
The dimension-5 operators in Eq.~{\ref{eq:dim5SFF}} are loop-generated by exchanging $X$. 
We focus on the case in which $X$ transforms like $({\bf 1}, \, {\bf 3})_{Q_X}$ under $(SU(2)_L,\, SU(3)_C)_{U(1)_Y}$.  The partial decay widths in this simple toy model are given by
\begin{equation}
\begin{split}
\Gamma_{\gamma\gamma} &= \frac{\alpha^2}{16 (4\pi)^3} c_{s\gamma\gamma}^2 M~, \\
\Gamma_{gg} &= \frac{\alpha_s^2}{2 (4\pi)^3} c_{sgg}^2 M~.
\end{split}
\end{equation}
The coefficients, $c_{s\gamma\gamma}$, $c_{sgg}$ of the effective operators in Eq.~\ref{eq:dim5SFF} are
\begin{equation}\label{eq:loopft}
\begin{split}
c_{s\gamma\gamma} &= 6\, Q_X^2 [y_X 2 \sqrt{\tau} A_{1/2}(\tau)] = 6\, Q_X^2 c_{sgg}~,
\end{split}
\end{equation}
where the loop function $A_{1/2}(\tau)$ (with $\tau = M^2/(4 m^2_X)$) can be found in~\cite{Spira:1995rr}. Assuming $\Gamma = \Gamma_{gg} + \Gamma_{\gamma\gamma}$ in Eq.~\ref{eq:xsec:Cgg}, the cross section has the following parametric dependence,
\begin{equation}\label{eq:xsec:Gammaggaa}
 \sigma(pp\rightarrow S\rightarrow \gamma\gamma) \approx 
\frac{9}{4}\frac{C_{gg}}{s}\frac{c_0^2(\tau)}{(4\pi)^5} g^4_s e^4 \frac{y_X^2 Q_X^4}{g_s^4 + (9/2)\, e^4 Q_X^4}~,
\end{equation}
where $c_0(\tau) = 2 \sqrt{\tau} A_{1/2}(\tau)$ and $c_0(\tau)$ converges to $(4/3)(M/m_X)$ in the limit $\tau \ll 1$. On the other hand, if 
the total decay width is set to a constant value $\Gamma = \Gamma_0$,\footnote{This assumption is suitable for the case that total width dominates over $\Gamma_{gg}+\Gamma_{\gamma\gamma}$, and $\Gamma_0 - (\Gamma_{gg}+\Gamma_{\gamma\gamma})$ is much less sensitive to the $y_X, Q_X$ (as well as the fermion multiplicity, $N_X$, that we will discuss below) than those appearing explicitly in the signal rate.} (for instance, $\sim$ 45 GeV as was indicated by ATLAS data~\cite{ATLAS-CONF-2015-081}) then the cross section scales like
\begin{equation}\label{eq:xsec:Gammaexp}
 \sigma(pp\rightarrow S\rightarrow \gamma\gamma) \approx 
\frac{9}{8}\frac{C_{gg}}{s}\frac{M}{\Gamma_0}\frac{c_0^4(\tau)}{(4\pi)^{10}} g^4_s e^4 y_X^4 Q_X^4~.
\end{equation}
The sum of two partial decay widths from the decay channels to gluons and photons is given by
\begin{equation}\label{eq:par:Gammaggaa}
 \Gamma_{gg}+\Gamma_{\gamma\gamma} = \frac{M}{2(4\pi)^5} c_0^2(\tau) y_X^2 \big [ g_s^4 + \frac{9}{2}e^4 Q_X^4 \big ]~.
\end{equation}
\begin{figure}[tp]
\begin{center}
\textbf{\footnotesize \hspace{8mm} $(\Gamma_{gg}+\Gamma_{\gamma\gamma})/45{\rm GeV}:$ $Q_X=8/3,5/3$}\\
\includegraphics[width=0.45\linewidth]{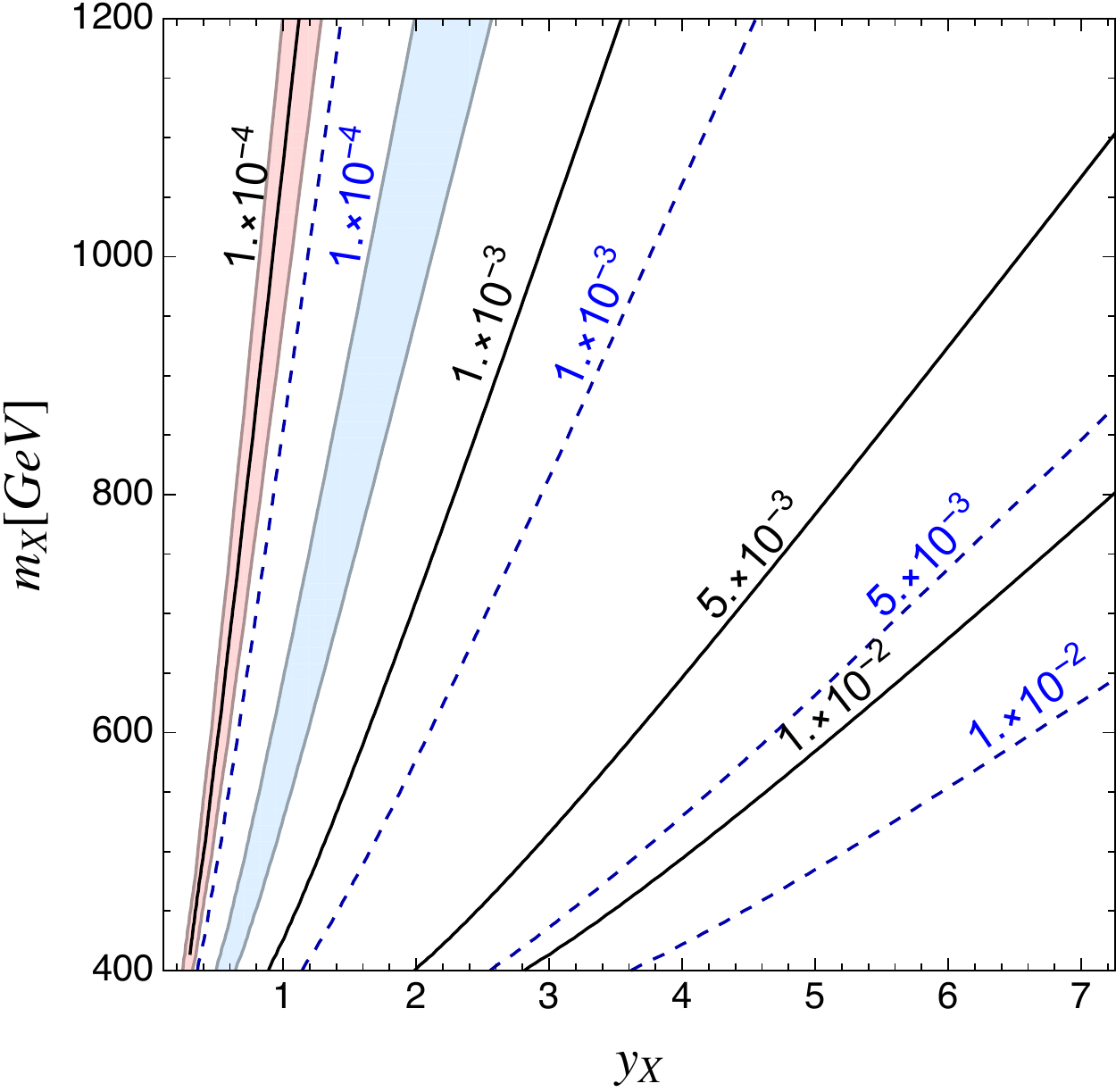}
\caption{The isocontours of the ratio of $(\Gamma_{\gamma\gamma} + \Gamma_{gg})$ to 45 GeV, assuming a new vector-like fermion with  electric charge 8/3 (black solid) and 5/3 (blue dashed). The shaded regions correspond to the $\sigma(pp\rightarrow S \rightarrow \gamma\gamma) = 6-10$ fb for the cases with $Q_X=8/3$ (light red) and $Q_X=5/3$ (light blue). The upper bound of the displayed $y_X$ value corresponds to the maximal value of the Yukawa coupling, $4\pi/\sqrt{N_C}$ ($N_C$ =3). 
}
\label{fig:GammaRatio}
\end{center}
\end{figure}
In the presence of multiple vector-like fermions, the loop function is rescaled by this multiplicity, denoted by $N_X$. This will introduce $N_X$ dependence\footnote{Here, we are assuming vector-like quarks that carry both colour and the electric charge. As a variant, one may consider two different types of vector-like fermions: one type with only colour and the other type with only the electric charge. We will not consider this option in this work.}
in the diphoton signal rate as well as in the partial decay widths from the decay channels to gluons and photons. Other important factors that can affect a New Physics interpretation are the $k$-factor in the gluon PDF, denoted by $k_{gg}$, and the rescaling factor of the overall observed signal rate in diphoton excess.\footnote{Since the diphoton excess suffers from low statistics, we take into account a possibility of a fluctuation in the observed signal events.}

The cross sections in Eq.~\ref{eq:xsec:Gammaggaa},~\ref{eq:xsec:Gammaexp} and the decay width in Eq.~\ref{eq:par:Gammaggaa} have different parametric dependences on $y_X$, $Q_X$, and $N_X$ as well as on the other couplings. 
Fitting them to the measured cross sections and the total width will shape the possible structure of New Physics. 
Interestingly, the current total decay width indicated by ATLAS, $\Gamma/M \sim 0.06$ (which translates into $\Gamma \sim 45$ GeV for 750 GeV resonance), appears very difficult to be explained by $\Gamma_{gg}+\Gamma_{\gamma\gamma}$ alone, as shown
 in Fig.~\ref{fig:GammaRatio}, while keeping the Yukawa couplings within a perturbative regime:
 The bigger the ratio $(\Gamma_{gg}+\Gamma_{\gamma\gamma})/(45\,{\rm GeV})$,  the stronger the involved Yukawa coupling.
 Since $\Gamma_{gg}+\Gamma_{\gamma\gamma}$ scales like $(N_X y_X)^2 (g^2_s + 9/2\, e^4 Q_X^4)$, a very large multiplicity, $N_X$, (or/and an unrealistically large electric charge, $Q_X$) is necessary to explain a bigger fraction of the total width by means of $\Gamma_{gg}$ and $\Gamma_{\gamma\gamma}$ while staying in a weakly coupled region.
\begin{figure}[tp]
\begin{center}
\textbf{\tiny \hspace{3mm} Assuming $\Gamma=\Gamma_{gg}+\Gamma_{\gamma\gamma}$: $N_X=1$ \hspace{10mm} Assuming $\Gamma=1$GeV: $N_X=1$ \hspace{1.5cm} Assuming $\Gamma=45$GeV: $N_X=1$}\\
\includegraphics[width=0.32\linewidth]{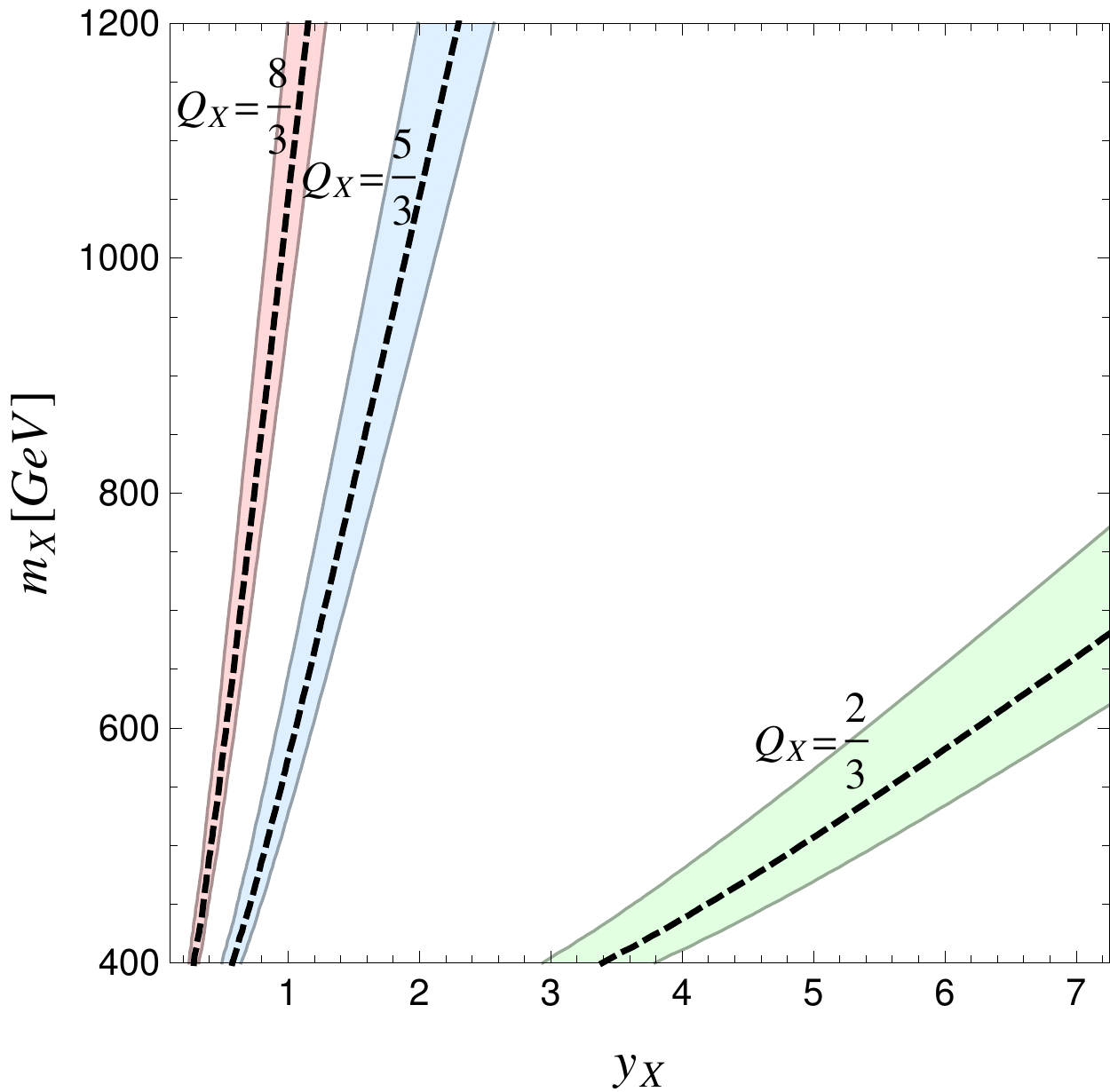}
\includegraphics[width=0.32\linewidth]{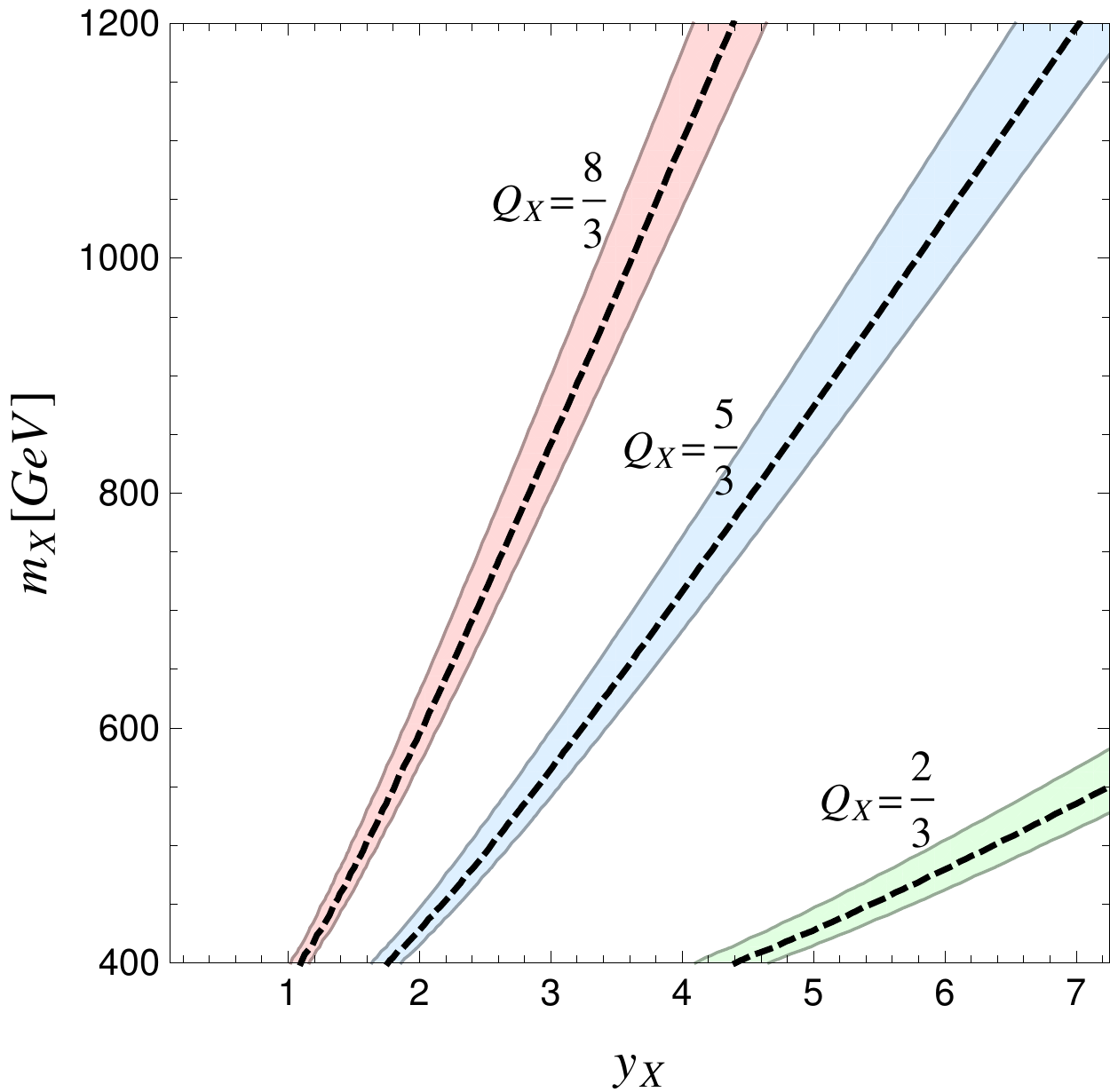}
\includegraphics[width=0.32\linewidth]{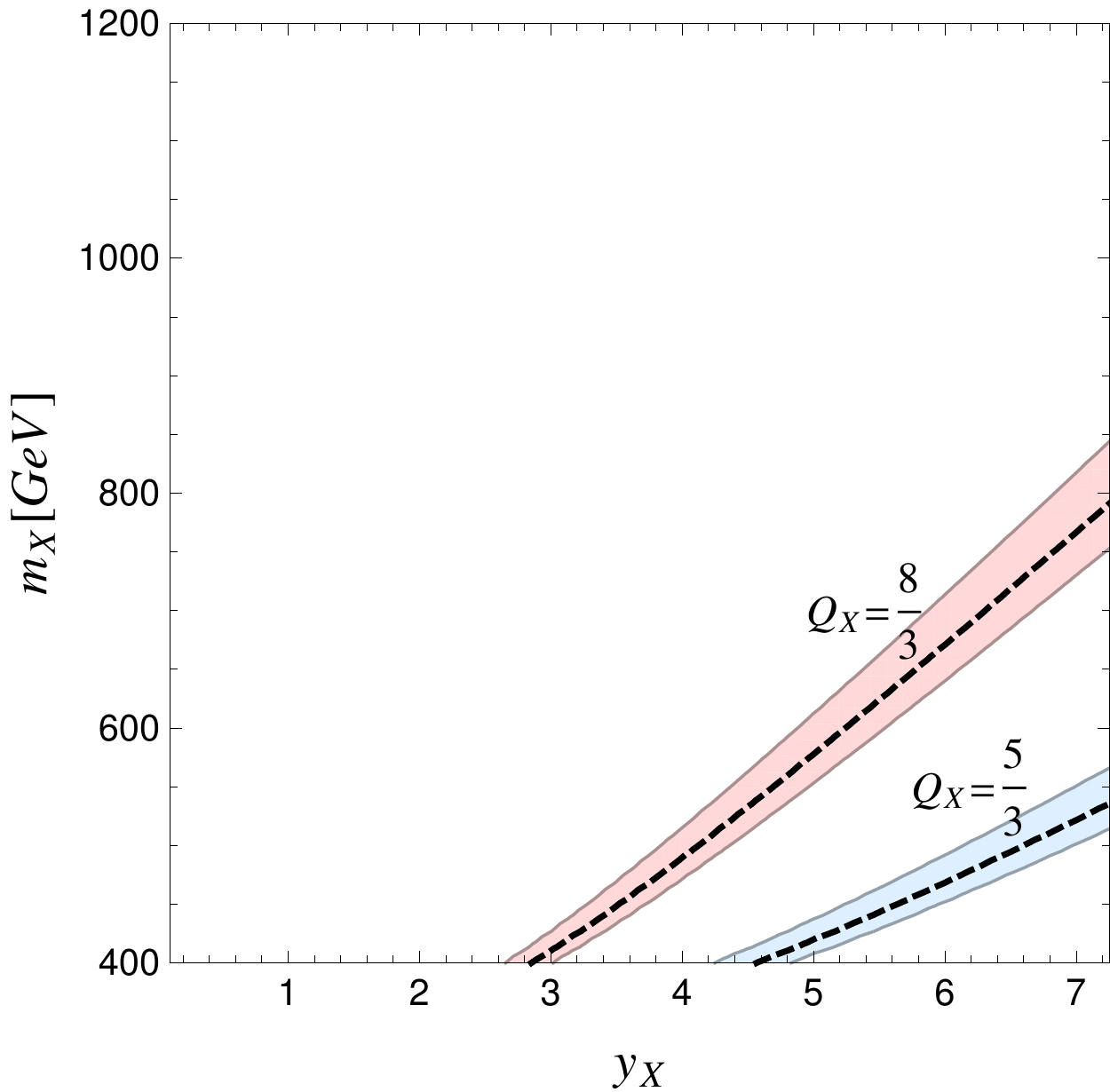}
\caption{Regions that conrrespond to $\sigma(pp\rightarrow S\rightarrow \gamma\gamma)$ = 6$-$10 fb with three different assumptions on the total decay width: $\Gamma=\Gamma_{gg}+\Gamma_{\gamma\gamma}$ (left), $\Gamma = 1$ GeV (middle) and $\Gamma = 45$ GeV (right). The dashed lines corresponds to 8 fb. 1 (45) GeV corresponds to 0.13\% (6\%) total decay width of 750 GeV resonance. 
}
\label{fig:xsecyxmx:threeGamma}
\end{center}
\end{figure}

\begin{figure}[tp]
\begin{center}
\textbf{\tiny \hspace{1mm} Assuming $\Gamma=\Gamma_{gg}+\Gamma_{\gamma\gamma}$: $N_X=5$ \hspace{10mm} Assuming $\Gamma=1$GeV: $N_X=5$ \hspace{1.5cm} Assuming $\Gamma=45$GeV: $N_X=5$\hspace{15mm}}\\
\includegraphics[width=0.32\linewidth]{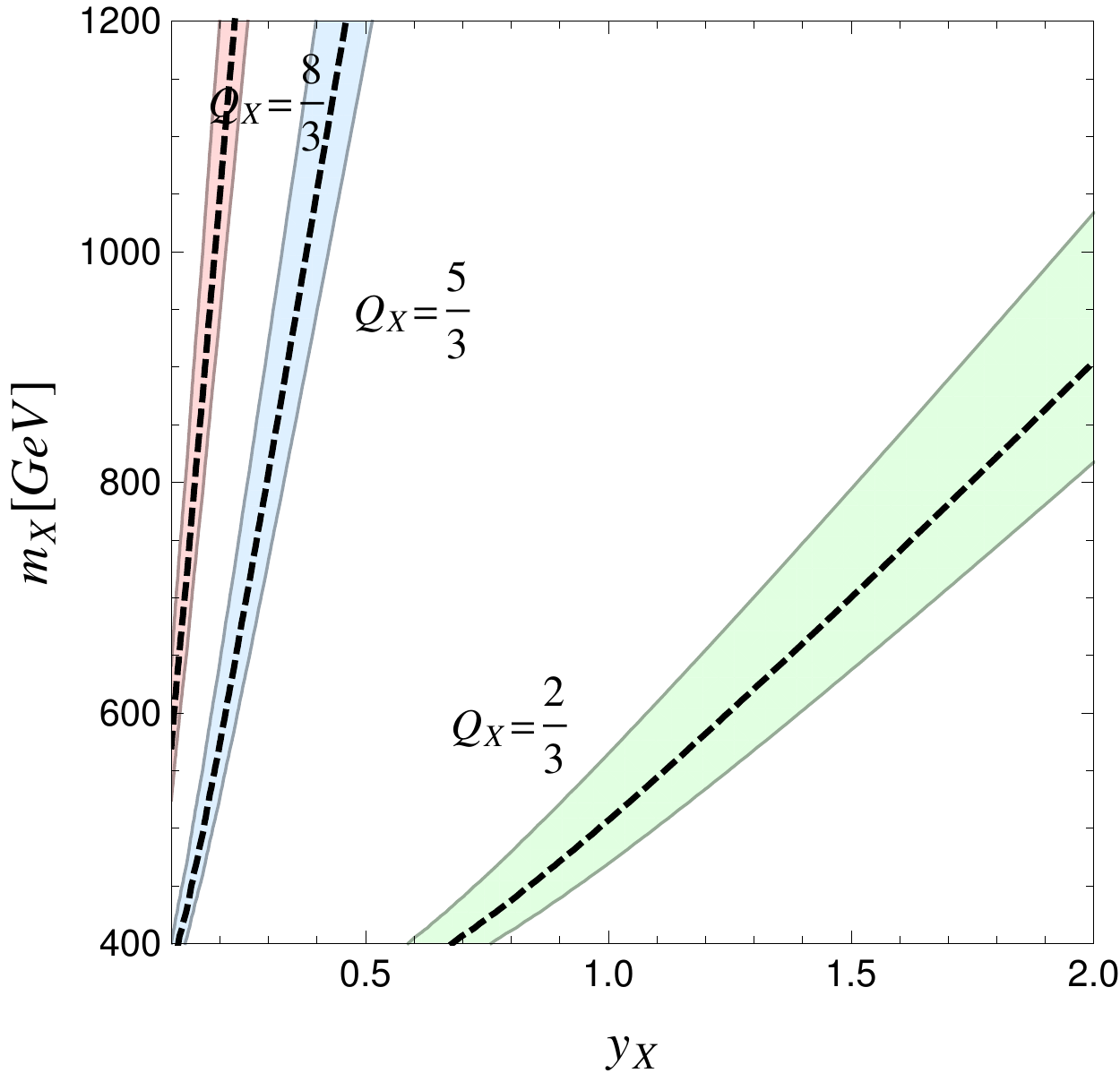}
\includegraphics[width=0.32\linewidth]{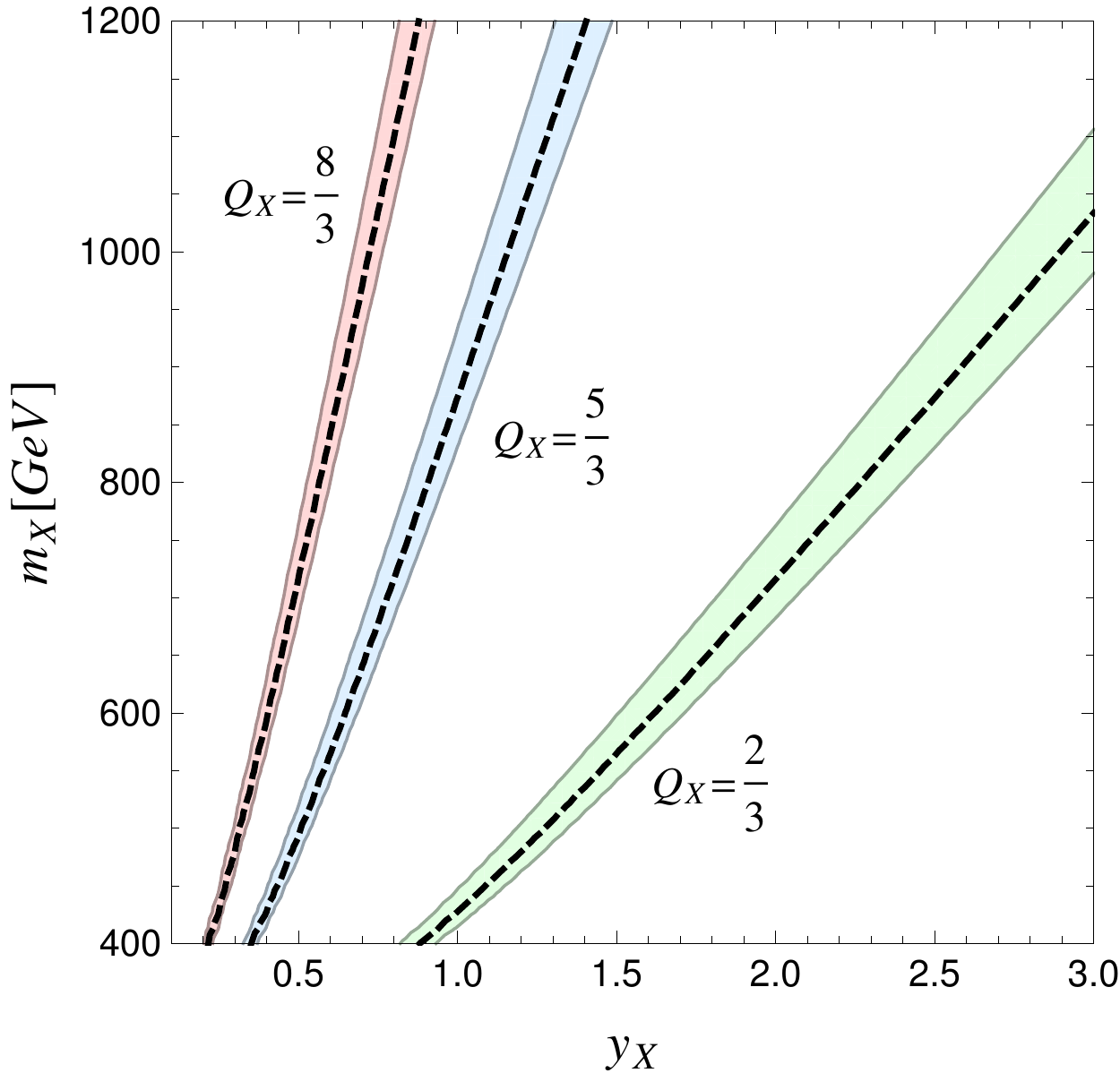}
\includegraphics[width=0.32\linewidth]{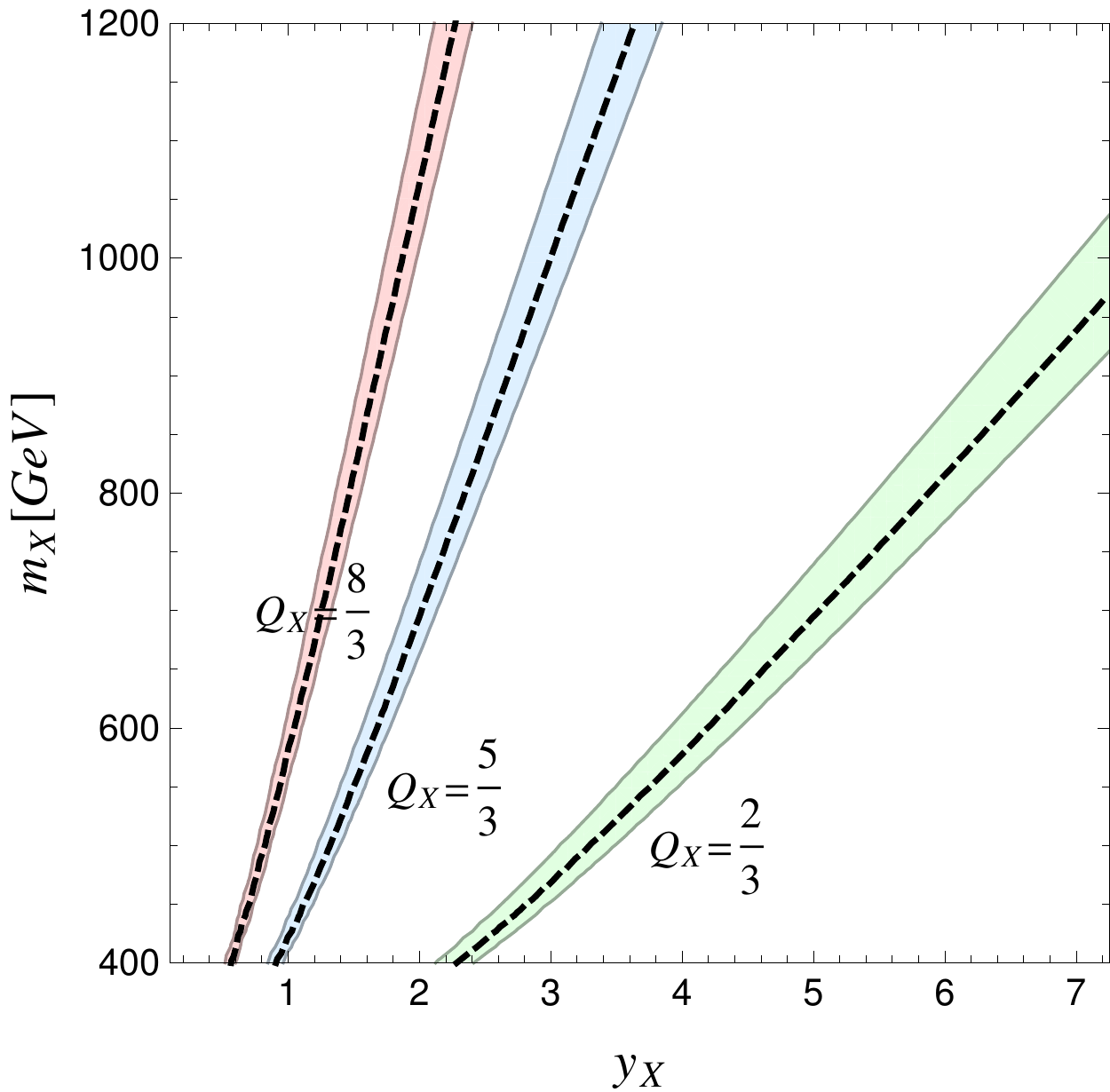}
\\
\textbf{\tiny \hspace{4mm} Assuming $\Gamma=\Gamma_{gg}+\Gamma_{\gamma\gamma}$: $N_X=1,k_{gg}=3$ \hspace{1mm} Assuming $\Gamma=1$GeV: $N_X=1,k_{gg}=3$ \hspace{2mm} Assuming $\Gamma=45$GeV: $N_X=1,k_{gg}=3$}\\
\includegraphics[width=0.32\linewidth]{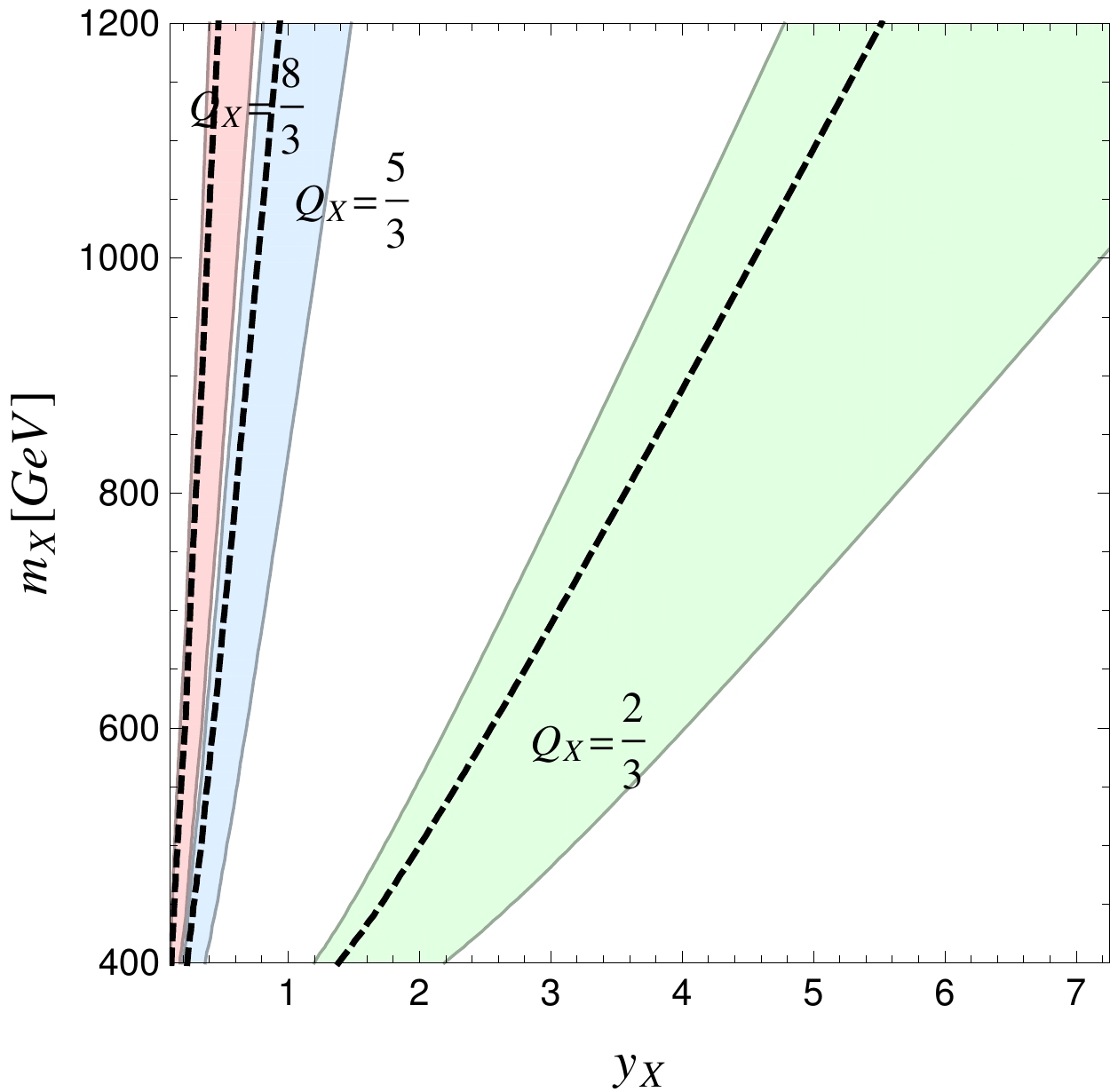}
\includegraphics[width=0.32\linewidth]{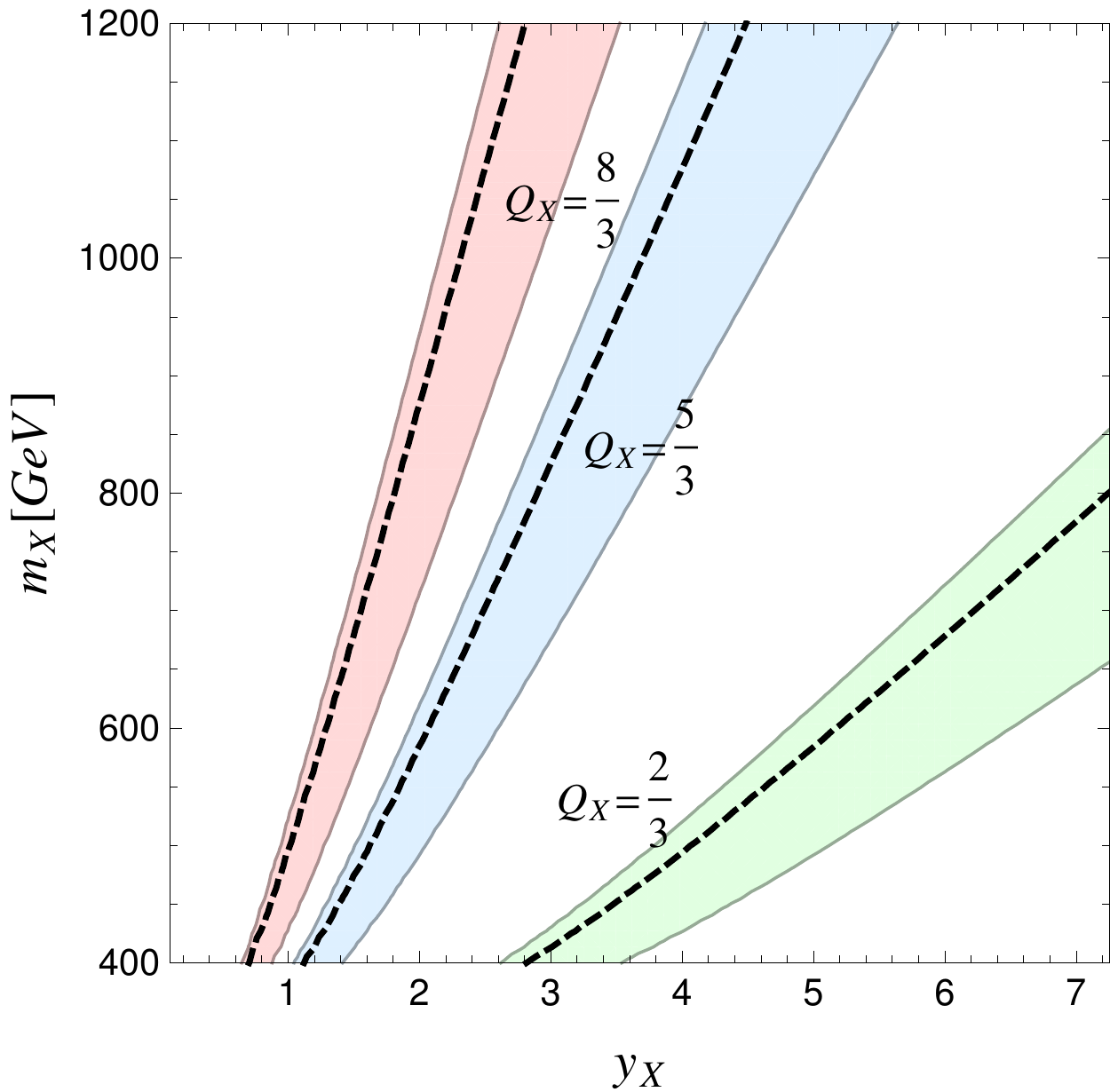}
\includegraphics[width=0.32\linewidth]{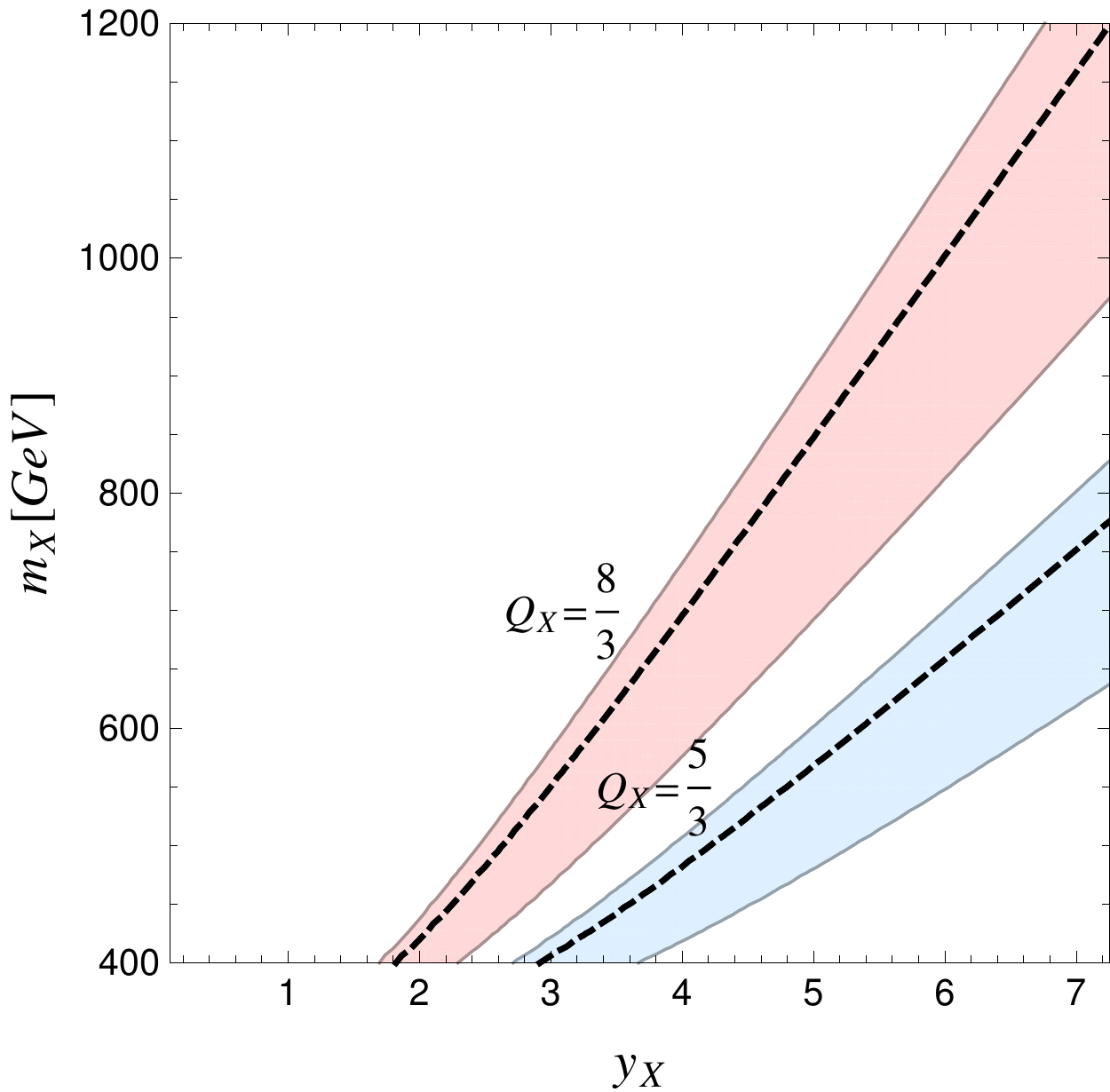}
\caption{Regions that conrrespond to $\sigma(pp\rightarrow S\rightarrow \gamma\gamma)$ = 6$-$10 fb (upper) and 3$-$10 fb (lower) with three different assumptions on total decay width: $\Gamma=\Gamma_{gg}+\Gamma_{\gamma\gamma}$ (left), $\Gamma = 1$ GeV (middle) and $\Gamma = 45$ GeV (right). The dashed lines corresponds to 8 fb (upper) and 4 fb (lower). $\Gamma = 1,\,45$ GeV corresponds to 0.13\%, 6\% total decay width of a $750$ GeV resonance. In the upper panels, a large particle multiplicity $N_X=5$ is considered. In the lower panels, we assume an overall $k$-factor of 3, and reduce the signal rate by a factor of 2.
}
\label{fig:xsecyxmx:threeGamma:Varying}
\end{center}
\end{figure}

A similar conclusion can be drawn by considering the signal rate in the $(y_X,\ m_X)$ parameter space. The situation is illustrated in Fig.~\ref{fig:xsecyxmx:threeGamma}. If $\Gamma = \Gamma_{gg}+\Gamma_{\gamma\gamma}$ (left panel of Fig.~\ref{fig:xsecyxmx:threeGamma}), the claimed signal rate can be obtained  with small Yukawa couplings, namely $y_X \lesssim 1$, 
only assuming large electric charges. The total width, normalized to 45 GeV, in that region is much smaller than $10^{-3}$ according to Fig.~\ref{fig:GammaRatio}. Forcing the total width $\Gamma_{gg}+\Gamma_{\gamma\gamma}$ towards 
the indicated value $\sim 45$ GeV requires strong Yukawa couplings even for  very large electric charges. The middle and right panels of Fig.~\ref{fig:xsecyxmx:threeGamma} illustrate the situation for two cases with $\Gamma =\Gamma_0 =$ 1 (45) GeV which correspond to $\Gamma/M$ for a 750 GeV resonance of 0.13\% (6\%). In these instances, according to Eq.~\ref{eq:xsec:Gammaexp}, the diphoton signal rate scales like 
\begin{equation}\label{eq:LargeWidth:Scaling}
\sim (N_X Q_X y_X)^4~,
\end{equation}
when multiple vector-like fermions with nearly degenerate masses exist. 
Na\"{\i}vely,  it is possible to play with $N_X$ and $Q_X$\footnote{This is different w.r.t. the scaling $\sim (N_X Q_X^2 y_X)^2/(g^4_s + 9/2\, e^4 Q_X^4)$ in Eq.~\ref{eq:xsec:Gammaggaa} where increasing $Q_X$ will have no effect when $Q_X$ becomes large enough for $\Gamma_{\gamma\gamma}$ to dominate over $\Gamma_{gg}$.} to bring large Yukawa couplings back to a weakly coupled regime. For instance, consider a strongly coupled model with $(y_X,\, m_X,\, Q_X)\sim(5,\, 900\, {\rm GeV},\, 5/3)$ in the middle panel of Fig.~\ref{fig:xsecyxmx:threeGamma} assuming $\Gamma = 1$ GeV. The Yukawa coupling can be brought back to the weakly coupled region, $y_X \lesssim 1$, when a large multiplicity, as big as $N_X \gtrsim5$, is available for the same electric charge. The related situation is illustrated in the upper middle panel of Fig.~\ref{fig:xsecyxmx:threeGamma:Varying}. One may consider a large electric charge as big as $Q_X \gtrsim$ 25/3 as well, 
while keeping $N_X = 1$, to achieve $y_X \lesssim 1$. Another possibility is to change both $N_X$ and $Q_X$ such as $N_X \gtrsim 3$ and $Q_X \gtrsim 8/3$.

However, a large $N_X$ or $Q_X$ 
can potentially send the theory back to a strongly coupled regime via the rapid running of the couplings or cause an instability of the scalar potential. 
This point is the main goal in this work, and it will be carried out in the next Sections in great detail. 

Finally, the lower panels of Fig.~\ref{fig:xsecyxmx:threeGamma:Varying} takes into account the 
effect of a large overall $k$-factor, $k_{gg} = 3$, on
 the diphoton signal rate\footnote{Notice that the specific choice $k_{gg} = 3$ was taken 
 for illustration purposes rather than being rigorously derived.} 
 and a reduction of the observed signal rate by the factor $\kappa_\sigma = 2$  (see \cite{Falkowski:2015swt} for related discussion). This can relax the combination $N_X Q_X y_X$ by the factor $(k_{gg}\, \kappa_\sigma)^{1/4}$ for the cases in which 
 the signal rate scales like Eq.~\ref{eq:xsec:Gammaexp}. For the cases in which 
 the signal rate scales like Eq.~\ref{eq:xsec:Gammaggaa}, assuming $\Gamma = \Gamma_{gg}+\Gamma_{\gamma\gamma}$ (very narrow width), the combination $N_X y_X Q_X^2$ can 
 be relaxed by the factor $(k_{gg}\, \kappa_\sigma)^{1/2}$ as long as $Q_X$ is not very large.

\section{The doublet-singlet model}\label{sec:singlet}
\label{sec:ScalarPot}
Let us now add to the game the SM Higgs sector. 
It is reasonable to assume that the new gauge singlet $S$ couples to the SM Higgs doublet $H$ via a mixing term, 
thus affecting both Higgs physics and the stability of the Higgs potential.

We consider the following scalar potential,  
\begin{equation}
V(H,S) = \mu_H^2 |H|^2 + \lambda_H |H|^4 + \frac{\lambda_{HS}}{2} |H|^2S^2 + \frac{\mu_S^2}{2} S^2 +\frac{\lambda_S}{4} S^4~,
\end{equation}
where we assumed that $S$ is real and odd under $S\rightarrow - S$.
The potential in the unitary gauge is obtained via $H(x) = (0, h(x)/\sqrt{2})^T$,
\begin{equation}\label{eq:PotentialhS}
V(h,S) = \frac{\mu_H^2}{2} h^2 + \frac{\lambda_H}{4} h^4 +  \frac{\lambda_{HS}}{4} h^2 S^2 + \frac{\mu_S^2}{2} S^2 +\frac{\lambda_S}{4} S^4~.
\end{equation}
We consider the most general situation in which both scalar fields take a vacuum expectation value (VEV), $\langle h\rangle = v$, $\langle S\rangle = u$.
The $\lambda_{HS}$ and VEVs of the scalar fields induce the mixing between $h$ and $S$,
\begin{equation}\label{eq:massbasis}
  h = \cos\theta H_1 + \sin\theta H_2~, \quad S = - \sin\theta H_1 + \cos\theta H_2~,
\end{equation}
where $H_{1}$, $H_{2}$ are mass eigenstates with masses of $m_{H_1}$, $m_{H_2}$ (see Appendix~\ref{app:ScalarPot} for details). $H_1$ is identified with the physical Higgs boson with $m_{H_1} = 125.09$ GeV, whereas $H_2$ with the new scalar resonance with $m_{H_2} \simeq 750$ GeV. We will use the short-hand notations $\cos\theta \equiv c_\theta$, $\sin\theta \equiv s_\theta$, and $\tan\theta \equiv t_\theta$ in the next Sections.

\subsection{Phenomenological implications}\label{sec:NewResonance}
%
%
\begin{figure}[tp]
\begin{center}
\includegraphics[width=0.455\linewidth]{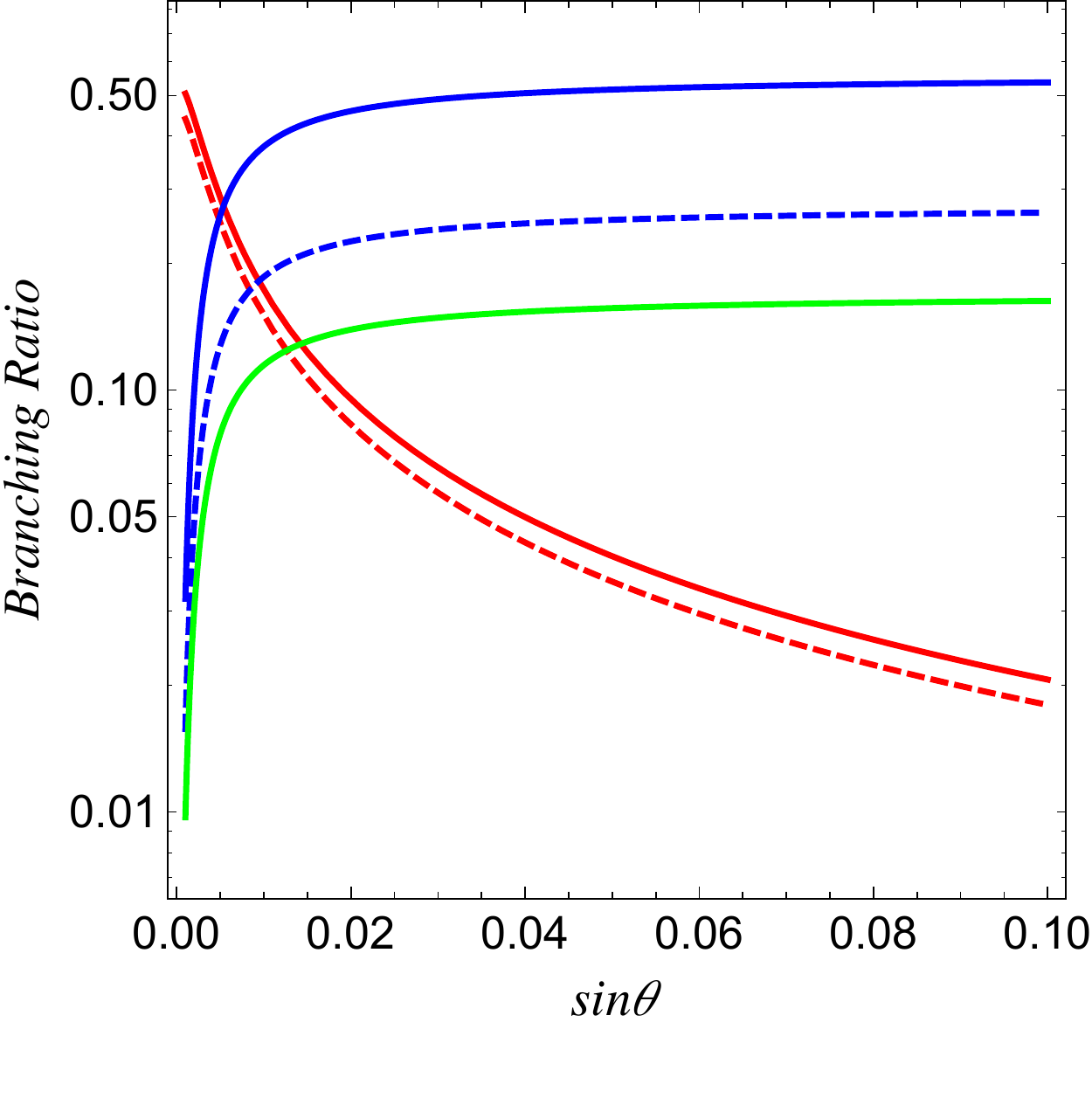}
\hspace{10mm}
\includegraphics[width=0.46\linewidth]{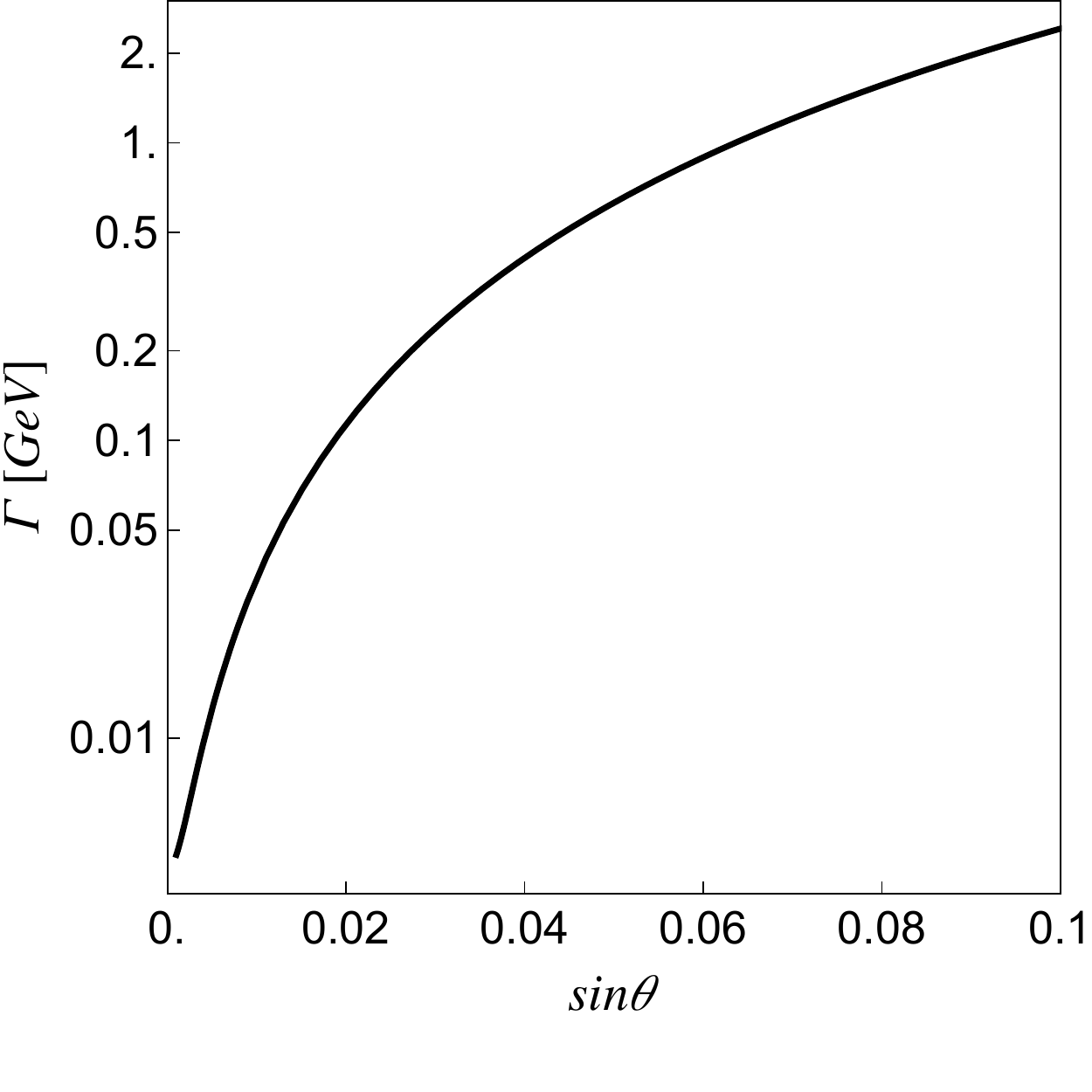}
\caption{Branching ratios (left), assuming $\Gamma =\sum_{i=gg,\gamma\gamma,WW,ZZ,t\bar{t},b\bar{b},hh} \Gamma_i$, and total width, $\Gamma$, (right) for the benchmark model with $m_X=900$ GeV, $Q_X= 8/3$, $N_X=1$, and $\lambda_{HS} = 0.02$. The Yukawa coupling $y_X$ is varied as the function of $\sin\theta$ to be compatible with the signal rate of 8 fb.  In left panel, $WW$ (solid blue), $ZZ$ (dashed blue), $gg$ (solid red), $\gamma\gamma$ (dashed red), $t\bar{t}$ (solid green). The branching ratios of other channels are not significant for the selected benchmark model and they are not shown on the plot in the left panel. Lowering the electric charge from $Q_X = 8/3$ makes $\Gamma_{gg}$ bigger than $\Gamma_{\gamma\gamma}$ in the left panel.
}
\label{fig:BR:width:WWZZ:Q85}
\end{center}
\end{figure}
A large mixing between the SM Higgs doublet and the new singlet can be phenomenologically dangerous as it changes the Higgs physics. 
In the language of the effective operators in Eq.~\ref{eq:dim5SFF}, 
the mixing induces an additional coupling of the SM Higgs to photons and gluons
\begin{equation}\label{eq:dim5SFFmixing}
\begin{split}
& \frac{e^2}{16\pi^2} \frac{c_{s\gamma\gamma}}{M} c_{\theta} H_2F_{\mu\nu}F^{\mu\nu} 
+ \frac{g_s^2}{16\pi^2}\frac{c_{sgg}}{M} c_{\theta} H_2 G_{\mu\nu}^a G^{a\,\mu\nu}  \\
& -\frac{e^2}{16\pi^2} \frac{c_{s\gamma\gamma}}{M} s_{\theta} H_1F_{\mu\nu}F^{\mu\nu} 
  - \frac{g_s^2}{16\pi^2}\frac{c_{sgg}}{M} s_{\theta} H_1 G_{\mu\nu}^aG^{a\,\mu\nu}~.
\end{split}
\end{equation}
On the other hand, the mixing introduces  a coupling 
of the heavy singlet to a pair of SM gauge bosons and fermions 
\begin{equation}\label{eq:Mixing}
\frac{1}{v}\left( c_{\theta} H_1 + s_{\theta}H_2 \right)\left( 2m_W^2W_{\mu}^+W^{-\, \mu}  +m_Z^2Z_{\mu}Z^{\mu}  - \sum_f m_f\bar{f}f \right)~,
\end{equation}
and it alters the corresponding SM Higgs couplings.
The decay channel $H_2 \to H_1\,H_1$ is kinematically allowed (since $m_{H_2} > 2m_{H_1}$) via the interaction,
\begin{equation}
-\frac{\kappa_{112}vs_{\theta}}{2}H_1^2 H_2~,
\end{equation}
where the induced coupling is given by
\begin{equation}
\kappa_{112} \equiv  \frac{2m_{H_2}^2 + m_{H_1}^2}{v^2}\left(
s_{\theta}^2 + \frac{\lambda_{HS}v^2}{m_{H_1}^2 - m_{H_2}^2}
\right)~.
\end{equation}
%
\begin{figure}[tp]
\begin{center}
\includegraphics[width=0.45\linewidth]{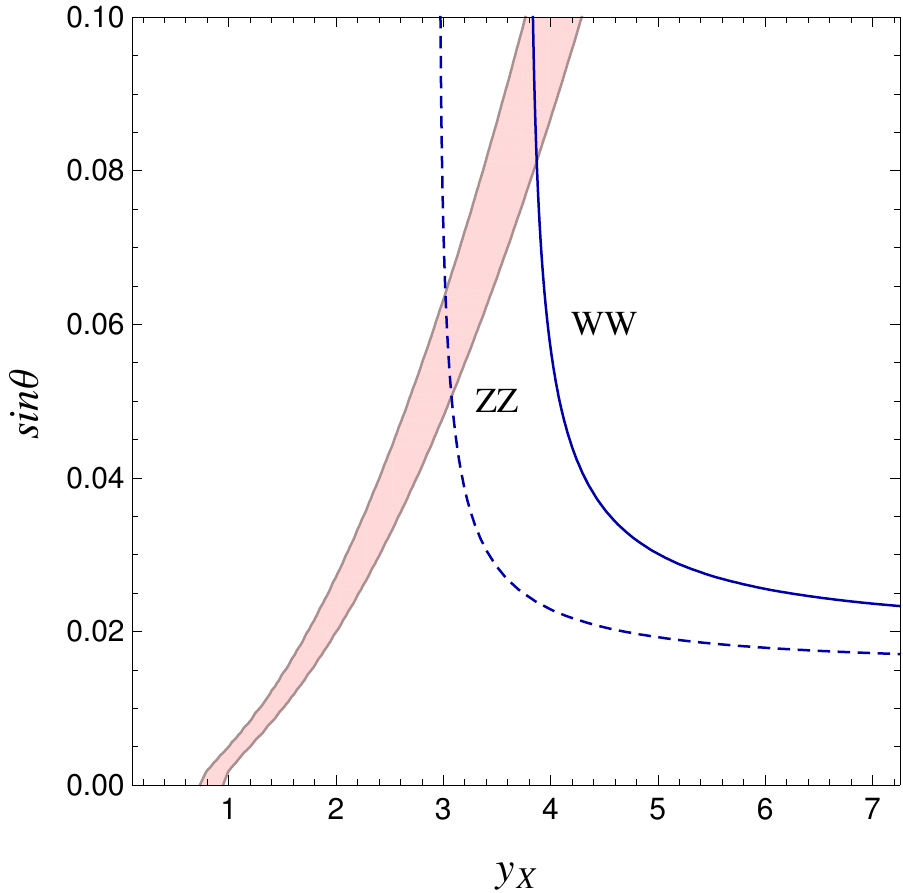}
\caption{The region that corresponds to 6$-$10 fb of $\sigma(pp\rightarrow S\rightarrow \gamma\gamma)$ for the benchmark point with $m_X=900$ GeV, $Q_X= 8/3$, $N_X=1$, and $y_X (\sin\theta = 0)=0.8$. The regions above solid blue (dashed blue) line are excluded by $WW$ ($ZZ$) channels.
}
\label{fig:WWZZ:Q85}
\end{center}
\end{figure}
Once the new resonance is linked to the SM Higgs sector via the mixing, the total width gets contributions from various decay channels, in addition to those from gluons and photons that we discussed in Section~\ref{sec:VectorLikeFrmions}. 
The relative size among various partial decay widths varies a lot over the parameter space as they have different scaling behavior. 
Fig.~\ref{fig:BR:width:WWZZ:Q85} illustrates the situation in the presence of new colored vector-like fermions with $m_X =900$ GeV, $Q_X =8/3$, $N_X = 1$, and varying Yukawa coupling $y_X = y_X(\sin\theta)$ to maintain the same signal rate of 8 fb (see the left panel in Fig.~\ref{fig:xsecyxmx:threeGamma}). 
We vary the mixing angle in Fig.~\ref{fig:BR:width:WWZZ:Q85} up to 0.1. 
The most stringent constraint on the mixing angle comes from the searches for heavy scalars in diboson decay channel~\cite{Falkowski:2015swt},
and $s_{\theta} \lesssim 0.1$ is the biggest allowed value. 
When the mixing is turned off, the total width is just $\Gamma_{gg}+\Gamma_{\gamma\gamma}$ that corresponds to the left panel of Fig.~\ref{fig:xsecyxmx:threeGamma} (very narrow width scenario).
 Turning it on, increases the contributions from $WW, ZZ, t\bar{t}$. The total decay width eventually reaches 
 $\Gamma \sim O(1 {\rm GeV})$ at around $s_\theta \sim 0.06$, the maximum value that is not excluded in $WW, ZZ$ search channels (see Fig.~\ref{fig:WWZZ:Q85})~\cite{Aad:2015kna,Khachatryan:2015cwa,Aad:2015agg}. 
 The resulting situation is exactly what we discussed in the middle panel of Fig.~\ref{fig:xsecyxmx:threeGamma} 
 where we set total width to 1 GeV. 
 The doublet-singlet model can interpolate between two cases, namely $\Gamma=\Gamma_{gg}+\Gamma_{\gamma\gamma}$ and $\Gamma =1$ GeV, via the mixing angle. 
 The Yukawa coupling, $y_X$, that produces the right signal rate increases with increasing mixing angle as in Fig.~\ref{fig:WWZZ:Q85}. It is because the signal rate scales, in presence of the mixing, roughly as $(c_\theta y_X)^n$ ($n \sim$ 2, 4 for two extreme cases in Eq.~\ref{eq:xsec:Gammaggaa} and \ref{eq:xsec:Gammaexp}). 
 The decreased $c_\theta$ is compensated by a larger $y_X$ to maintain the same signal rate (note that two boundary values, $y_X \sim$ 0.8 and 3 for the mixing angles $\sin\theta = 0$ and $\sim$0.06 in Fig.~\ref{fig:WWZZ:Q85} match those in Fig.~\ref{fig:xsecyxmx:threeGamma}).

\subsection{Vacuum (in)stability}\label{sec:Instability}
We discuss under which conditions the potential in Eq.~\ref{eq:PotentialhS} describes a consistent weakly interacting theory. The existence of a local minimum already provides a set of constraints on the couplings in the scalar potential (see Appendix~\ref{app:ScalarPot} for details),
\begin{equation}\label{eq:LocalMin}
\lambda_{HS}\mu_S^2 - 2\lambda_S\mu_H^2 >0~,~~~ \lambda_{HS}\mu_H^2 - 2\lambda_H^2\mu_S^2> 0~,~~~4\lambda_H\lambda_S - \lambda_{HS}^2 > 0~.
\end{equation}
Especially the third condition in Eq.~\ref{eq:LocalMin} restricts the relation among three dimensionless couplings, $\lambda_H>  \lambda_{HS}^2/(4\lambda_S)$. However, this constraint needs to be respected only at low scale, typically of the order of the mass scale of the scalar fields. On the contrary, imposing the positivity of the scalar potential leads to stronger constraints.
 
The potential in Eq.~\ref{eq:PotentialhS} can be written, neglecting quadratic terms, in the following form
\begin{equation}\label{eq:HighPotential}
\begin{split}
V(h,S) &\approx \frac{\lambda_H}{4} h^4 +  \frac{\lambda_{HS}}{4} h^2 S^2
+\frac{\lambda_S}{4} S^4 \\
&= \frac{1}{4}\left[
\left(
\sqrt{\lambda_H}h^2 - \sqrt{\lambda_S}S^2
\right)^2 + h^2S^2 \left(
\lambda_{HS} + 2\sqrt{\lambda_H\lambda_S}
\right)
\right]~.
\end{split}
\end{equation}
We distinguish between two cases, depending on the sign of $\lambda_{HS}$.

\begin{itemize}
\item[$\circ$] {\underline{$\lambda_{HS}>0$}}.  From the first line in eq.~(\ref{eq:HighPotential}) 
it is clear that in order to ensure the positivity of the potential the conditions
$\lambda_H(\Lambda) > 0$ and $\lambda_S(\Lambda) > 0$ must be respected all the way 
up to some high-energy scale $\Lambda$ defining the limit of validity of the theory.
If either $\lambda_H(\Lambda) < 0$ or $\lambda_S(\Lambda) < 0$ (for moderately   low scale $\Lambda$ not too far
 away from the TeV scale, i.e. the mass scale of the new particles), the model can not be considered as a consistent theory.

\item[$\circ$] {\underline{$\lambda_{HS} < 0$}}. In this case, as is clear from the second line in eq.~(\ref{eq:HighPotential}), 
the conditions $\lambda_H(\Lambda) > 0$ and $\lambda_S(\Lambda) > 0$ are not enough to ensure the positivity of the potential, and 
we need to impose $\lambda_S(\Lambda) > 0$ together with
$\lambda_H(\Lambda) > \lambda_{HS}^2(\Lambda)/4\lambda_S(\Lambda)$. 

\end{itemize} 

In addition to the vacuum stability, the condition of perturbativity requires $|\lambda_i|,|y_X| \lesssim 4\pi$ during the RG evolution.\footnote{Strictly speaking, $|y_X| \lesssim 4\pi/\sqrt{N_C}$ with $N_C = 3$ for the Yukawa coupling.}

%

\section{Peering at high scales using the Renormalization Group Equations}\label{sec:RGE}

We extrapolate the model discussed in Section~\ref{sec:singlet} at high scales using the RGEs.
In Section~\ref{sec:RGETheory} we set the ground for our discussion by introducing one-loop $\beta$ functions and matching conditions.
After a qualitative overview, in Section~\ref{eq:RGEresults} we numerically solve the RGEs focusing 
our attention on the parameter space of the model in which---as explained in Section~\ref{sec:VectorLikeFrmions}---the diphoton excess can be reproduced. 
The aim of this Section is to investigate whether a weakly coupled realization stays within the perturbative regime once the running is taken into account.

\subsection{Theoretical setup: one-loop beta functions and matching}\label{sec:RGETheory}

The $\beta$ functions for a generic coupling $g$ are defined as
\begin{equation}
\beta_g = \mu \frac{d g}{d \mu} = \frac{1}{(4 \pi)^2} \beta_g^{(1)} + \frac{1}{(4 \pi)^4} \beta^{(2)}_g + \ldots \,,
\end{equation}
where $\mu$ is the renormalization scale.
We consider the case with $N_X$ copies of vector-like fermions in the same representation.
For simplicity, we consider the Yukawa matrix $\hat{y}_X = y_X\,\mathrm{1}_{N_X\times N_X}$.
In the $\overline{{\rm MS}}$ scheme the one-loop $\beta$-functions of the gauge couplings\footnote{We use the hypercharge gauge coupling in GUT normalization $g_1^2 = 5g_Y^2/3$.
} are given by (see also~\cite{Falkowski:2015iwa,Xiao:2014kba})
\begin{equation}
\begin{split}
\beta_{g_1}^{(1)} =& \left(\frac{41}{10} + N_XQ_X^2 \frac{12}{5} \right)g_1^3~,\label{eq:RGEGauge}\\
\beta_{g_2}^{(1)} =& -\frac{19}{6}g_2^3~, \\ 
\beta_{g_3}^{(1)} =&  \left(
-7 + N_X\frac{2}{3}
\right)
g_3^3~.
\end{split}
\end{equation}
Those of the Yukawa couplings are 
\begin{equation}
\begin{split}
\beta_{y_t}^{(1)} =&\ y_t\left(
\frac{9}{2} y_t^2 - \frac{17}{20}g_1^2 - \frac{9}{4}g_2^2 - 8g_3^2
\right)~,\label{eq:RGEYukawa}\\
\beta_{y_X}^{(1)} =&\ y_X\left[
3(2N_X + 1) y_X^2 - \frac{18}{5}Q_X^2g_1^2 - 8g_3^2
\right]~,
\end{split}
\end{equation}
The one-loop $\beta$-functions for the scalar couplings in the potential are
\begin{equation}
\begin{split}
\beta_{\lambda_H}^{(1)} =&\ \left[
\lambda_{H}\left(
12 y_t^2 - \frac{9}{5}g_1^2 -9g_2^2 + 24\lambda_H
\right)
\right.\\
&-	\left. 
6y_t^4 +\frac{9}{20} g_1^2g_2^2 + \frac{27}{200}g_1^4 +\frac{9}{8}g_2^4 + \frac{1}{2}\lambda_{HS}^2
\right]~,\label{eq:RGEScalar}\\
\beta_{\lambda_{HS}}^{(1)} =&\
\lambda_{HS}\left(
6y_t^2 -\frac{9}{10}g_1^2 -\frac{9}{2}g_2^2 +12\lambda_H + 6\lambda_S + 12 N_X y_X^2 + 4 \lambda_{HS} 
\right)~,\\
\beta_{\lambda_{S}}^{(1)} =&\ 2\lambda_{HS}^2 + 24 N_X\lambda_S y_X^2 + 18\lambda_S^2 - 24 N_X y_X^4~.
\end{split}
\end{equation}
Let us now discuss these RGEs in a qualitative way. First of all, the vector-like fermions alter the running of the hypercharge gauge coupling (see Eq.~\ref{eq:RGEGauge}). They enter with a positive sign, proportional to the parametric combination $N_X Q_X^2$, thus worsening the problem of the hypercharge Landau pole in the SM. This plays an important role in particular when considering models of vector-like fermions with large $N_X$. In full generality, Eq.~\ref{eq:RGEGauge} can be solved analytically, and we find 
\begin{equation}\label{eq:HyperRunning}
g_1(\mu) = \frac{4\sqrt{5}\pi g_{1}(\mu_0)}{\sqrt{
80\pi^2 - g_{1}^2(\mu_0)(24 N_X Q_X^2 + 41)\ln(\mu/\mu_0)
}}~.
\end{equation}
\begin{figure}[!htb!]
\minipage{0.325\textwidth}
  \includegraphics[width=1.\linewidth]{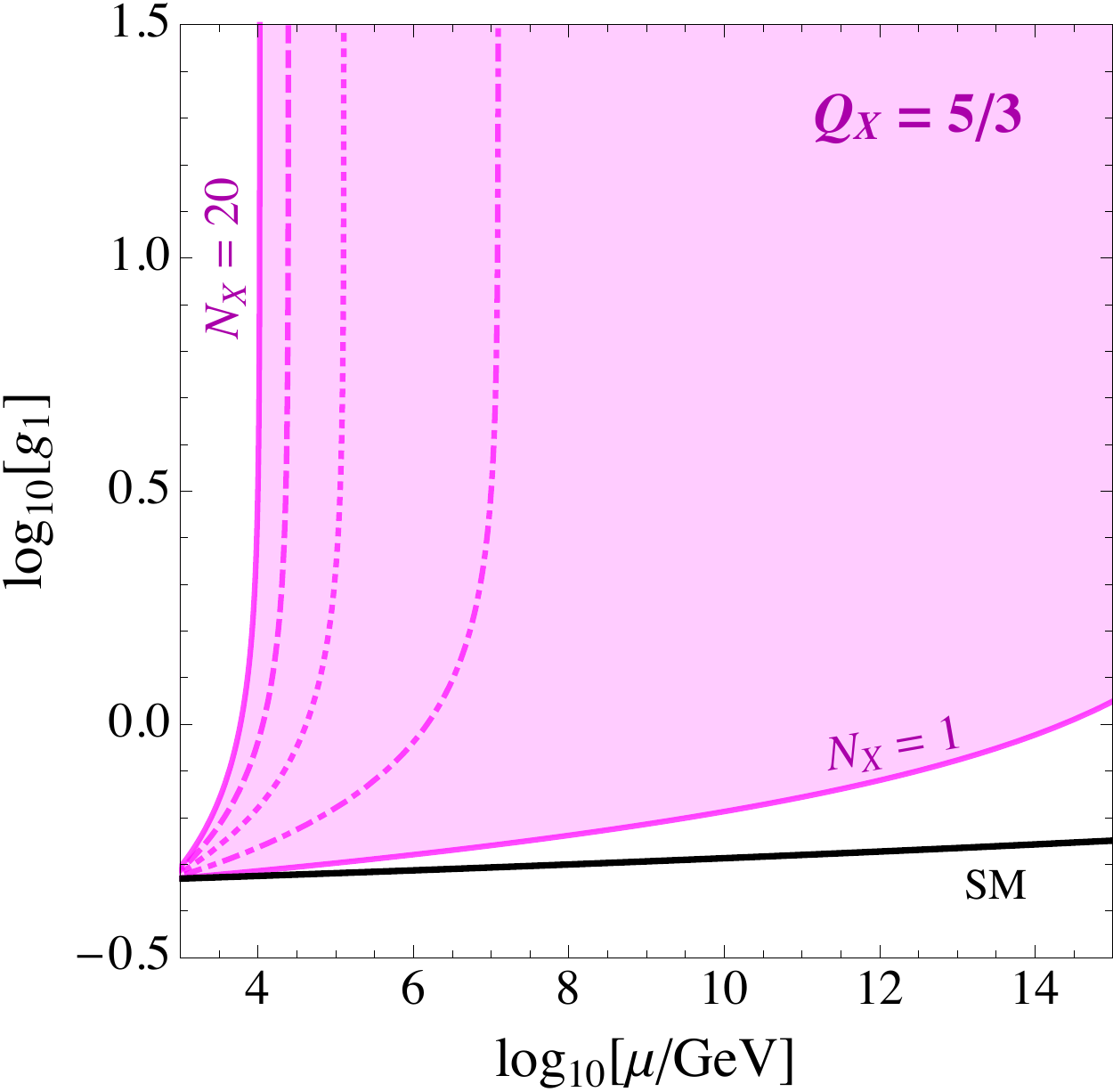}
\endminipage\hfill
\minipage{0.325\textwidth}
  \includegraphics[width=1.\linewidth]{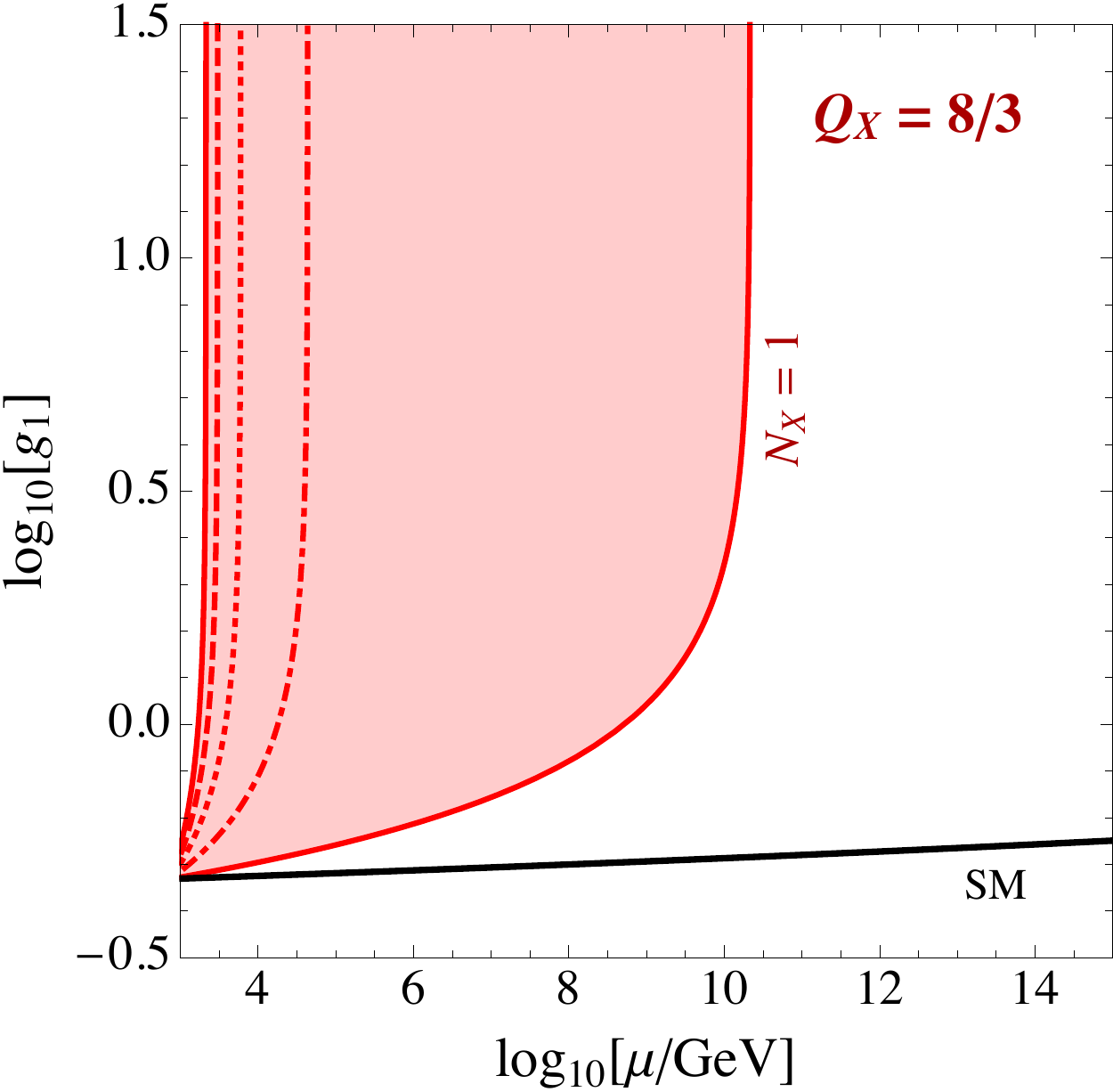}
\endminipage\hfill
\minipage{0.325\textwidth}
  \includegraphics[width=1.\linewidth]{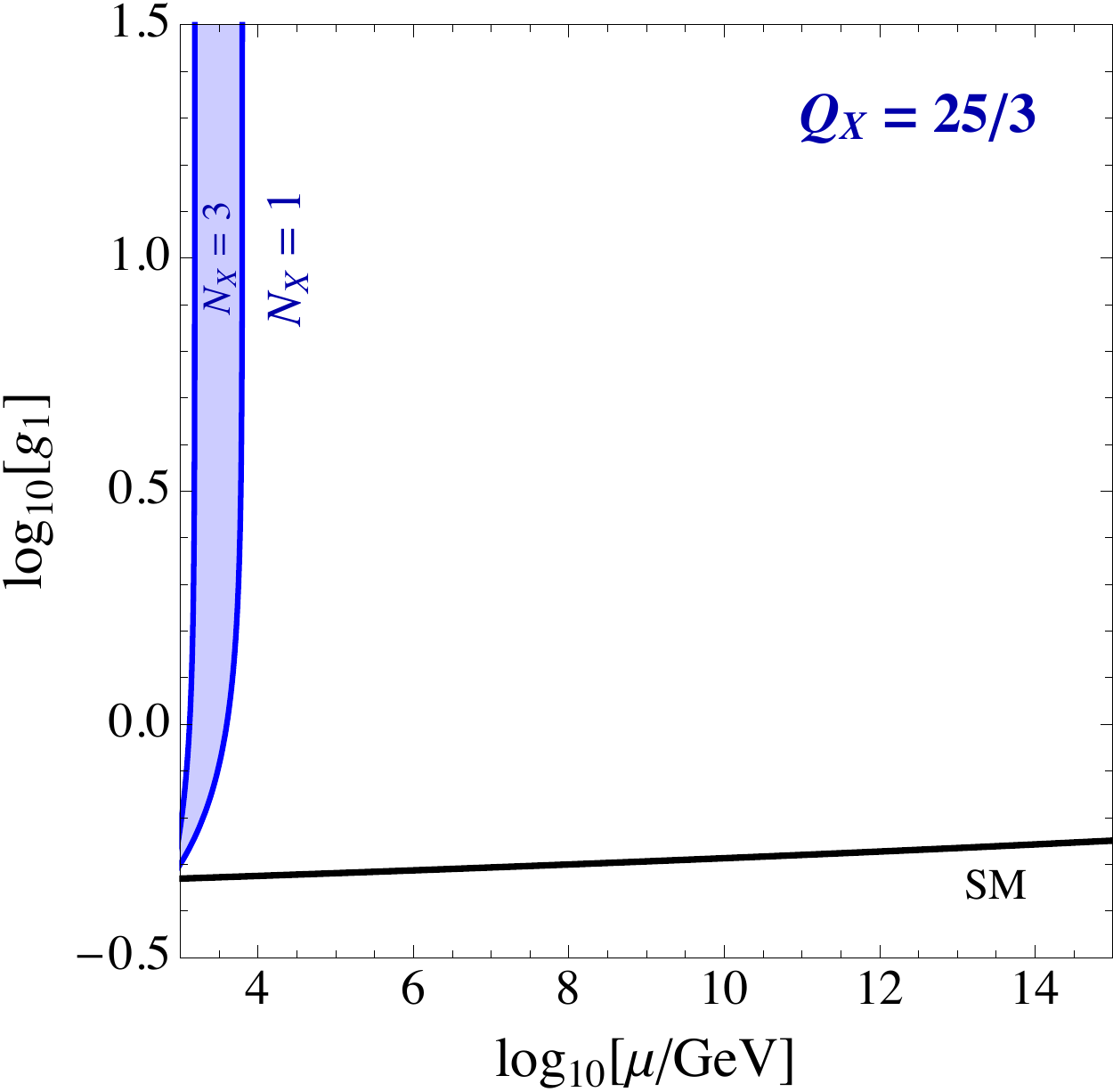}
\endminipage
 \caption{
Running of the hypercharge gauge coupling in Eq.~\ref{eq:HyperRunning} for different electric charges: $Q_X = 5/3$  (left panel),  $Q_X = 8/3$ (central panel),  $Q_X = 25/3$ (right panel). For $Q_X = 5/3$ and $Q_X = 8/3$ the shaded area covers multiplicities between $N_X = 1$ (on the far right-hand side) and  $N_X = 20$ (on the far left-hand side). Dashed, dotted and dot-dashed intermediate lines correspond, respectively, to $N_X = 15, 10, 5$. For $Q_X = 25/3$ we show $N_X = 1$ (on the far right-hand side) and $N_X = 3$ (on the far left-hand side). The solid black line corresponds to the running in the SM at one loop. }
\label{fig:HyperchargeRunning}
\end{figure}
In Fig.~\ref{fig:HyperchargeRunning} we show the running of $g_1(\mu)$ for different electric charge $Q_X = 5/3,8/3,25/3$ and different multiplicities (see caption for details). For $Q_X = 2/3$ (not shown in the plots) the impact on the hypercharge running is very limited, and only at very large multiplicities ($N_X \gtrsim 20$) the hypercharge Landau pole is pushed below $10^8$ GeV. For larger $Q_X$, the impact on the running of $g_1$ can be very significant pushing the hypercharge Landau pole---in particular for large $N_X$---towards unrealistically low scales. 
 
The second important feature is related to the running of the Yukawa coupling $y_X$ in Eq.~\ref{eq:RGEYukawa}.
The scale at which $y_X$ becomes strong indicates the limit of validity of the theory (see~\cite{Gupta:2015zzs} for similar discussion). The running of $y_X$ is dominated by two opposing effects: On the one hand, $y_X$ is pushed towards larger values by the wave-function renormalization term $\beta_{y_X}^{(1)} \propto 3(2N_X + 1)y_X^3$, on the other hand, it is pushed towards smaller values by the vertex correction $\beta_{y_X}^{(1)} \propto - [(18/5)Q_X g_1^2 + 8 g_3^2]y_X$. Understanding which effect dominates is a matter of numerical coefficients. To give some idea, for $N_X = 1$, $Q_X = 5/3$, and taking $g_3 \simeq 1$, $g_1 \simeq 0.5$
we find that the wave function renormalization term starts dominating if $y_X \gtrsim 1.2$.
From this value on, $y_X$ increases following its RG evolution.
Notice that, as shown in the left panel of Fig.~\ref{fig:xsecyxmx:threeGamma}, for $Q_X = 5/3$ and $N_X = 1$ 
the Yukawa coupling needed to explain the observed excess falls exactly in the ballpark estimated above.
This very simple argument tells us that the running of $y_X$ is controlled by a delicate numerical 
interplay between two terms of opposite sign whose net effect depends on the specific choice of the parameters $N_X$ and $Q_X$. 

Finally, let us comment about the running of $\lambda_S$, Eq.~\ref{eq:RGEScalar}. As discussed in Section~\ref{sec:Instability}, negative values of $\lambda_S$ indicates an instability in the scalar potential.
The running of $\lambda_S$ is, again, the consequence of two counterbalancing effects: 
There is a positive contribution, $\beta_{\lambda_S}^{(1)} \propto 2\lambda_{HS}^2+24 N_X \lambda_S y_X^2 + 18\lambda_S^2$, and a negative term $\beta_{\lambda_S}^{(1)} \propto -24N_X y_X^4$. The positive contribution dominates if
\begin{equation}\label{eq:LambdaTh}
\lambda_S > \lambda_S^{\rm th}~,~~~~\lambda_S^{\rm th} = \frac{1}{3}\left(
\sqrt{-\lambda_{HS}^2 + 4 N_X^2 y_X^4 + 12 N_X y_X^4} - 2N_X y_X^2
\right)~.
\end{equation}
For $N_X = 1$, $y_X = 1$, $\lambda_{HS} = 0$ we have $\lambda_S^{\rm th} = 2/3$. Increasing $y_X$, the threshold value of $\lambda_S$ increases too, and for instance we have $\lambda_S^{\rm th} = 6$ if $y_X =3$. If $\lambda_S > \lambda_S^{\rm th}$ the problem of negative $\lambda_S$ is avoided. However, the perturbativity bound sets an important upper limit on the largest values allowed. Notice that Eq.~\ref{eq:LambdaTh} is a rough estimate that was obtained ignoring the $\mu$-dependence; in Section~\ref{eq:RGEresults} we will return to this point from a more quantitative perspective.

\begin{figure}[!htb!]
\begin{center}
  \includegraphics[width=.6\linewidth]{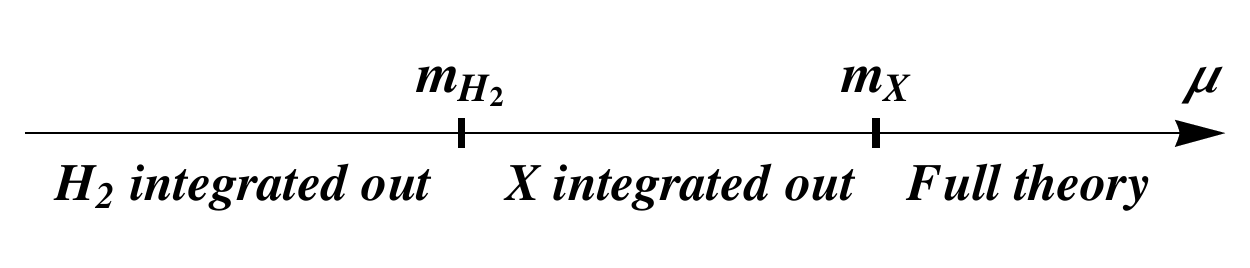}
\end{center}
\caption{Schematic representation of the RG running. Above the mass scale $\mu > m_X$ we use the full RGEs in Eqs.~\ref{eq:RGEGauge},\ref{eq:RGEYukawa},\ref{eq:RGEScalar}. For $m_{H_2} \leqslant \mu \leqslant m_X$ we have an effective field theory in which the vector-like quark is integrated out. Finally, for $\mu < m_{H_2}$, the heavy scalar is integrated out and we are left with the standard model field content. Notice that the choice $m_{H_2} \leqslant m_X$ adopted in our analysis is mere convention, and if $m_X \leqslant m_{H_2}$ everything remains unchanged once the proper hierarchy of masses is taken into account in the running.
}
\label{fig:residuals}
\end{figure}

Before proceeding, let us close this section with a few technical remarks related to the solution of Eqs.~\ref{eq:RGEGauge},\ref{eq:RGEYukawa},\ref{eq:RGEScalar}. The RGEs in Eqs.~\ref{eq:RGEGauge},\ref{eq:RGEYukawa},\ref{eq:RGEScalar} are valid only for $\mu > m_X > m_{H_2}$. The situation is sketched in Fig.~\ref{fig:residuals}.
Running down using the RG flow, heavy fields are integrated out and their contributions disappear from the $\beta$ functions. At each threshold the matching procedure between the full theory above the threshold and the effective theory below produces some threshold corrections. In our setup, this is true both for the vector-like fermion and the heavy scalar. As far as the vector-like fermion is concerned, the threshold corrections generated at $\mu = m_X$ can be computed as follows~\cite{Casas:1999cd,Rose:2015fua}. First, the contribution of $X$ to the effective potential $V(h,S)$ is
\begin{equation}
\Delta V_{\rm eff}^X(h,S,\mu) = -\frac{3}{16\pi^2} m_X^4(s) \left[
\log\frac{m_X^2(s)}{\mu^2} - \frac{3}{2}
\right]~,
\end{equation}
where $m_X(s) = y_X s + m_X^{(B)}$ is the vector-like quark mass as a function of the background field $s$.\footnote{Here $m_X^{(B)}$ is the bare mass of the vector-like fermion in the original Lagrangian 
$\mathcal{L}_X = -y_X S\overline{X}X - m_X^{(B)}\overline{X}X$. As discussed in Appendix~\ref{app:ScalarPot}, in fact, we are considering the most general situation in which $S$ takes a VEV.}
At $\mu = m_X$ the 
threshold $\Delta V_{\rm th}^X(h,S) = \Delta V_{\rm eff}^X(h,S,\mu = m_X)$ is generated. 
The corresponding contribution to the quartic $\lambda_S/4\, S^4$ in Eq.~\ref{eq:PotentialhS} is given by
\begin{equation}\label{eq:Threshold}
\Delta \lambda_S = \frac{1}{6}\left.
\frac{\partial^4\Delta V_{\rm th}^X(h,S)}{\partial s^4}
\right|_{s=0}~.
\end{equation}
At $\mu = m_{H_2}$ a threshold correction for the quartic coupling $(\lambda_H/4)\, h^4$ is generated~\cite{EliasMiro:2012ay}.
This threshold correction corresponds to a tree-level shift of the Higgs quartic coupling $\lambda_H$, and its value can be extracted from Eq.~\ref{app:eq:LargeVev}
\begin{equation}
\Delta\lambda_H = -\frac{\lambda_{HS}^2}{2\lambda_S}~.
\end{equation}

The solution of 
the RGEs requires suitable matching conditions in order to relate the running $\overline{{\rm MS}}$ parameters 
with on-shell observables. Since we are working with one-loop $\beta$ functions, we need to impose only tree level matching conditions.
\begin{figure}[!htb!]
\begin{center}
\fbox{\footnotesize $\lambda_{HS} = 0.01$, $s_{\theta} = 0.01$, $y_X = 0.75$, $m_X = 1.5$ TeV, $Q_X = 2/3$, $N_X = 1$}
\end{center}
\minipage{0.325\textwidth}
  \includegraphics[width=1.\linewidth]{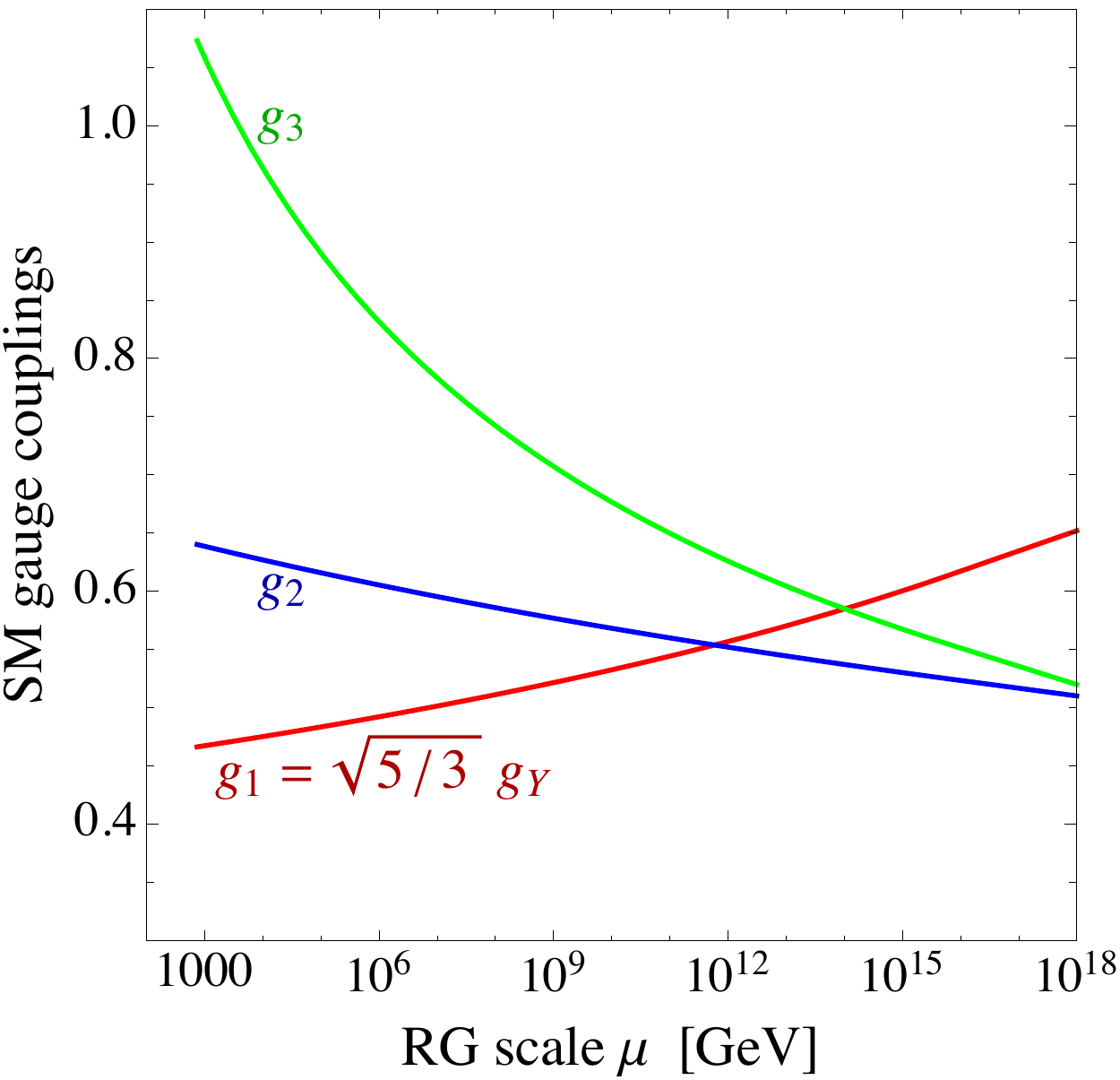}
\endminipage\hfill
\minipage{0.325\textwidth}
  \includegraphics[width=1.\linewidth]{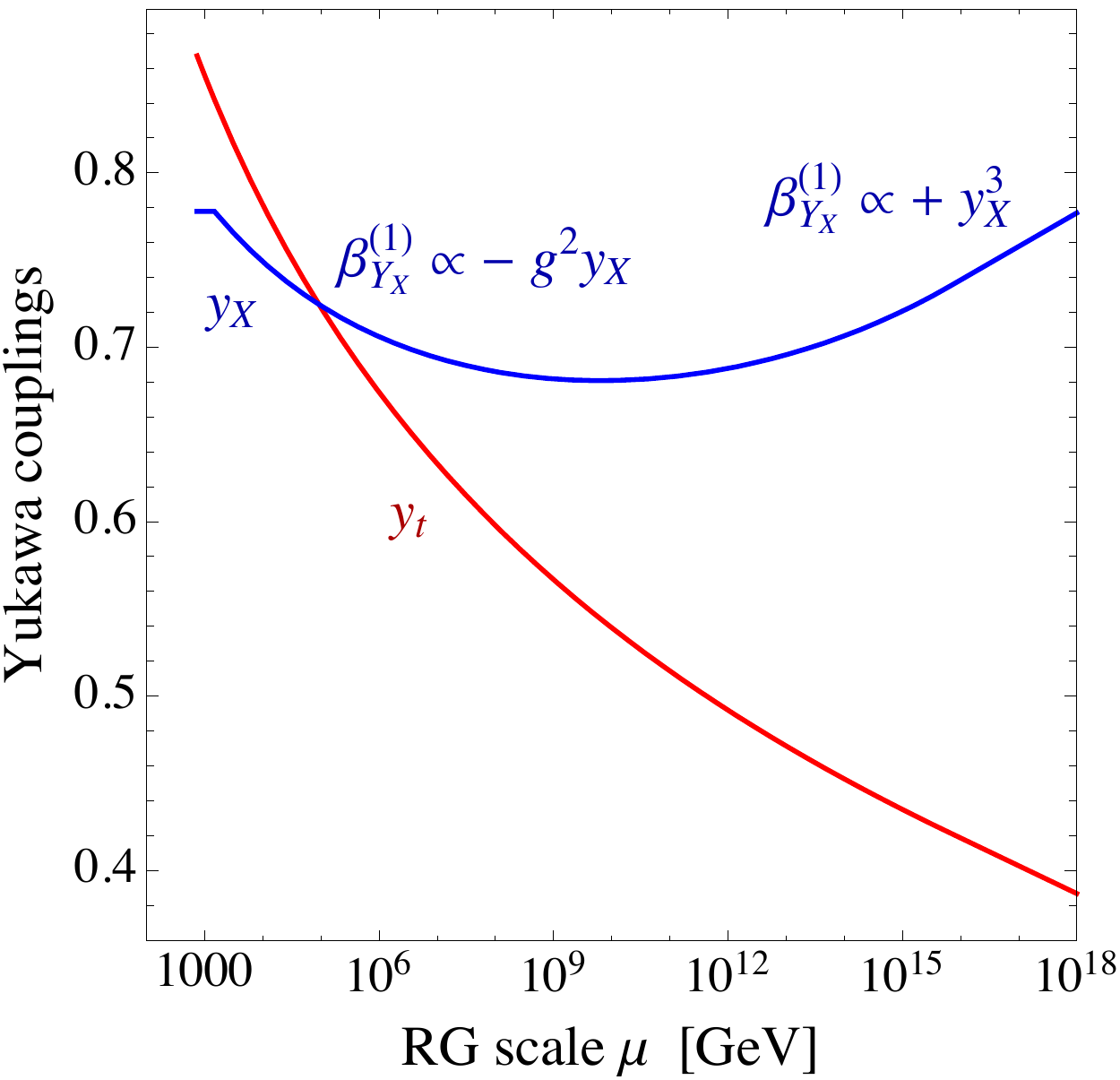}
\endminipage\hfill
\minipage{0.345\textwidth}
  \includegraphics[width=1.\linewidth]{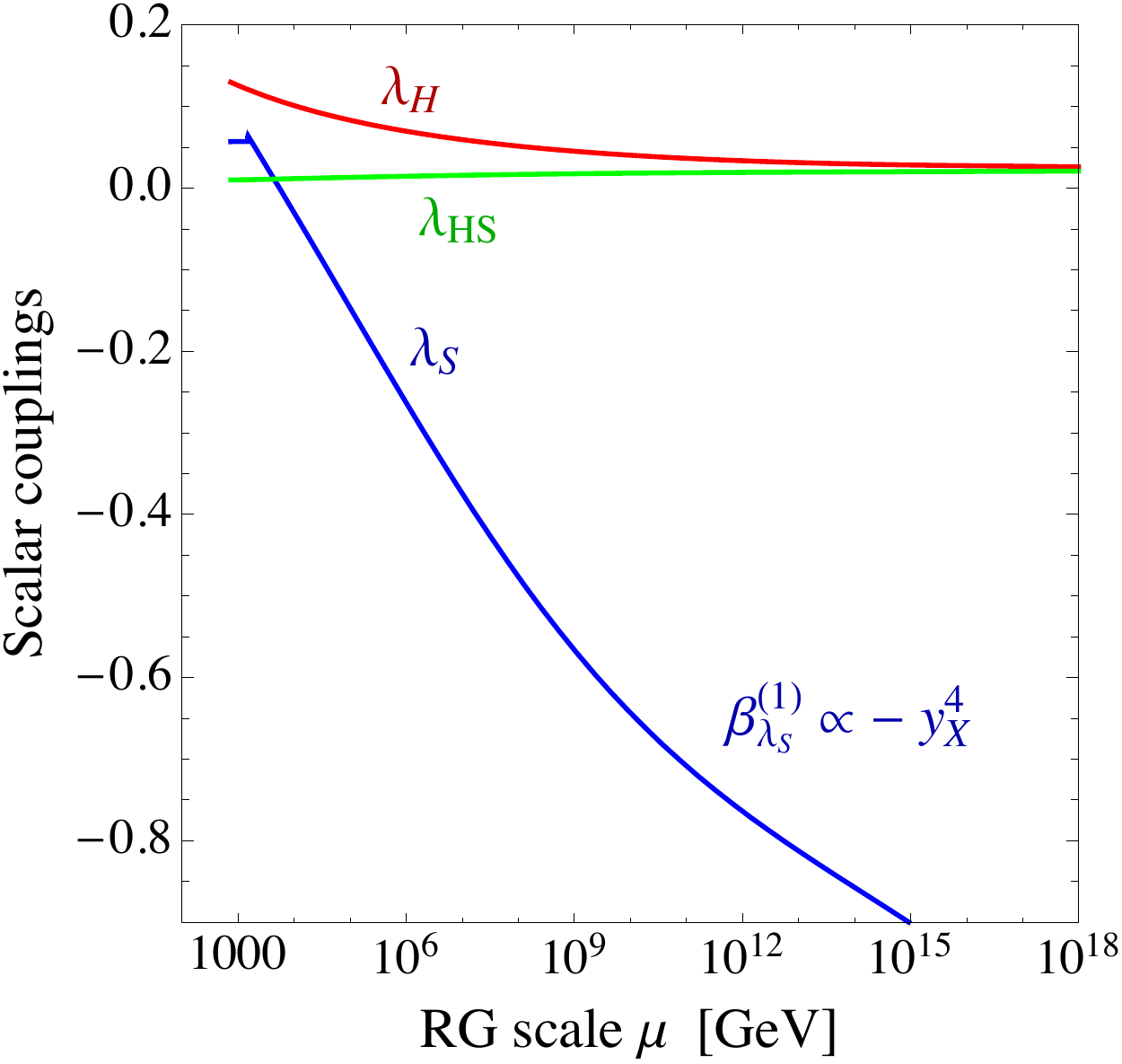}
\endminipage
 \caption{
 RG evolution. We show the gauge couplings (left panel), Yukawa couplings (central panel) and scalar couplings (right panel).
  }
 \label{fig:RGPlot}
\end{figure}
The potential has $5$ parameters, $(\lambda_H, \lambda_S, \lambda_{HS}, v, u)$ that are related to the physical parameters $(m_{H_1}^2,m_{H_2}^2,s_{\theta},v,\lambda_{HS})$. As already stated before, $m_{H_1} = 125.09$ GeV is the Higgs boson mass while, following the  discussion in Section~\ref{sec:NewResonance}, $m_{H_2} = 750$ GeV is the mass of the new putative scalar resonance; $s_{\theta}$ describes the mixing in the scalar sector, and its value is an external parameter controlled by the fit in  Section~\ref{sec:NewResonance}. Furthermore, $v = 246$ GeV since it enters in the definition of the gauge boson masses. We treat $\lambda_{HS}$ as an external free parameter. In order to relate the internal parameters $(\lambda_H, \lambda_S, \lambda_{HS}, v, u)$ to the observable ones $(m_{H_1}^2,m_{H_2}^2,s_{\theta},v,\lambda_{HS})$, we work out explicit expressions for $\lambda_H$, $\lambda_S$ and $u$. They are 
\begin{eqnarray}
\lambda_H &=& \frac{m_{H_1}^2}{2v^2} + \frac{s_{\theta}^2(m_{H_2}^2 - m_{H_1}^2)}{2v^2}~,\\
\lambda_S &=& \frac{2\lambda_{HS}^2 v^2}{s_{2\theta}^2(m_{H_2}^2 - m_{H_1}^2)}
\left[
\frac{m_{H_2}^2}{(m_{H_2}^2 - m_{H_1}^2)} - s_{\theta}^2
\right]~,\label{eq:LambdaS}\\
u &=& \frac{s_{\theta}c_{\theta}(m_{H_2}^2 - m_{H_1}^2)}{\lambda_{HS}v}~.
\end{eqnarray}
As far as the other SM parameters---namely $g_1$, $g_2$, $g_3$, $y_t$---are concerned, at the electroweak scale they can be related to the $W$ and $Z$ boson pole masses, the top quark pole mass, and the $\overline{{\rm MS}}$ QCD structure constant at the $Z$ pole~\cite{Buttazzo:2013uya}.

\subsection{Phenomenological analysis: on the importance of running couplings}
\label{eq:RGEresults}
Let us now discuss our results.
For simplicity, the starting point in the RG running is chosen at the scale, $\mu_0 \equiv m_{H_2} = 750$ GeV. Only the threshold in Eq.~\ref{eq:Threshold}, therefore, is included in our analysis. We use the following initial values $g_Y(\mu_0) = 0.361$, $g_2(\mu_0) = 0.640$, $g_3(\mu_0) = 1.073$, $y_t(\mu_0) = 0.867$ at $\mu = \mu_0$. 
For illustrative purposes, let us start our discussion from the benchmark values $\lambda_{HS} = 0.01$, 
 $s_{\theta} = 0.01$, $y_X = 0.75$, $m_{X} = 1.5$ TeV, $N_X = 1$, $Q_X = 2/3$. 
Notice that, using Eq.~\ref{eq:LambdaS}, we have in this case $\lambda_S \simeq 0.06$. This benchmark point is far from the values of $Q_X$, $m_X$, and $y_X$ singled out in Fig.~\ref{fig:xsecyxmx:threeGamma} as good candidates for explaining the diphoton excess. However, we believe that this choice provides a good starting point to illustrate, on the quantitative level, the properties of the RGEs outlined qualitatively in Section~\ref{sec:RGETheory}. Our results are shown in Fig.~\ref{fig:RGPlot} for the running of the gauge couplings, the Yukawa couplings, and the couplings in the scalar potential. Three observations can be made.

\begin{itemize}

\item[${\it i)}$] The presence of the vector-like fermions modify the running of $U(1)_Y$ and $SU(3)_C$ gauge couplings, as was expected due to their quantum numbers, namely $(\textbf{1},\textbf{3})_{2/3}$ under the SM gauge group. They give an additional positive contribution both to the running of $g_1$ and $g_3$. However, our selected values, $N_X = 1$ and $Q_X = 2/3$, do not cause any evident problem, and the running of $g_1$ increases with the renormalization group scale with the rate similar to the SM one (see also Fig.~\ref{fig:HyperchargeRunning}).

\item[${\it ii)}$]  
The Yukawa coupling $y_X$ is frozen at the input value $y_X = 0.75$ below $\mu = m_X$, 
and it starts running above $\mu = m_X$. The running is driven by two distinct contributions. At low scale, the dominant contribution to the one-loop $\beta$ function comes from the term with the QCD coupling, $\beta_{y_X}^{(1)} \propto -g_3^2 y_X$.
It has a negative sign, and it pushes $y_X$ towards smaller values. As $\mu$ increases, $g_3$ gets smaller (see left panel in Fig.~\ref{fig:RGPlot}), and the dominant contribution to the one-loop $\beta$ function becomes $\beta_{y_X}^{(1)} \propto + 9 y_X^3$. From this point on, $y_X$ increases and it eventually violates the perturbative bound. However, notice that---at least for the specific values chosen in Fig.~\ref{fig:RGPlot}---the running of $y_X$ at high RG scale  values is not dramatically fast, and $y_X$ stays within the validity of perturbation theory all the way up to the Planck scale.

\item[${\it iii)}$] The running of the scalar couplings reveals a pathology in the model. As soon as the Yukawa coupling $y_X$ enters in the RG running, it easily dominates the running of the scalar coupling $\lambda_S$ via the term, $\beta_{\lambda_S}^{(1)} \propto -24 y_X^4$. As a consequence, $\lambda_S$ is dragged towards negative values already at low scale (not far away $m_X$). As explained in Section~\ref{sec:Instability}, the condition $\lambda_S < 0$ generates a dangerous run-away direction in the scalar potential.

\end{itemize} 

The numerical example analyzed in Fig.~\ref{fig:RGPlot} shows the presence of a dangerous pathology in the model. The presence 
of the Yukawa coupling between the vector-like fermion $X$ and the scalar field $S$ generates an instability in the scalar potential of the theory already at a low scale. The problem is already evident in Fig.~\ref{fig:RGPlot}  even if we decided to work, for illustrative purposes, with a moderately small Yukawa coupling ($y_X = 0.75$). A larger Yukawa $y_X$ will exacerbate the problem further. Notice that the peculiar behavior of $\lambda_S$ highlighted in this example agrees nicely with the discussion in Section~\ref{sec:RGETheory}: The initial value of $\lambda_S$ (that is $\lambda_S = 0.06$) is much smaller than the estimated threshold value by Eq.~\ref{eq:LambdaTh} ($\lambda_{S}^{\rm th} \simeq 0.3$ in this case). Starting from larger values of $\lambda_S$ would reverse the situation, pushing $\lambda_S$ towards increasing positive values.
\begin{figure}[!htb!]
\minipage{0.5\textwidth}
  \includegraphics[width= .88 \linewidth]{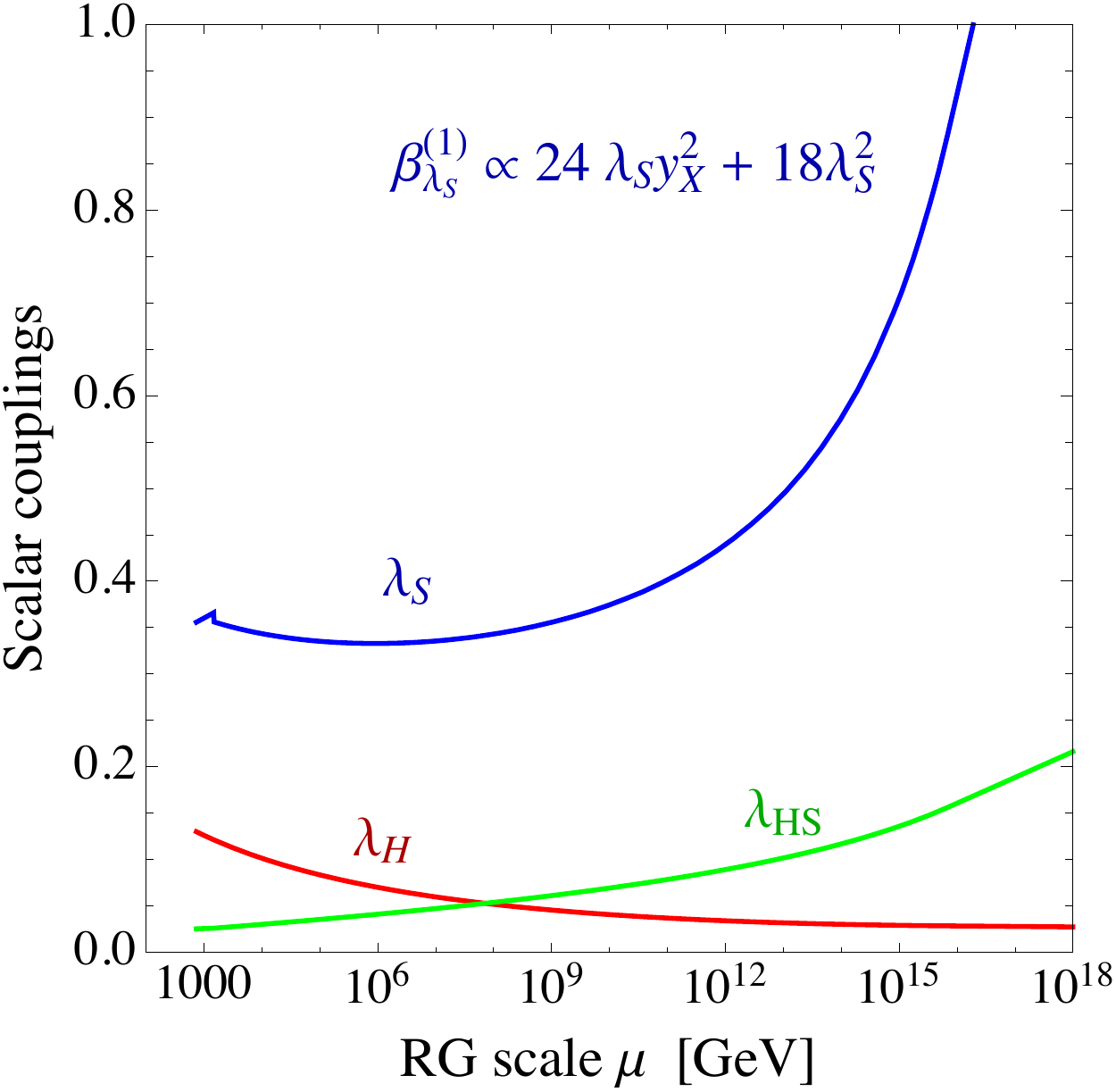}
\endminipage\hfill
\minipage{0.5\textwidth}
  \includegraphics[width= .91 \linewidth]{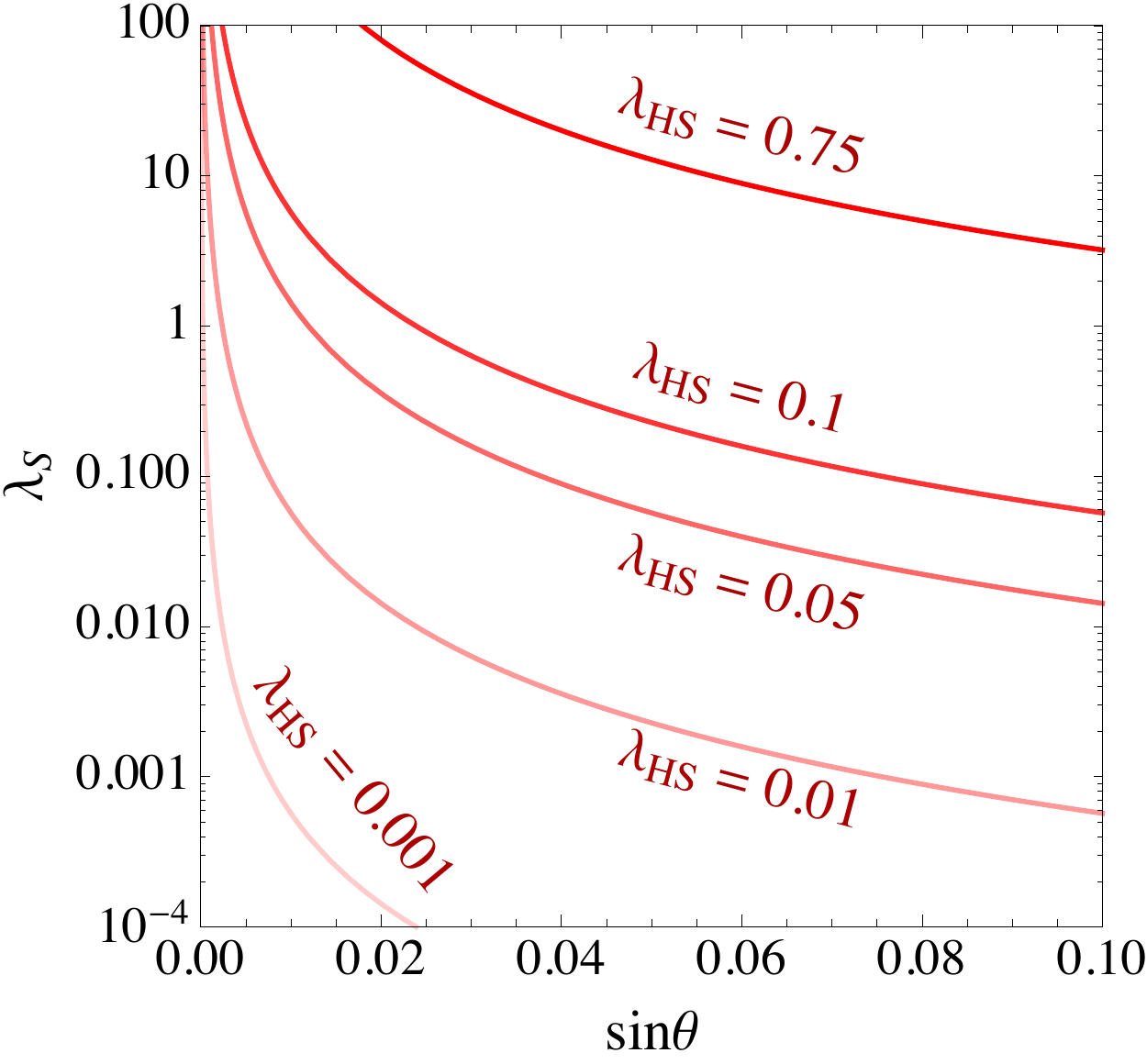}
\endminipage\\
\caption{ 
Left panel. Running of the scalar couplings with $\lambda_S(\mu_0) \simeq 0.35$. Right panel. 
Mixing angle dependence of $\lambda_S$, as described by Eq.~\ref{eq:LambdaS}.
}
\label{fig:RunningLambda}
\end{figure}
This is illustrated in Fig.~\ref{fig:RunningLambda}, left panel, where we started from $\lambda_S \simeq 0.35$.
As shown in the plot, $\lambda_S$ is pushed large and positive by the positive terms in its one-loop $\beta$ function.

In presense of the mixing, the value of $\lambda_S$ is fixed by two internal parameters, $\lambda_{HS}$ and $s_\theta$ (see Eq.~\ref{eq:LambdaS} and related discussion in Section~\ref{sec:RGETheory}). In the right panel of Fig.~\ref{fig:RunningLambda}, we show the value of $\lambda_S$ as a function of the mixing angle for different values of $\lambda_{HS}$. As is evident from this plot, a large range of values for both $\lambda_S$ and $\lambda_{HS}$ is allowed. For instance if $s_{\theta} = 0.01$, we have $\lambda_S \simeq 5.7$ for $\lambda_{HS} = 0.1$, and $\lambda_S \simeq 5.7 \times 10^{-4}$ for $\lambda_{HS} = 10^{-3}$. At larger mixing angles, larger values for $\lambda_{HS}$ are allowed.\footnote{We checked that all the values in the right panel of Fig.~\ref{fig:RunningLambda} satisfy the local minimum condition in Eq.~\ref{eq:LocalMin}.} For instance, considering $s_{\theta} = 0.1$, we have $\lambda_S \simeq 3.2$ for $\lambda_{HS} = 0.75$, and  $\lambda_S \simeq 5.7 \times 10^{-4}$ for $\lambda_{HS} = 10^{-2}$. For simplicity, we will ignore the mixing between the Higgs and the singlet scalar in the discussion of the next Section
\footnote{The limit with no mixing in Eq.~\ref{eq:LambdaS} can be understood by noticing that $s_{\theta}$ is not a free parameter, namely $s_{2\theta} = 2\lambda_{HS}uv/(m_{H_2}^2- m_{H_1}^2)$ (see Appendix~\ref{app:ScalarPot}). By taking $\lambda_{HS} \to 0$, we have $\lambda_S \to m_{H_2}^2/2u^2$ where $u$ is the vacuum expectation value of the singlet $S$, or $u\equiv \langle S\rangle$.}. 
In Appendix~\ref{app:MixingAngle}, we comment about the generalization of our result to the case with non-zero mixing.


Equipped with these results, we can now move to discuss few cases numerically more similar to the ones  suggested by the fit outlined 
in Section~\ref{sec:NewResonance}.

\subsubsection{$m_X = 900$ GeV, $N_X = 1$, $Q_X = 5/3$, $y_X = 5$.}\label{sec:BechmarkCase}

According to the central panel in Fig.~\ref{fig:xsecyxmx:threeGamma}, this case provides a good fit of the diphoton excess, assuming $\Gamma = 1$ GeV. Notice that these values are very realistic, since they correspond to a hypothetical top partner quark not yet ruled out by direct searches.
Without a mixing, $\lambda_S$ is a free parameter, and we vary its initial value  in the interval $\lambda_S(\mu_0) \in [10^{-3}, 5]$.\footnote{Notice that in the limit $\lambda_{HS} \to 0$ 
the one-loop $\beta$ function 
for the Higgs quartic coupling does not receive additional contributions. 
On the quantitative level,  
the only effect on the running of $\lambda_H$ is indirectly induced by the different running of $g_1$ and $g_3$, 
and therefore it does not change much if compared with the running in the SM~\cite{Buttazzo:2013uya}. In our analysis we mainly focus on the running
of $\lambda_S$ and $y_X$, since they are the most sensitive parameters to the RG evolution.}
\begin{figure}[!htb!]
\begin{center}
\fbox{\footnotesize $y_X = 5$, $m_X = 900$ GeV, $Q_X = 5/3$, $N_X = 1$ ($\lambda_{HS} = 0$)}
\end{center}
\minipage{0.5\textwidth}
  \includegraphics[width= .92 \linewidth]{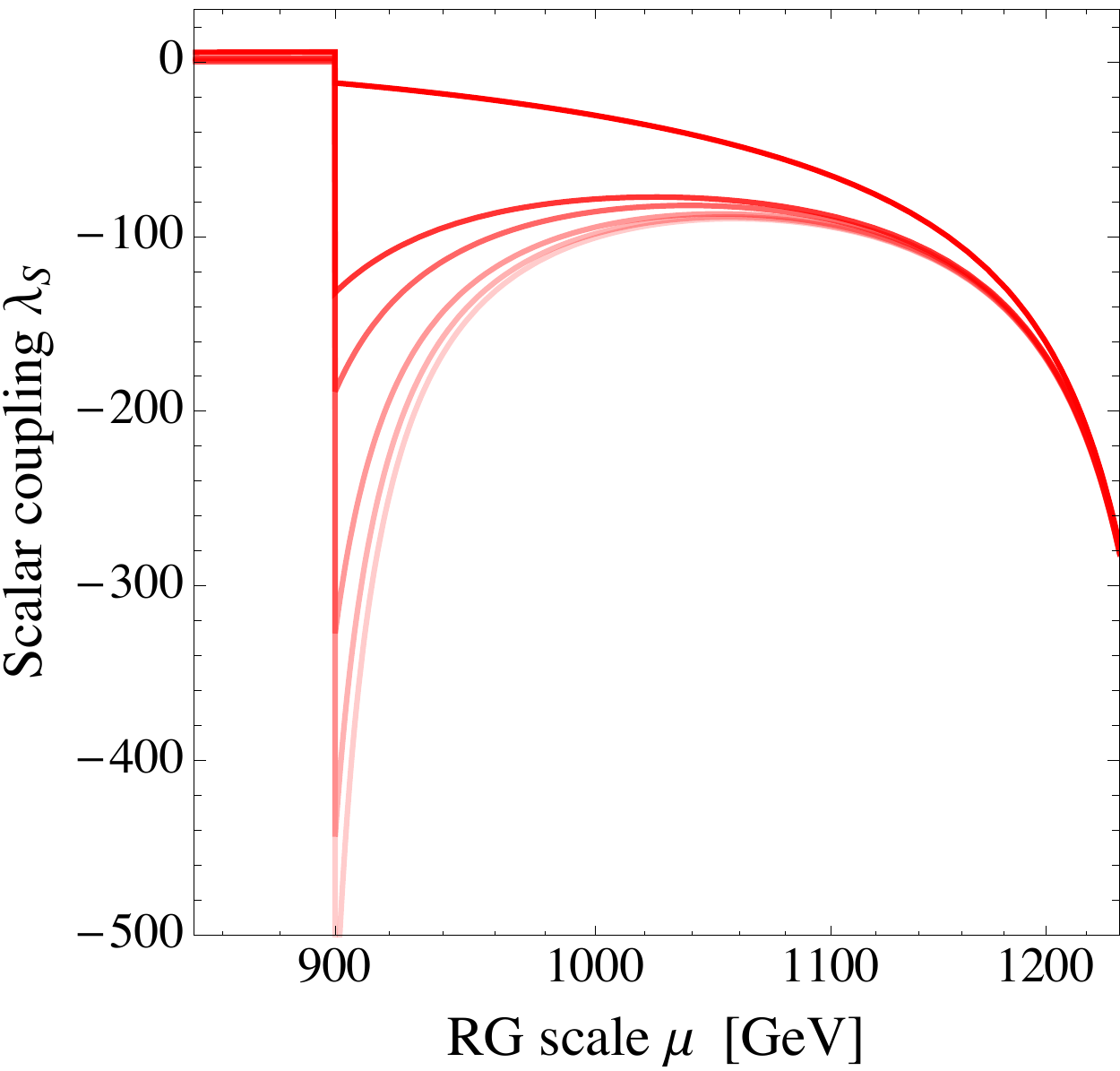}
\endminipage\hfill
\minipage{0.5\textwidth}
  \includegraphics[width= .86 \linewidth]{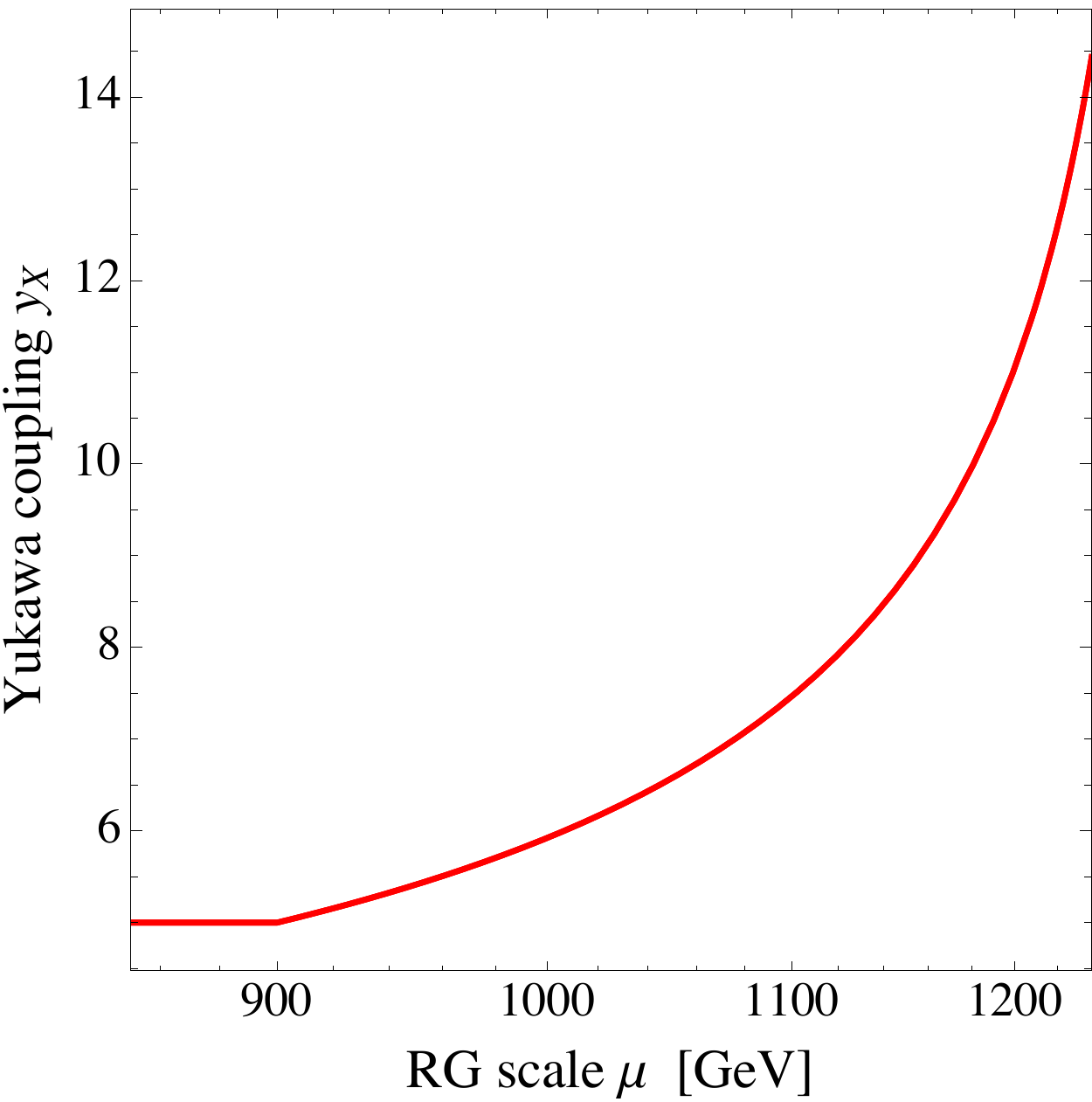}
\endminipage\\
\caption{ 
Left panel. Running of $\lambda_S$ in the model with $y_X = 5$, $m_X = 900$ GeV, $Q_X = 5/3$, $N_X = 1$.
From lighter to darker, the running corresponds to the initial values $\lambda_S(\mu_0) = 10^{-3}, 10^{-2}, 10^{-1}, 1,2, 5$.
Right panel. Running of $y_X$.
}
\label{fig:RunningLambdaSPositive}
\end{figure}
We show our results in Fig.~\ref{fig:RunningLambdaSPositive}, and we focus on the running of $\lambda_S$ (left panel) and $y_X$ (right panel).
The impact on the hypercharge gauge coupling is very limited, see Fig.~\ref{fig:HyperchargeRunning}.
As is clear from the plot, the very large initial value of the Yukawa coupling $y_X$ has dramatic effect on the running.
As far as $\lambda_S$ is concerned, the threshold correction, given by Eq.~\ref{eq:Threshold}, is extremely large (being proportional to $y_X^4$), and the contribution $\beta_{\lambda_S}^{(1)} \propto -24\, y_X^4$ largely dominates. As a result, $\lambda_S$ always rapidly runs towards negative values, thus generating an instability in the scalar potential. Notice that a large initial value of $\lambda_S$ will be difficult to fix the problem. In the running of $y_X$, the term  $\beta_{y_X}^{(1)} \propto 9\, y_X^3$ dominates, and $y_X$ eventually violates perturbativity at the TeV scale. We therefore conclude that this case is unrealistic as a candidate for a weakly coupled model due to the large Yukawa $y_X$.

However, as explained in Section~\ref{sec:VectorLikeFrmions}, thanks to a degeneracy in the scaling of the diphoton signal rate
it is possible to alleviate the problem of large $y_X$ in different ways as long as the combination $ y_X N_X Q_X \sim 25/3$ is kept fixed.
For fixed $Q_X$, it is possible to decrease the value of the Yukawa coupling 
by changing the multiplicity of the vector-like fermions $N_X$.
For fixed $N_X$, it is possible to decrease the value of the Yukawa coupling 
by changing the electric charge of the vector-like fermions $Q_X$.
Finally, both $N_X$ and $Q_X$ can be changed.
We discuss now these possibilities from the perspective of the RGEs.

\subsubsection{$m_X = 900$ GeV, $N_X = 5$, $Q_X = 5/3$, $y_X = 1$: $y_X N_X Q_X = 25/3$.}

In this scenario, a small Yukawa coupling  $y_X = 1$ is obtained by increasing the value of $N_X$.
We take $N_X = 5$ while $Q_X = 5/3$ is fixed as the case in~\ref{sec:BechmarkCase}.
\begin{figure}[!htb!]
\begin{center}
\fbox{\footnotesize  $m_X = 900$ GeV, $Q_X = 5/3$, $N_X = 5$, $y_X = 1$ ($\lambda_{HS} = 0$)}
\end{center}
\minipage{0.5\textwidth}
  \includegraphics[width= .84 \linewidth]{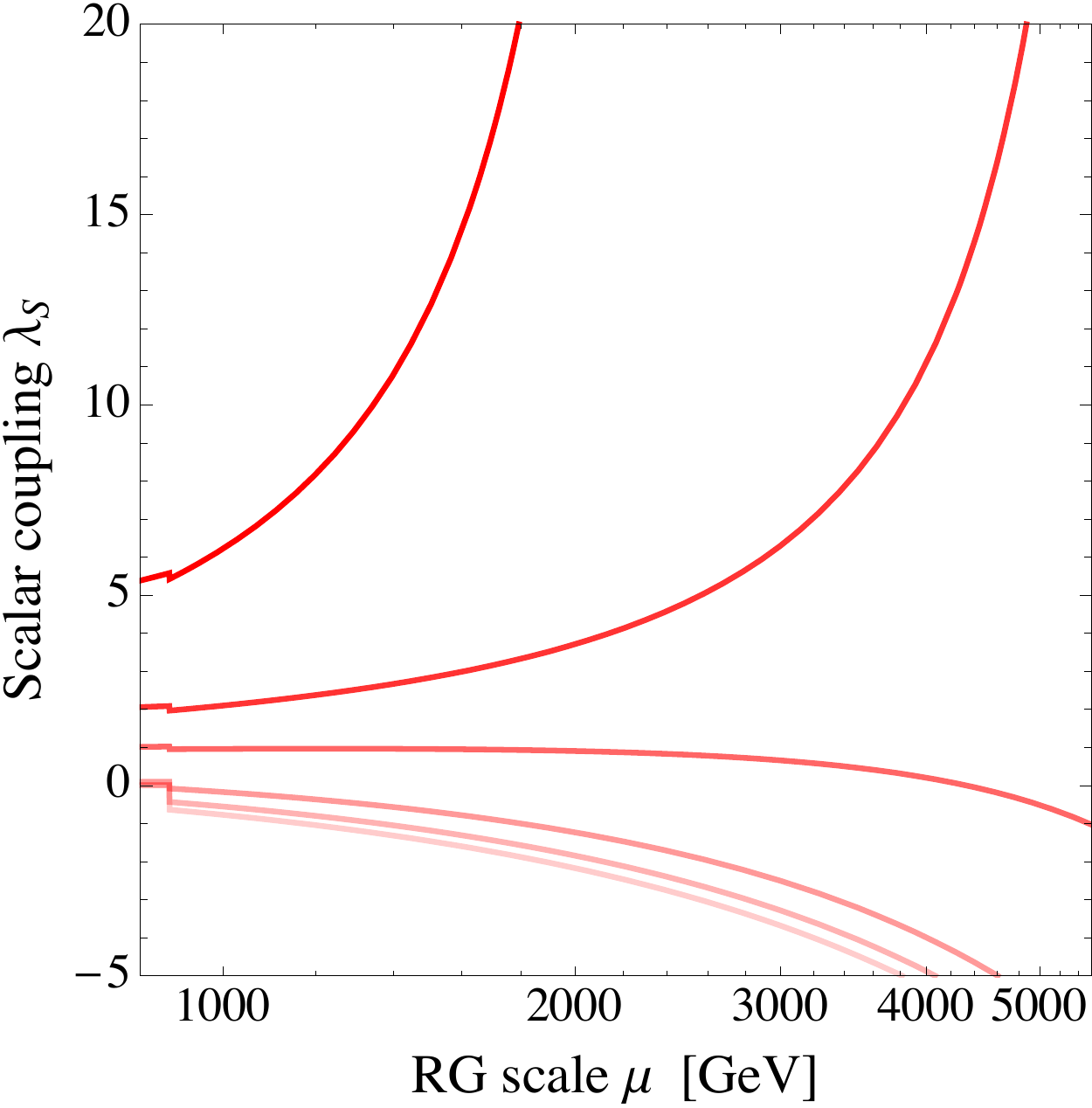}
\endminipage\hfill
\minipage{0.5\textwidth}
  \includegraphics[width= .86 \linewidth]{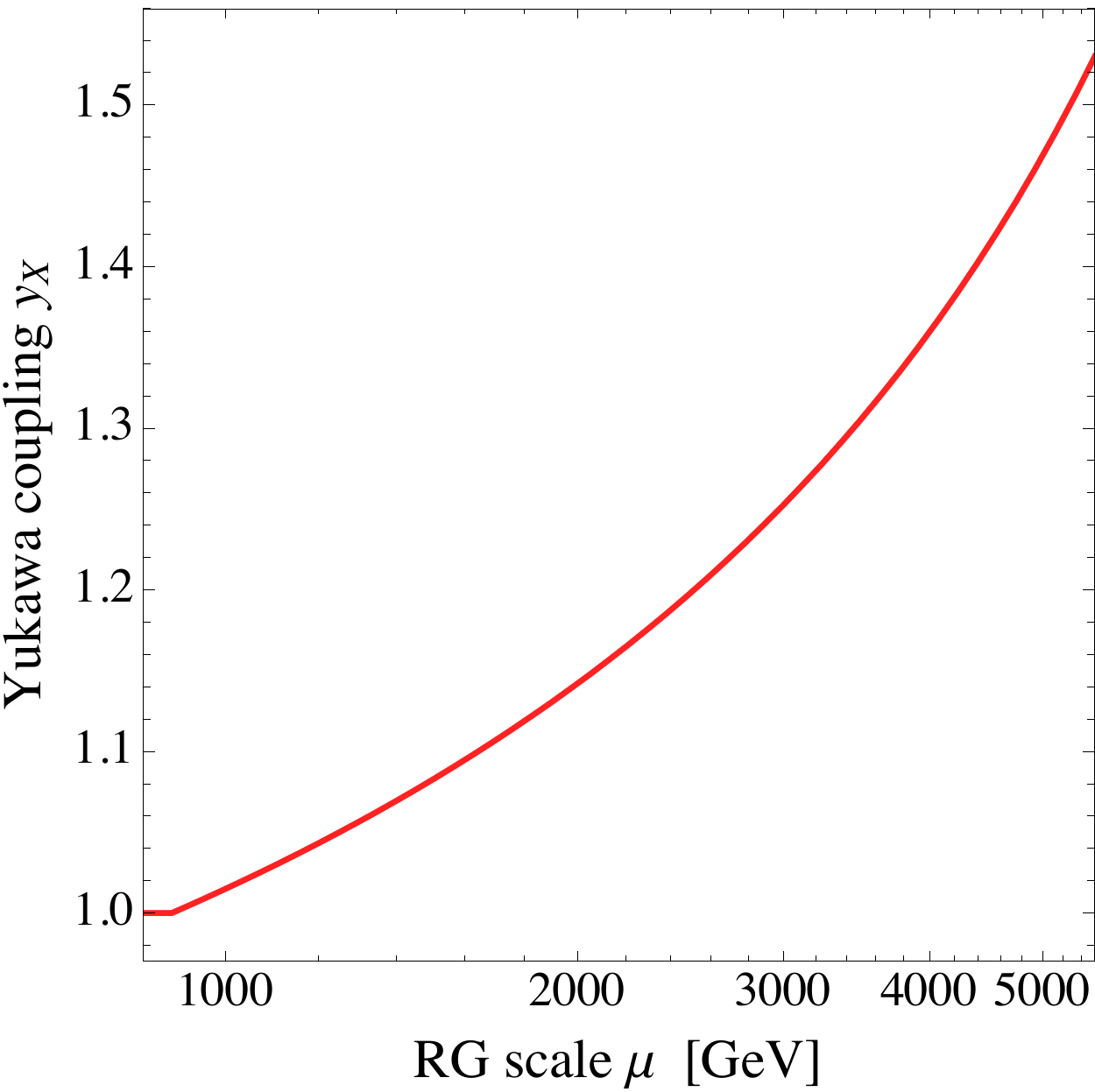}
\endminipage\\
\caption{ 
The same as in Fig.~\ref{fig:RunningLambdaSPositive} but for different values of $N_X$, $y_X$ (see label plot).
}
\label{fig:RunningLambdaS2Positive}
\end{figure}
We present our results in Fig.~\ref{fig:RunningLambdaS2Positive}.
In the left panel, we show the running of the quartic coupling, $\lambda_S$. The main difference w.r.t. the previous case with $y_X = 5$ is the smaller Yukawa coupling. 
Regarding the running of $\lambda_S$, we recover what we already discussed in the first part of Section~\ref{eq:RGEresults}.
If $\lambda_S$ is large enough, the positive term 
$\beta_{\lambda_S}^{(1)} \propto 120\, \lambda_S y_X^2 + 18\, \lambda_S^2$ dominates and $\lambda_S$ increases along the RG flow. In Fig.~\ref{fig:RunningLambdaS2Positive}, this behavior is reflected by the initial values $\lambda_S(\mu_0) = 2,5$. For these choices, $\lambda_S$ violates perturbativity at few TeV. If the initial values $\lambda_S(\mu_0)$ are small enough, the negative term $\beta_{\lambda_S}^{(1)} \propto -120\, y_X^4$ dominates, and $\lambda_S$ is dragged towards negative values.  This is always true for $\lambda_S(\mu_0) \lesssim 1$. For $\lambda_S(\mu_0) = 1$, we find that $\lambda_S(\Lambda) <0$ at $\Lambda \sim 5$ TeV. As far as the running of the Yukawa coupling $y_X$ is concerned, it runs now very slowly, staying within the perturbation regime up to a very high scale.

We argue that also in this case the theory reveals an instability at a scale not far away $O(\rm TeV)$ for generic values of the couplings. While the left panel in Fig.~\ref{fig:RunningLambdaS2Positive} shows that there exists a very fine-tuned initial value, $1 \lesssim \lambda_S(\mu_0) \lesssim 2$ which are almost unaffected by  RG effects, this particular point does not correspond to a special property of the theory.

\subsubsection{$m_X = 900$ GeV, $N_X = 1$, $Q_X = 25/3$, $y_X = 1$: $y_X N_X Q_X = 25/3$.}

In this scenario, a small Yukawa coupling $y_X = 1$ is obtained by increasing the value of $Q_X$.
We take $Q_X =25/3$ while $N_X = 1$ is fixed as the case in~\ref{sec:BechmarkCase}.
\begin{figure}[!htb!]
\begin{center}
\fbox{\footnotesize $m_X = 900$ GeV, $Q_X = 25/3$, $N_X = 1$, $y_X = 1$ ($\lambda_{HS} = 0$)}
\end{center}
\minipage{0.5\textwidth}
  \includegraphics[width= .84 \linewidth]{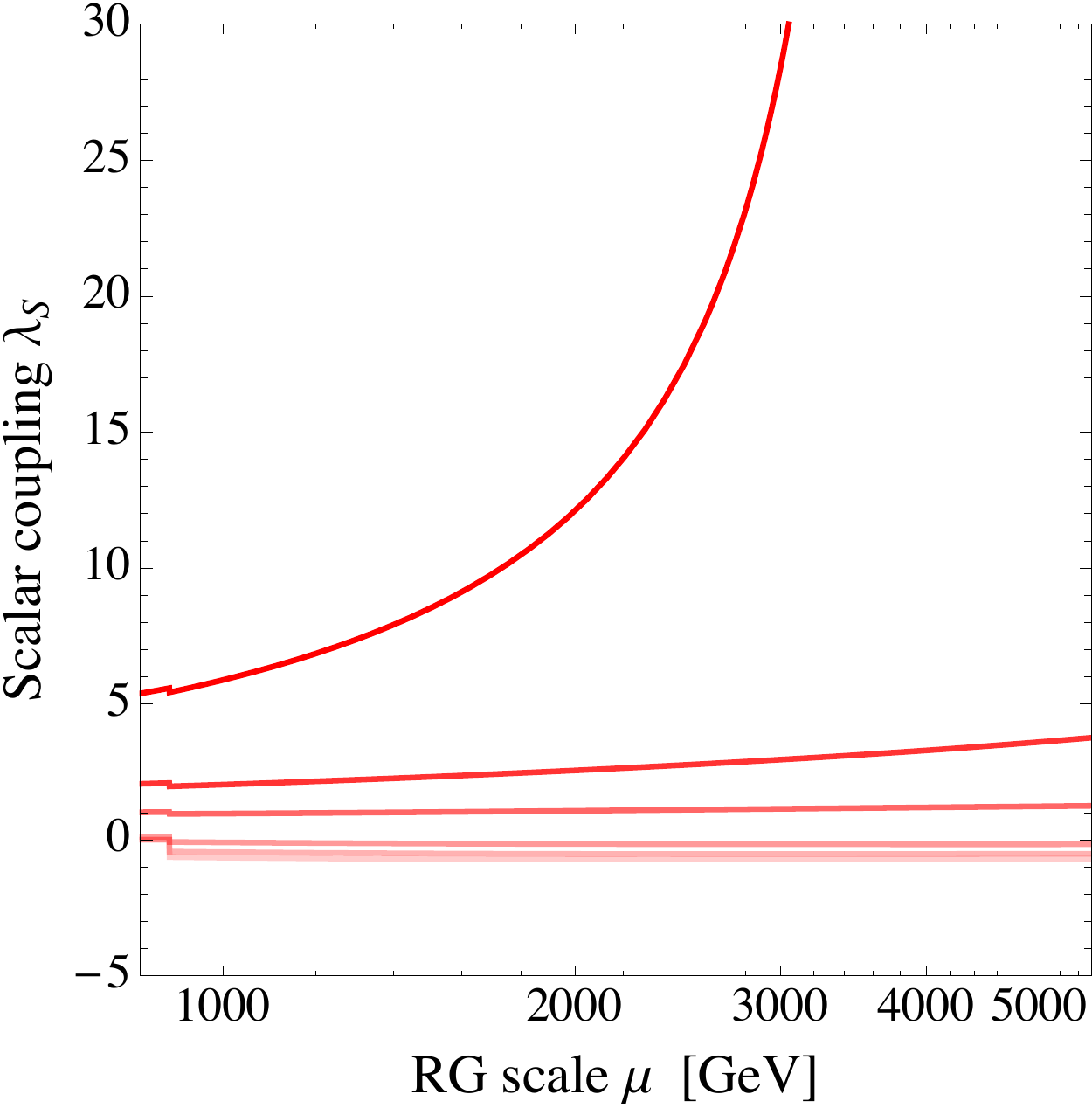}
\endminipage\hfill
\minipage{0.5\textwidth}
  \includegraphics[width= .86 \linewidth]{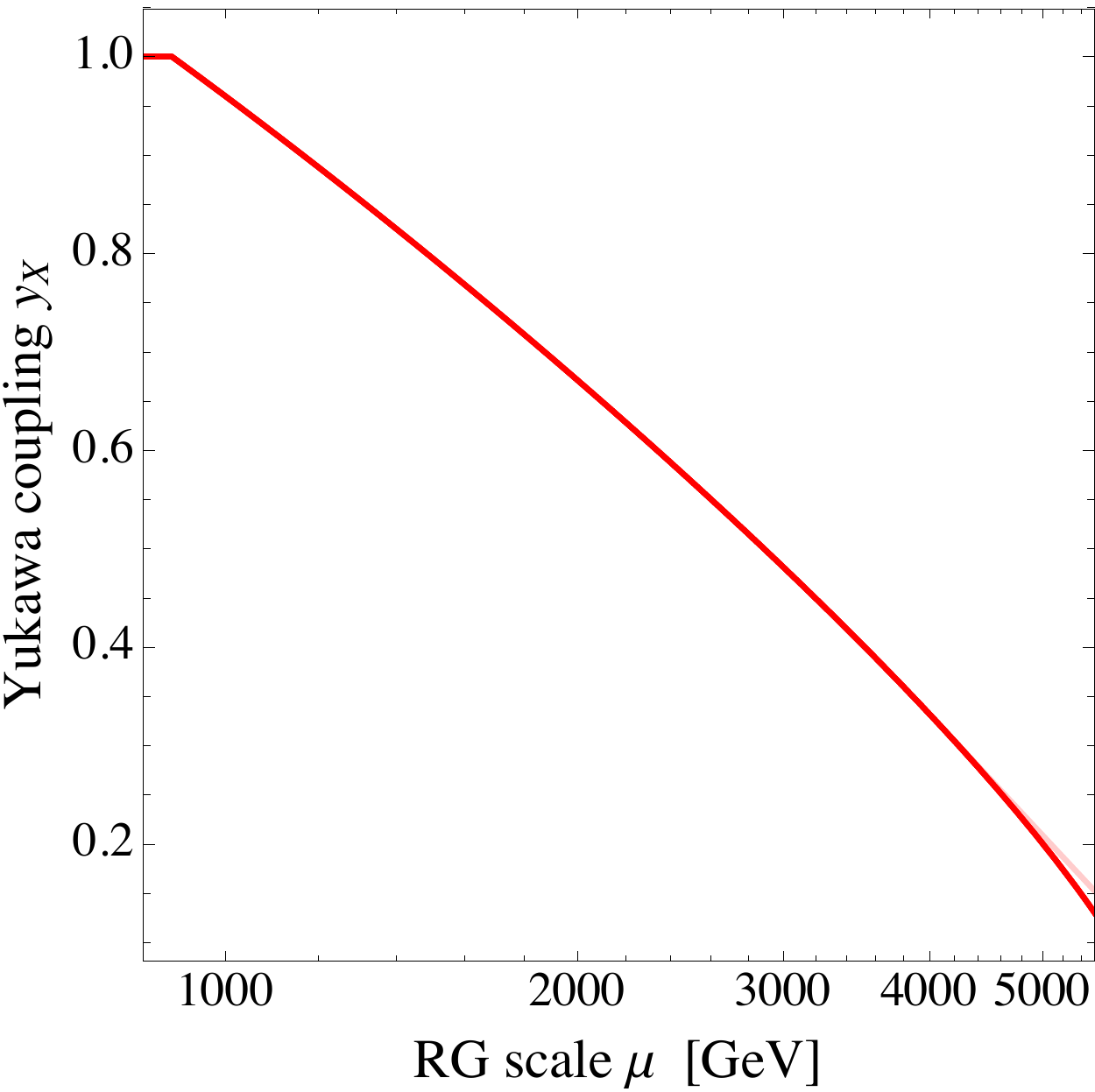}
\endminipage\\
\caption{ 
The same as in Fig.~\ref{fig:RunningLambdaSPositive} but for different values of $Q_X$, $y_X$ (see plot label).
}
\label{fig:RunningLambdaS3Positive}
\end{figure}
Our results are illustrated in Fig.~\ref{fig:RunningLambdaS3Positive}. In this scenario, the Yukawa coupling $y_X$ decreases as the large electric charge $Q_X = 25/3$ makes the negative contribution to its RG running, 
$\beta_{y_X}^{(1)} \propto - Q_X^2 g_1^2 y_X$, dominant.
Regarding the running of $\lambda_S$, it is possible to find some acceptable trajectories with $\lambda_S(\mu_0) \sim O(1)$ in the RG space.
However, as is clear from Fig.~\ref{fig:HyperchargeRunning}, this case is very
unrealistic from the point of view of the hypercharge gauge coupling. We discard this solution as a candidate for the weakly coupled realization of the diphoton excess.

\subsubsection{$m_X = 900$ GeV, $N_X = 3$, $Q_X = 8/3$, $y_X = 1$: $y_X N_X Q_X \sim 25/3$.}

\begin{figure}[!htb!]
\begin{center}
\fbox{\footnotesize $y_X = 1$, $m_X = 900$ GeV, $Q_X = 8/3$, $N_X = 3$ ($\lambda_{HS} = 0$)}
\end{center}
\minipage{0.5\textwidth}
  \includegraphics[width= .84 \linewidth]{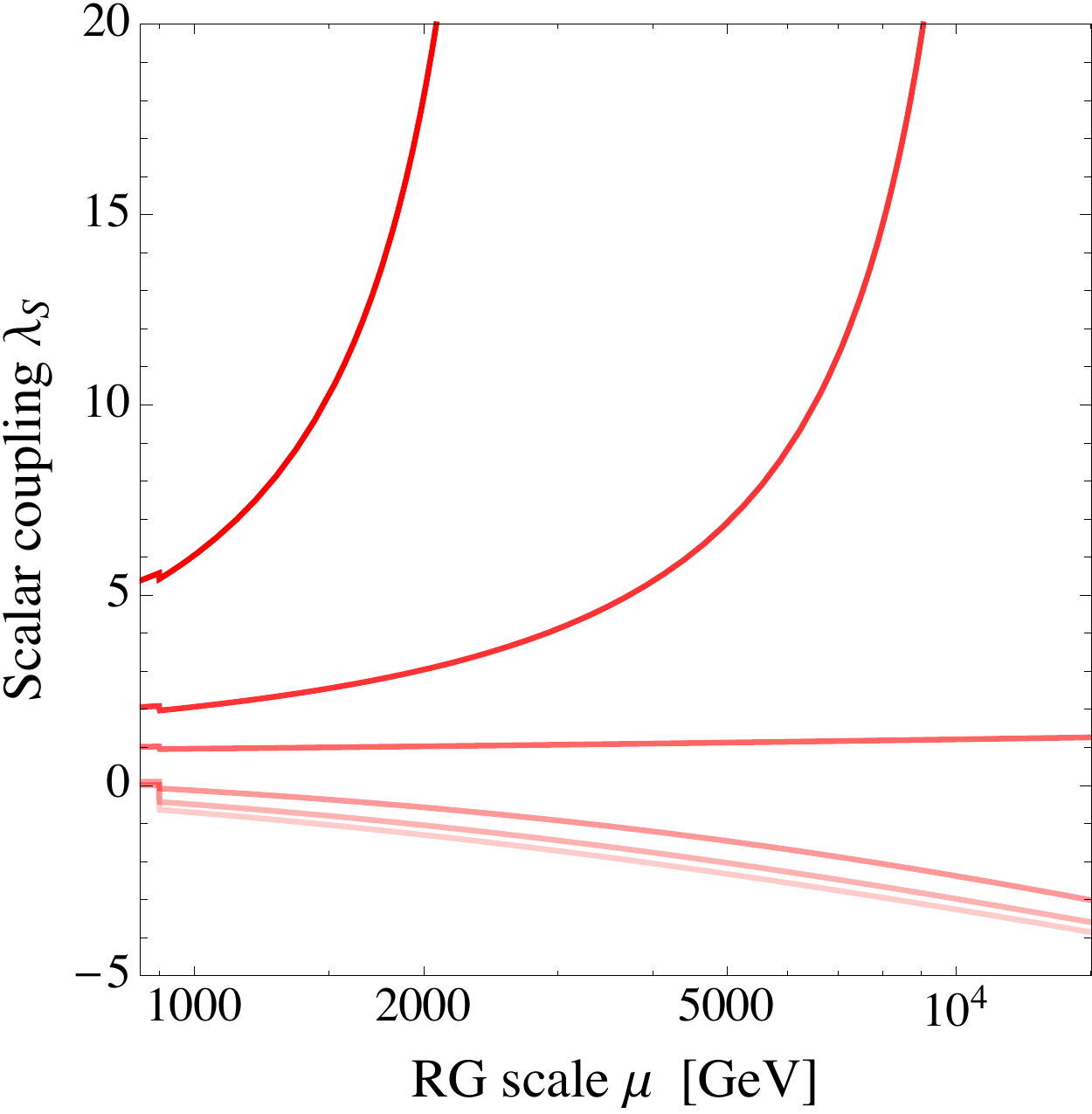}
\endminipage\hfill
\minipage{0.5\textwidth}
  \includegraphics[width= .86 \linewidth]{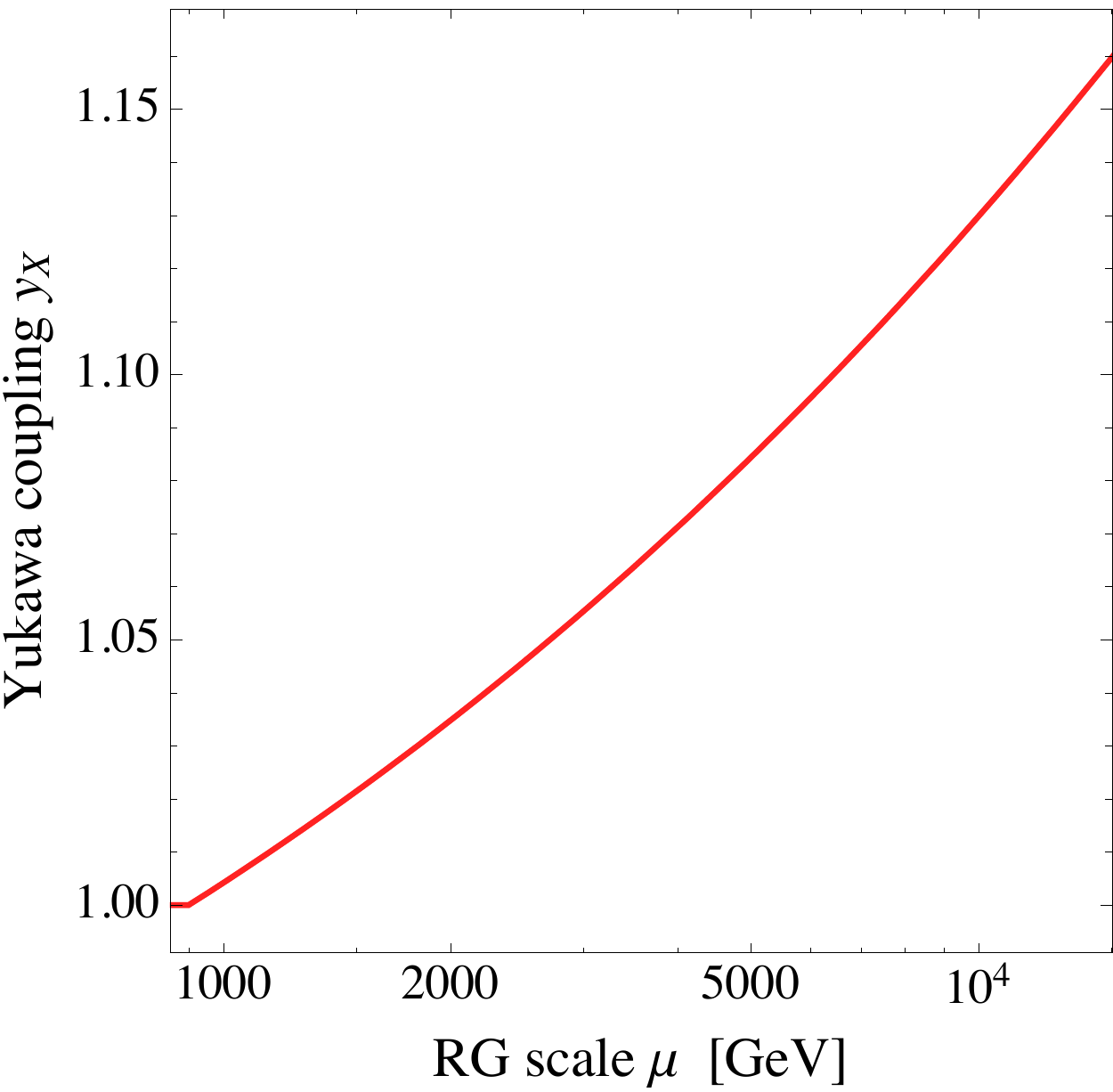}
\endminipage\\
\caption{ 
The same as in Fig.~\ref{fig:RunningLambdaSPositive} but for different values of $Q_X$, $N_X$, $y_X$ (see label plot).
}
\label{fig:RunningLambdaS4Positive}
\end{figure}

Let us move to discuss an intermediate situation in which both the multiplicity $N_X$ and the electric charge $Q_X$ 
are modified in such a way that $y_X \sim 1$.
We present our results in Fig.~\ref{fig:RunningLambdaS4Positive}, where we focus on $Q_X = 8/3$ and $N_X = 3$.
The running of the Yukawa coupling is again dominated by the positive term $\beta_{y_X}^{(1)} \propto 21\, y_X^3$, and $y_X$ always increases along the RG flow. The running of $\lambda_S$ depends on its initial value. If $\lambda_S(\mu_0) \simeq 1$, a stable solution exists, almost unaffected by the RG flow. As for the case with $Q_X = 5/3$, $N_X = 5$ that was discussed above, this solution looks like the result of a fine tuning rather than a special point in the parameter space. However, it is undeniable that in this scenario a weakly coupled theory valid to a high energy scale can be constructed. From the central panel in Fig.~\ref{fig:HyperchargeRunning}, we see that the choice $Q_X = 8/3$ and $N_X = 3$ sizably alter the hypercharge running. We find that  the corresponding Landau pole is lowered to $\Lambda \sim 300$ TeV. We can therefore identify this scale as the upper limit of the validity for this theory.\\

We close this Section by briefly discussing the case with $\Gamma = 45$ GeV. If $N_X = 1$ a very large Yukawa coupling ($y_X \gtrsim 6.5$ if $m_X \simeq 700$ GeV
\footnote{For the same mass of the vector-like fermions, $N_X Q_X y_X$ needs to increase by factor of $(45)^{1/4} \sim 2.6$ to maintain the same signal rate (see Eq.~\ref{eq:LargeWidth:Scaling}) when the total width changes from $\Gamma = 1$ GeV to $\Gamma = 45$ GeV.} and $Q_X = 8/3$) is needed in order to fit the excess. As illustrated in the upper-right panel of Fig.~\ref{fig:xsecyxmx:threeGamma:Varying}, it is possible to bring the Yukawa back to a perturbative value by increasing the value of $N_X$, and we find $y_X \simeq 1.5$ with $N_X = 5$ (taking fixed $m_X \simeq 700$ GeV and $Q_X = 8/3$). In this case, the biggest obstruction is represented by the running of the hypercharge gauge coupling since the Landau pole is lowered to $\sim 10$ TeV (see Fig.~\ref{fig:HyperchargeRunning}). It would be interesting to investigate the case with $\Gamma = 45$ GeV in more detail following the strategy outlined in this paper, even if such a large value of total width is very difficult to be realistic from the point of view of a weakly coupled theory. If experimentally confirmed, it will give a strong indication in favor of a strongly coupled interpretation of the excess.

\subsection{Concluding remarks and summary plots}
\label{eq:MoneyPlot}

\begin{figure}[!htb!]
\begin{center}
\minipage{0.5\textwidth}
  \includegraphics[width=.90\linewidth]{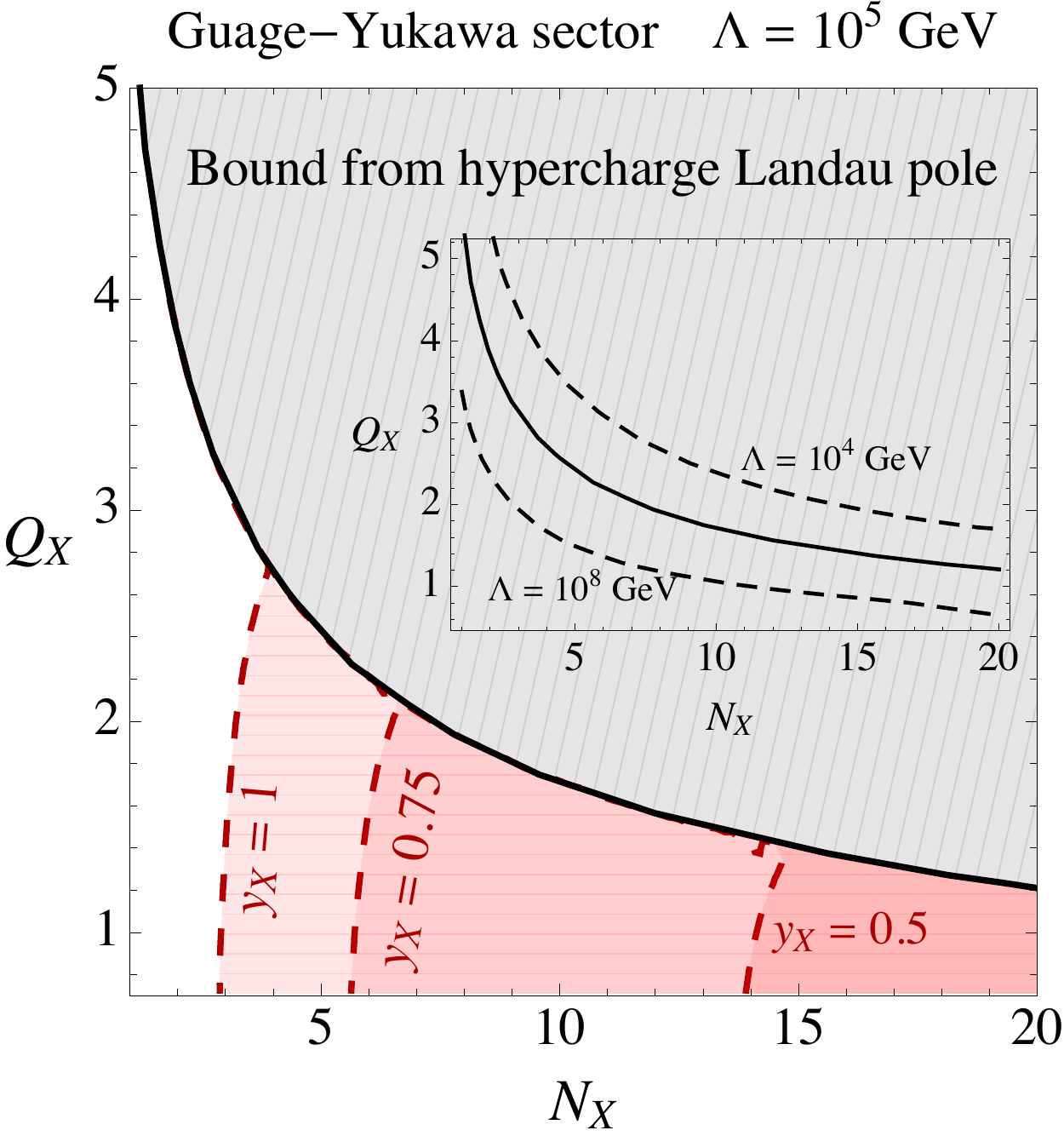}
\endminipage\hfill
\minipage{0.5\textwidth}
  \includegraphics[width=.90\linewidth]{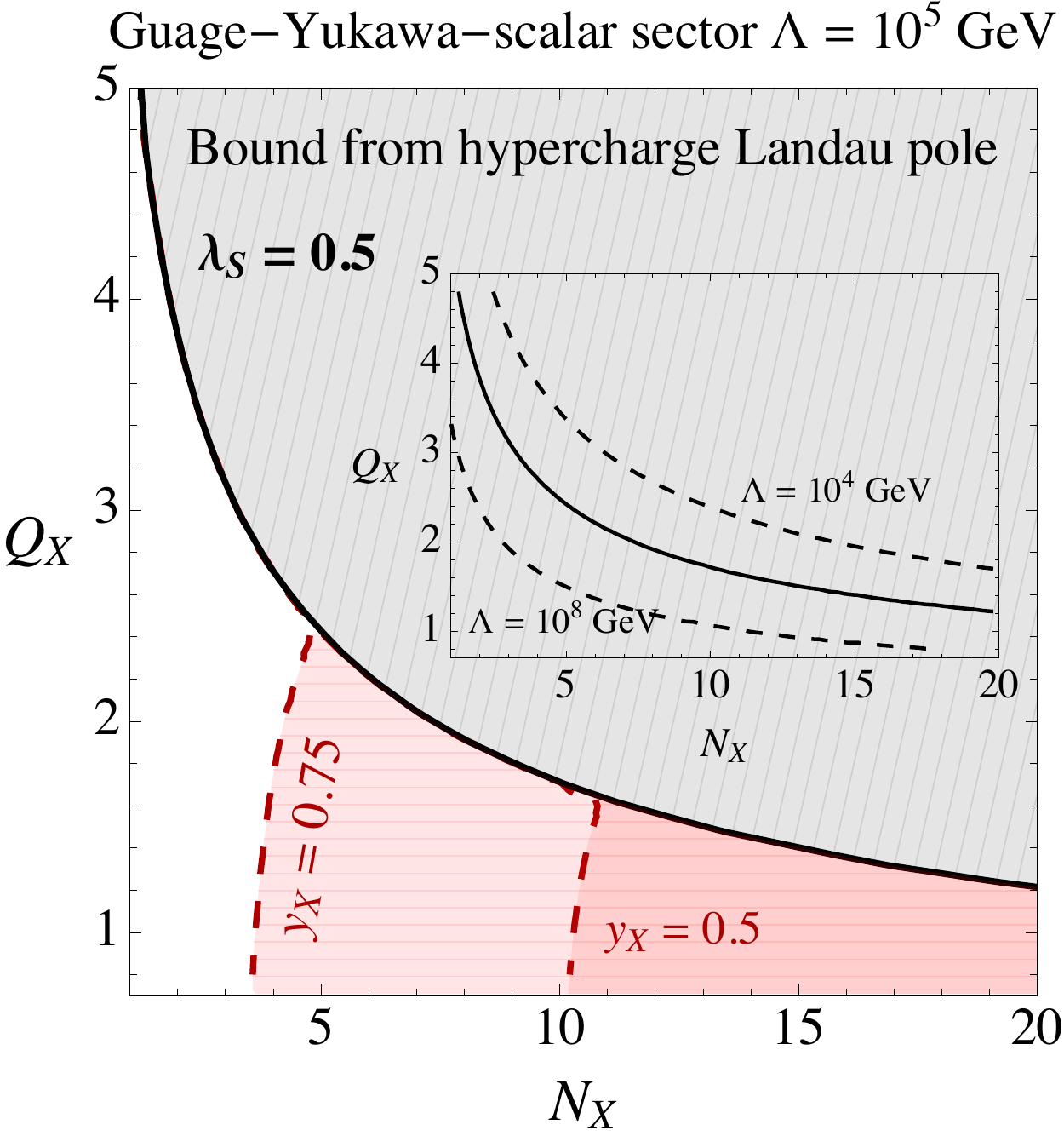}
\endminipage\hfill
\caption{Left panel. 
Bound on the parameter space $(N_X,\, Q_X)$ coming  
from the Landau pole of the hypercharge gauge coupling and perturbativity of the Yukawa coupling $y_X$. 
In the upper region of the solid line, the theory develops a Landau pole below $\Lambda =10^5$ GeV. Similarly for $\Lambda =10^4, 10^8$ GeV (black dashed lines in the inset plot).
The  red dashed lines correspond to the cases incorporating fixed Yukawa couplings.
Right panel. 
Bound on the parameter space $(N_X,\, Q_X)$ coming  
from the Landau pole of the hypercharge gauge coupling, perturbativity of the Yukawa coupling $y_X$, vacuum stability condition $\lambda_S > 0$, and 
perturbativity of the scalar coupling $\lambda_S$. For simplicity we consider a fixed value $\lambda_S = 0.5$ at the electroweak scale.}
\label{fig:gaugeYukawa}
\end{center}
\end{figure}

Before concluding, it is useful to summarize the main results of our paper.
In Fig.~\ref{fig:gaugeYukawa} we show the allowed parameter space in the plane 
$(N_X,\, Q_X)$ once the constraints coming from our RG analysis are imposed. For simplicity, we neglect the mixing ($\lambda_{HS} = 0$), and 
we fix the mass of the vector-like quarks ($m_X = 900$ GeV). 
We take $\Lambda = 10^5$ GeV as a reference cut-off scale but in the inset plots we show how the bound changes considering $\Lambda = [10^4 - 10^8]$ GeV.

In the left panel of  Fig.~\ref{fig:gaugeYukawa}, we focus on the impact of the RGEs related to the gauge-Yukawa sector of the theory, and to this end we put $\lambda_S = 0$. 
The theory suffers from an hypercharge Landau pole below $\Lambda = 10^5$ GeV in the region shaded in gray. 
In the region shaded in red, on the contrary,  the Yukawa coupling $y_X$ violates the perturbativity bound below the same cut-off scale.
As evident from the plot, the combination of the two bounds significantly reduces the allowed parameter space. This is in particular true if large Yukawa couplings are needed.
The compatibility of the parameter space in Fig.~\ref{fig:gaugeYukawa} with the signal region for $\Gamma =1$ GeV ($\Gamma =45$ GeV) 
can be figured out by comparing it to the middle panels (right-most panels) of Fig.~\ref{fig:xsecyxmx:threeGamma:Varying},\,\ref{fig:xsecyxmx:threeGamma}.

In the right panel of  Fig.~\ref{fig:gaugeYukawa}, we include the scalar sector of the theory. For simplicity, we consider 
$\lambda_S = 0.5$ as initial condition. 
The interplay with the Yukawa coupling, 
entering in the $\beta$ function through the combination $\beta_{\lambda_S} \propto 24 N_X y_X^2 (\lambda_S - y_X^2)$, 
has the net effect of reducing the allowed parameter space since $\lambda_S$ quickly runs---driven by $y_X$---towards negative values, thus destabilizing the vacuum. 
This is evident from the comparison between left and right panel of Fig.~\ref{fig:gaugeYukawa}. For instance the value $y_X = 1$, allowed in the left panel in the left-most corner of the parameter space, becomes completely forbidden once scalar couplings are included.

Note that from the informations encoded in Fig.~\ref{fig:gaugeYukawa} 
it is possible to conclude that the case with $\Gamma =45$ GeV is more disfavored by our perturbative analysis if compared with the assumption  $\Gamma =1$ GeV.
A large width implies a smaller branching ratio $S\rightarrow \gamma\gamma$, thus requiring a large Yukawa coupling  to compensate the suppression through 
the gluon fusion production of S. Large Yukawa couplings, however, do not fit in the parameter space represented in Fig.~\ref{fig:gaugeYukawa}.

\section{Summary and outlook}\label{sec:Summary}
Recently,  both the ATLAS and CMS collaborations reported an excess around $750$ GeV in the invariant mass distribution of the diphoton. The excess can be interpreted in terms of a weakly coupled theory of New Physics beyond the SM, and a simple such scenario consists of a scalar resonance coupled to photons and gluons via loops of $N_X$ vector-like quarks with electric charge $Q_X$, that are almost degenerate in mass. Alternatively, one can identify the excess as the imprint of a scalar resonance belonging to a strongly interacting sector. 

At the moment---given the small statistical significance of the excess, still compatible with a fluctuation of the background---the two interpretations, weakly versus strongly coupled, are more or less equally preferred by data from a phenomenological viewpoint. In this paper, we confronted these two possibilities from a more theoretical perspective and our approach was the following. On general grounds, by taking a weakly coupled theory custom-tailored to fit the observed properties of the excess at low energies, it is possible to extrapolate its structure to high energies by means of the RGEs. The logic is to check whether  the theory develops some pathology along the RG flow, thus indicating its inconsistency and the scale of the corresponding  breaking. We performed this exercise in the context of a simple weakly coupled theory able to explain the diphoton excess, in which the SM is enlarged by means of a new scalar singlet together with new  vector-like fermions responsible for its interactions with photons and gluons. In this simple setup, we showed that the theory quickly runs towards an instability of the scalar potential, already at a scale not far above the TeV scale.
This problematic behavior is shared by many variations of the simple setup mentioned above that we checked (i.e. introducing multiple vector-like fermions or changing their electric charge), and therefore it seems to point towards an inconsistency of the underlying weakly coupled theory.

Exceptions are possible. We showed that one can finely balance between $Q_X$ and $N_X$ such that the vacuum stability of the scalar potential and the perturbativity of all the dimensionless couplings is ensured up to high scales. However, we also showed that this particular direction corresponds to fine-tuned points in the parameter space rather than to natural realizations of the theory.

Note that the results obtained in this paper are even stronger in the case of resonant production via photon fusion. 
The production cross section from photon fusion is much smaller than the one originated from gluon fusion because of 
the smaller photon luminosity. 
In order to compensate the reduced signal rate, the partial decay widths need to be significantly increased, thus strengthening the perturbativity constraints.

Of course our analysis does not pretend to exclude all weakly coupled explanations of the diphoton excess. One can always engineer more complicated theoretical frameworks; for instance, it is possible to introduce both vector-like quarks and leptons in order to disentangle gluon production from diphoton decay and gain more freedom in the parameter space that could be used to keep the Yukawas in a perturbative regime. In any case, we argue that---even in the context of toy models---extrapolating the theory to high scales is an important exercise that should be carried out in order to fully reveal the actual strength of the dimensionless couplings.

Only time will tell us if the diphoton excess reported by the ATLAS and CMS collaborations corresponds to our first glimpse of New Physics beyond the SM, or just to another sneaky statistical fluctuation. Meanwhile, it costs nothing to speculate on possible theoretical implications of such a potential discovery. In this respect, the strategy outlined in this paper could be a valid guiding principle to check whether a weakly coupled explanation of the excess behaves properly as a good theory or hides some deeper inconsistency just above the energy scale at which it was tailored.\\

Note added:  While we were working on this paper, we noted~\cite{Chakraborty:2015jvs,Zhang:2015uuo,Dhuria:2015ufo} which address  similar issues. It is worth pointing out the main differences between these papers and our work. In~\cite{Chakraborty:2015jvs} the authors---motivated by the apparent large width of the resonance---focused on the existence of multiple real scalar gauge singlets $S_i$ almost degenerate in mass. On the contrary, they include only one single vector-like fermion with $Q_X = 2/3$.
They conclude that the model stays within the validity of perturbation theory only if a large number of singlet $N\gg 1$ is allowed. Clearly, this analysis follows an orthogonal direction if compared with our setup. The analysis of~\cite{Zhang:2015uuo} is, on the technical level, the closest w.r.t. ours (since only one scalar---real or complex---singlet field $S$, mixed with the Higgs doublet, was introduced). However, there are few important differences on which we would like to remark. 
First, the authors include only one vector-like quark, which transforms as a triplet under $SU(3)_C$, with generic electric charge $Q_X$. Second, and most important, the main goal of~\cite{Zhang:2015uuo} is to understand if the conditions of vacuum stability and perturbativity---explored up to three different cut-off scales, namely the Planck, GUT and see-saw scales---are compatible with the signal strength required to fit the diphoton excess. The answer is of course negative, since a good fit  of the diphoton excess can be obtained only  with a large Yukawa coupling. In our paper we offer a broader and deeper perspective on the issue. By increasing the multiplicity of the vector-like quarks, in fact, we proved that a good fit can be obtained with a moderate Yukawa coupling thus apparently solving the issue. However, by carefully studying the RGEs, we also proved that the theory---with the exception of very few fine-tuned directions---is brought back to the strong coupling regime once the running is taken into account. Finally, we comment on~\cite{Dhuria:2015ufo}. The authors claim that the model with $Q_X = 2$, $N_X = 2$, $y_X = 0.52$ stays within the validity of perturbation theory all the way up to the Planck scale, providing a good fit of the diphoton excess with a signal strength $\sigma = 30$ fb (alternatively, with $Q_X = 2$, $N_X = 1$, $y_X = 0.33$ if $\sigma \in [2-4]$ fb). According to our analysis in Section~\ref{sec:VectorLikeFrmions}, small Yukawa couplings like those considered in~\cite{Dhuria:2015ufo}  are allowed if one assumes $\Gamma = \Gamma_{gg} + \Gamma_{\gamma\gamma}$. In this case, good directions indeed exist in the parameter space (although they look a bit fine-tuned, similar to those highlighted in our Fig.~\ref{fig:RunningLambdaS3Positive}). In our paper we focus on the case $\Gamma = 1$ GeV (more preferred by data), and we provide a comprehensive description of the interplay between signal strength, decay width, and RGEs.

\acknowledgments
We thank Roberto Contino and Adam Falkowski for useful discussions and Rakhi Mahbubani for reading our manuscript. MS is grateful for the hospitality of the CERN Theory Group where this work has been initiated and done. 

\appendix
\section{Scalar Potential}\label{app:ScalarPot}
The generic scalar potential of the SM Higgs doublet and new singlet scalar can be written as 
\begin{equation}
V(H,S) = \mu_H^2 |H|^2 + \lambda_H |H|^4 + \frac{\lambda_{HS}}{2} |H|^2S^2 + \frac{\mu_S^2}{2} S^2 +\frac{\lambda_S}{4} S^4~,
\end{equation}
where we assumed that $S$ is real and odd under $S\rightarrow - S$. The potential in the unitary gauge is obtained via $H(x) = (1/\sqrt{2})\, U(x)(0, h(x))^T$,
\begin{equation}\label{app:eq:PotentialhS}
V(h,S) = \frac{\mu_H^2}{2} h^2 + \frac{\lambda_H}{4} h^4 +  \frac{\lambda_{HS}}{4} h^2 S^2 + \frac{\mu_S^2}{2} S^2 +\frac{\lambda_S}{4} S^4~.
\end{equation}
The potential has a minimum at VEVs, $\langle h\rangle = v$, $\langle S\rangle = u$ if the following conditions
involving first derivatives
\begin{equation}
\left.\frac{\partial V(h,S)}{\partial h}\right|_{\langle h\rangle = v,~\langle S\rangle = u}=0~,~~~~
\left.\frac{\partial V(h,S)}{\partial S}\right|_{\langle h\rangle = v,~\langle S\rangle = u}=0~,
\end{equation}
and the determinant of the Hessian matrix $\left.{\rm det}
(M_{hS}^2)
\right|_{\langle h\rangle = v,~\langle S\rangle = u}>0$ with
\begin{equation}\label{app:eq:MassMatrix}
\left.M_{hS}^2\right|_{\langle h\rangle = v,~\langle S\rangle = u} \equiv
\left.\left(
\begin{array}{cc}
\frac{\partial^2 V(h,S)}{\partial h^2}  &   \frac{\partial^2 V(h,S)}{\partial h\partial S}   \\
\frac{\partial^2 V(h,S)}{\partial S\partial h}  &  \frac{\partial^2 V(h,S)}{\partial S^2}
\end{array}
\right)\right|_{\langle h\rangle = v,~\langle S\rangle = u}=
\left(
\begin{array}{cc}
 2\lambda_H v^2  & \lambda_{HS} u v     \\
 \lambda_{HS}uv &   2\lambda_S u^2
\end{array}
\right)~,
\end{equation}
are satisfied. After simple algebra it follows
\begin{equation}
v^2 = \frac{2(\lambda_{HS}\mu_S^2 - 2\lambda_S^2\mu_H^2)}{4\lambda_H \lambda_S - \lambda_{HS}^2}~,~~~
u^2 = \frac{2(\lambda_{HS}\mu_H^2 - 2\lambda_H^2\mu_S^2)}{4\lambda_H \lambda_S - \lambda_{HS}^2}~,
\end{equation}
with  $4\lambda_H\lambda_S - \lambda_{HS}^2 > 0$.
The conditions for a local minimum therefore are
\begin{equation}\label{app:eq:LocalMin}
\lambda_{HS}\mu_S^2 - 2\lambda_S\mu_H^2 >0~,~~~ \lambda_{HS}\mu_H^2 - 2\lambda_H^2\mu_S^2> 0~,~~~4\lambda_H\lambda_S - \lambda_{HS}^2 > 0~.
\end{equation} 
Let us now turn to discuss the mass eigenstates. The mass matrix in Eq.~\ref{app:eq:MassMatrix}
can be easily diagonalized by means of an orthogonal transformation
\begin{equation}
\mathcal{O} \equiv
\left(
\begin{array}{cc}
 c_{\theta} & s_{\theta}  \\
-s_{\theta}  & c_{\theta}  
\end{array}
\right)~,~~\mathcal{O}^T\left.M_{hS}^2\right|_{\langle h\rangle = v,~\langle S\rangle = u} \mathcal{O} 
={\rm diag}(m_{H_1}^2, m_{H_2}^2)~,
\end{equation}
where we used the short-hand notations $c_{\theta}\equiv \cos\theta$, $s_{\theta}\equiv \sin\theta$, $t_{\theta}\equiv \tan\theta$. 
The eigenvalues and the mixing angle are given by
\begin{eqnarray}
m_{H_1}^2 &=& \lambda_Hv^2 +\lambda_Su^2 - \sqrt{
(\lambda_S u^2 - \lambda_H v^2)^2 + \lambda_{HS}^2 u^2 v^2
}~,\\
m_{H_2}^2 &=& \lambda_Hv^2 +\lambda_Su^2 + \sqrt{
(\lambda_S u^2 - \lambda_H v^2)^2 + \lambda_{HS}^2 u^2 v^2
}~,\\
t_{2\theta} &=& \frac{\lambda_{HS} u v}{\lambda_S u^2 - \lambda_H v^2}~,
\end{eqnarray}
while the mass eigenstates are
\begin{equation}
\left(
\begin{array}{c}
  H_1   \\
  H_2
\end{array}
\right)  = \mathcal{O}^T 
\left(
\begin{array}{c}
  h   \\
  S
\end{array}
\right)~.
\end{equation}
We identify $H_1$ with the physical Higgs boson with $m_{H_1} = 125.09$ GeV, while $H_2$  is the new scalar resonance with 
 $m_{H_2} \simeq 750$ GeV. In the large VEV limit $u^2 \gg v^2$ we have, neglecting terms $O(v^2/u^2)$
 \begin{equation}\label{app:eq:LargeVev}
 m_{H_1}^2 \approx 2\lambda_H v^2 - \frac{\lambda_{HS}^2 v^2}{2\lambda_S}~,~~
  m_{H_2}^2 \approx 2\lambda_S u^2 + \frac{\lambda_{HS}^2 v^2}{2\lambda_S}~,~~
  t_{2\theta}\approx \frac{\lambda_{HS}v}{\lambda_S u}~.
 \end{equation}

\section{On the impact of the mixing angle}\label{app:MixingAngle}

In this Appendix we briefly discuss the impact of a non-zero mixing angle on the results of our analysis.
For the sake of simplicity, we consider the case with small mixing angle $s_{\theta} = 0.01$ (consistent with the bound in~\cite{Falkowski:2015swt}),   
and we focus on the choice $N_X = 1$, $Q_X = 5/3$, $y_X = 5$, $m_X = 900$ GeV.
We stress that the purpose of this section is not to provide a comprehensive scan on the allowed values,
rather to show that 
the presence of a non-zero mixing does not alter, at the qualitative level, our results.
As explained in the right panel of Fig.~\ref{fig:RunningLambda},
for a fixed mixing angle we have the freedom to vary the free parameter $\lambda_{HS}(\mu_0)$. 
In turn, for each value of $\lambda_{HS}(\mu_0)$, the starting value of $\lambda_S$ is fixed via Eq.~\ref{eq:LambdaS}.
According to Fig.~\ref{fig:RunningLambda}, in the following we scan over a large range of $\lambda_{HS}(\mu_0)$, and in particular we 
choose $\lambda_{HS}(\mu_0) \in [10^{-3}, 10^{-1}]$.
\begin{figure}[!htb!]
\begin{center}
\fbox{\footnotesize $s_{\theta} = 0.01$, $y_X = 5$, $m_X = 900$ GeV, $Q_X = 5/3$, $N_X = 1$, $(\lambda_{HS} \neq 0$)}
\end{center}
\minipage{0.325\textwidth}
  \includegraphics[width=1.\linewidth]{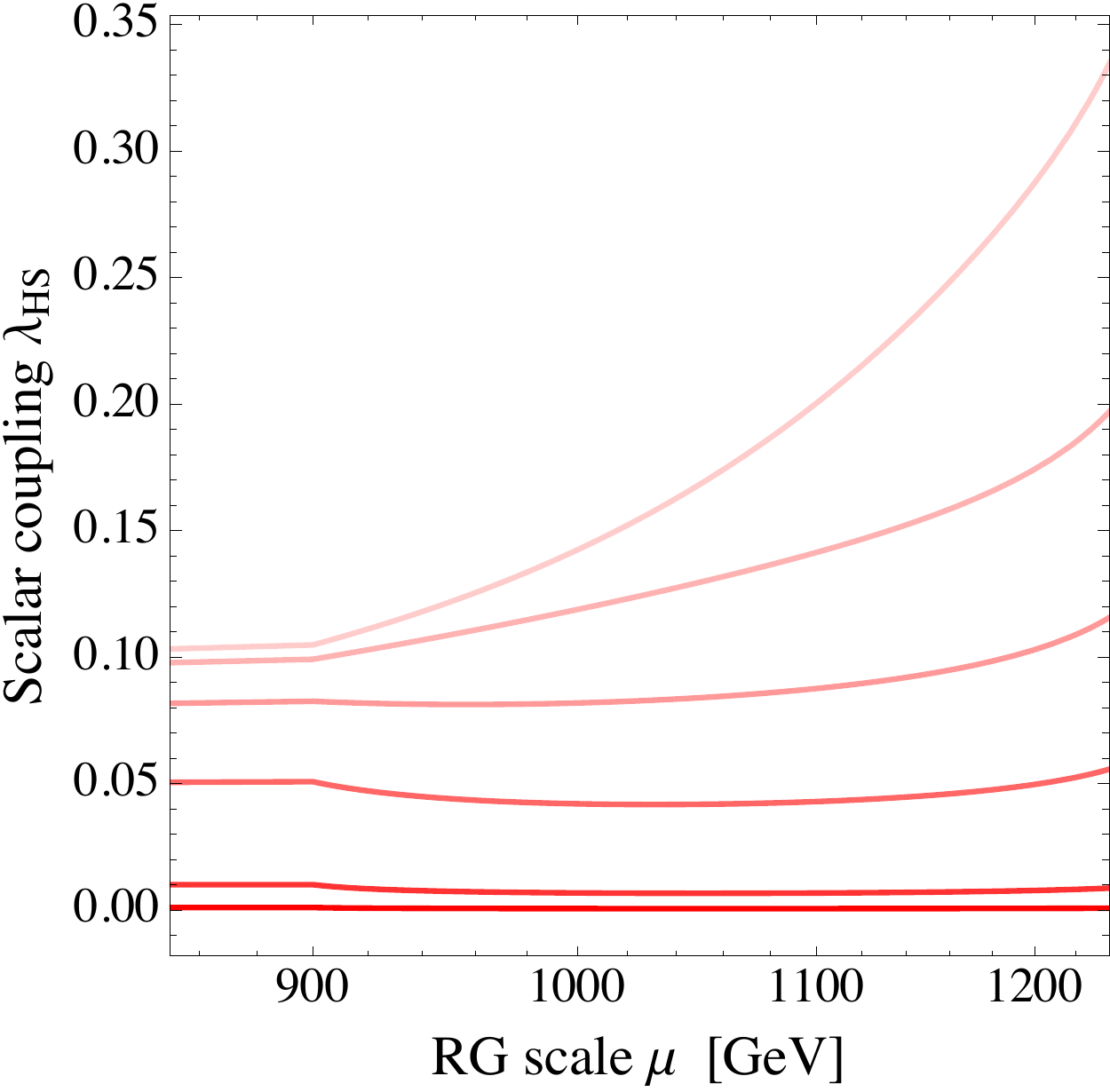}
\endminipage\hfill
\minipage{0.325\textwidth}
  \includegraphics[width=1.\linewidth]{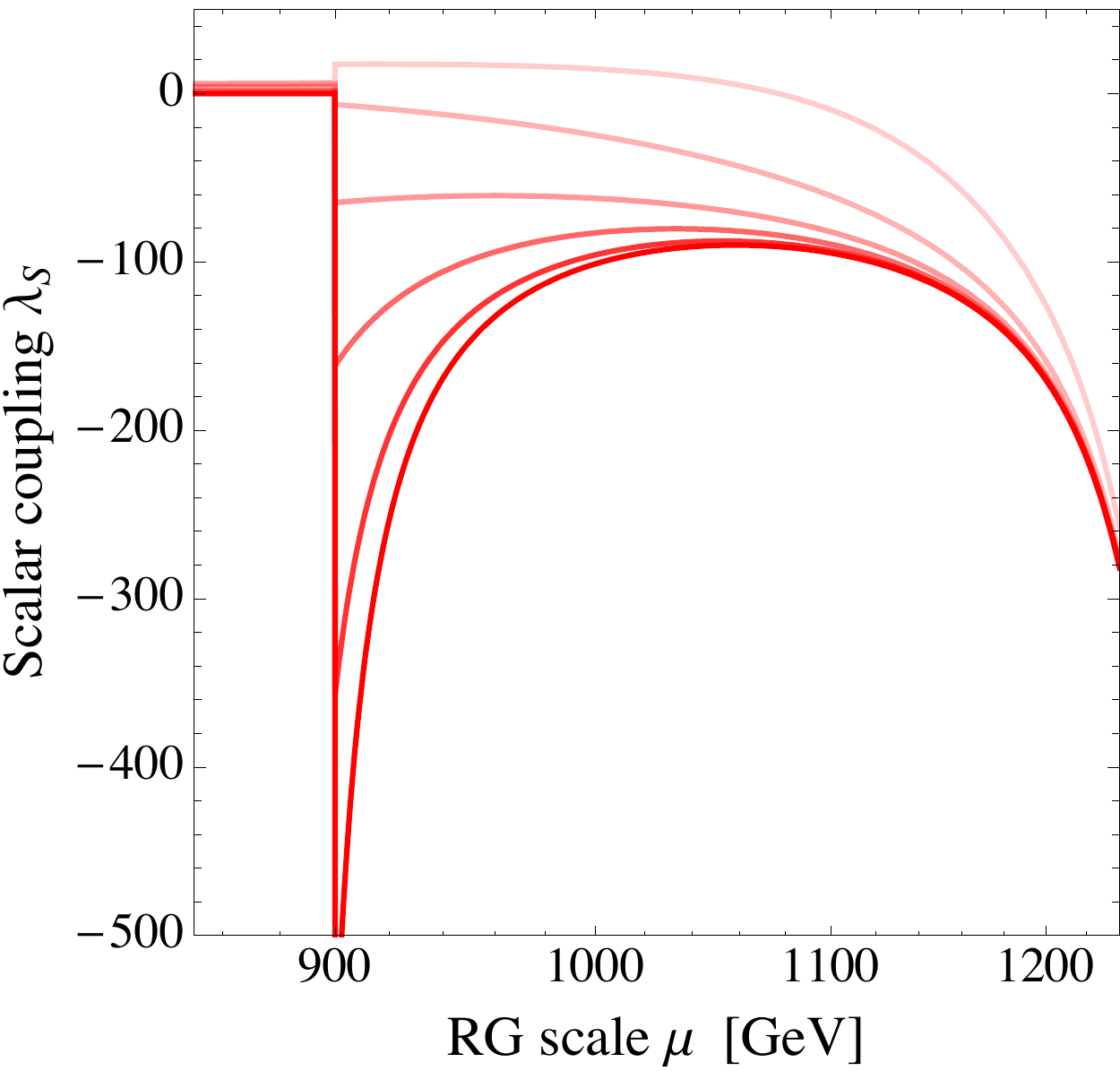}
\endminipage\hfill
\minipage{0.3\textwidth}
  \includegraphics[width=1.\linewidth]{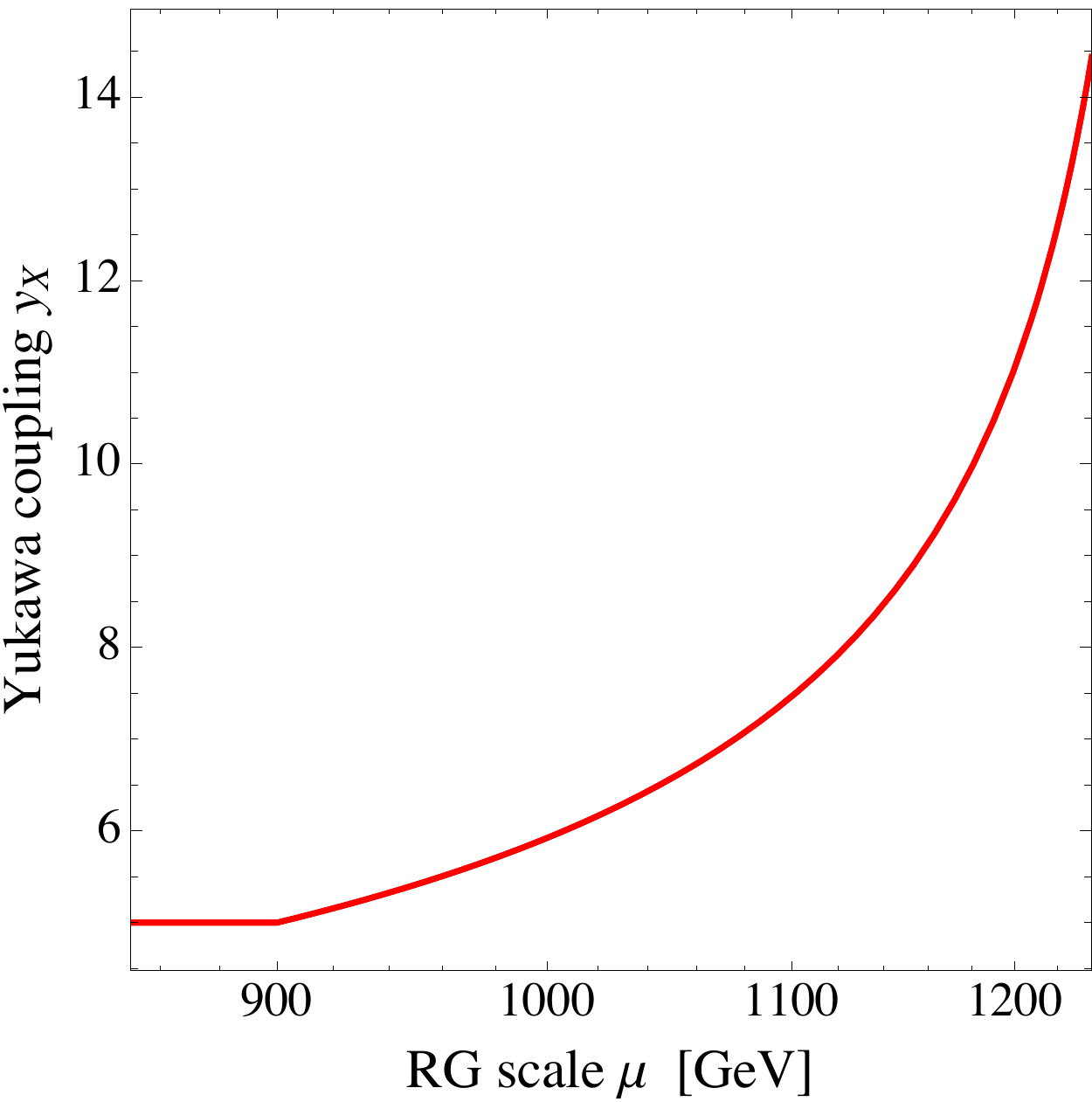}
\endminipage
 \caption{
 Left panel. Running of $\lambda_{HS}$ in the model with $y_X = 5$, $m_X = 900$ GeV, $Q_X = 5/3$, $N_X = 1$, $s_{\theta} = 0.01$.
From lighter to darker, the running corresponds to the initial values $\lambda_{HS}(\mu_0) = 10^{-3}, 10^{-2}, 5 \times10^{-2}, 8 \times10^{-2}, 9.5 \times10^{-2},10^{-1}$.
Central panel. Running of $\lambda_S$. Right panel. Running of $y_X$.
  }
 \label{fig:Mixing}
\end{figure}
We show our results in Fig.~\ref{fig:Mixing}, where we consider the running of $\lambda_{HS}$ (left), $\lambda_{S}$ (central), and $y_X$ (right). 
Fig.~\ref{fig:Mixing} should be compared with the running for the unmixed case described in Fig.~\ref{fig:RunningLambdaSPositive}.
Our conclusions obviously remain unchanged. 
We checked that, at the qualitative level, all the results obtained in Section~\ref{eq:RGEresults} are not altered by the presence of a non-zero mixing angle
if---as described in Fig.~\ref{fig:RunningLambdaSPositive}, and done explicitly in this Appendix---one scans over the allowed values of $\lambda_{HS}$.

For completeness, let us also discuss the case with negative $\lambda_{HS}$.
\begin{figure}[!htb!]
\begin{center}
\fbox{\footnotesize $s_{\theta} = 0.01$, $y_X = 5$, $m_X = 900$ GeV, $Q_X = 5/3$, $N_X = 1$, $(\lambda_{HS} < 0$)}
\end{center}
\minipage{0.325\textwidth}
  \includegraphics[width=1.\linewidth]{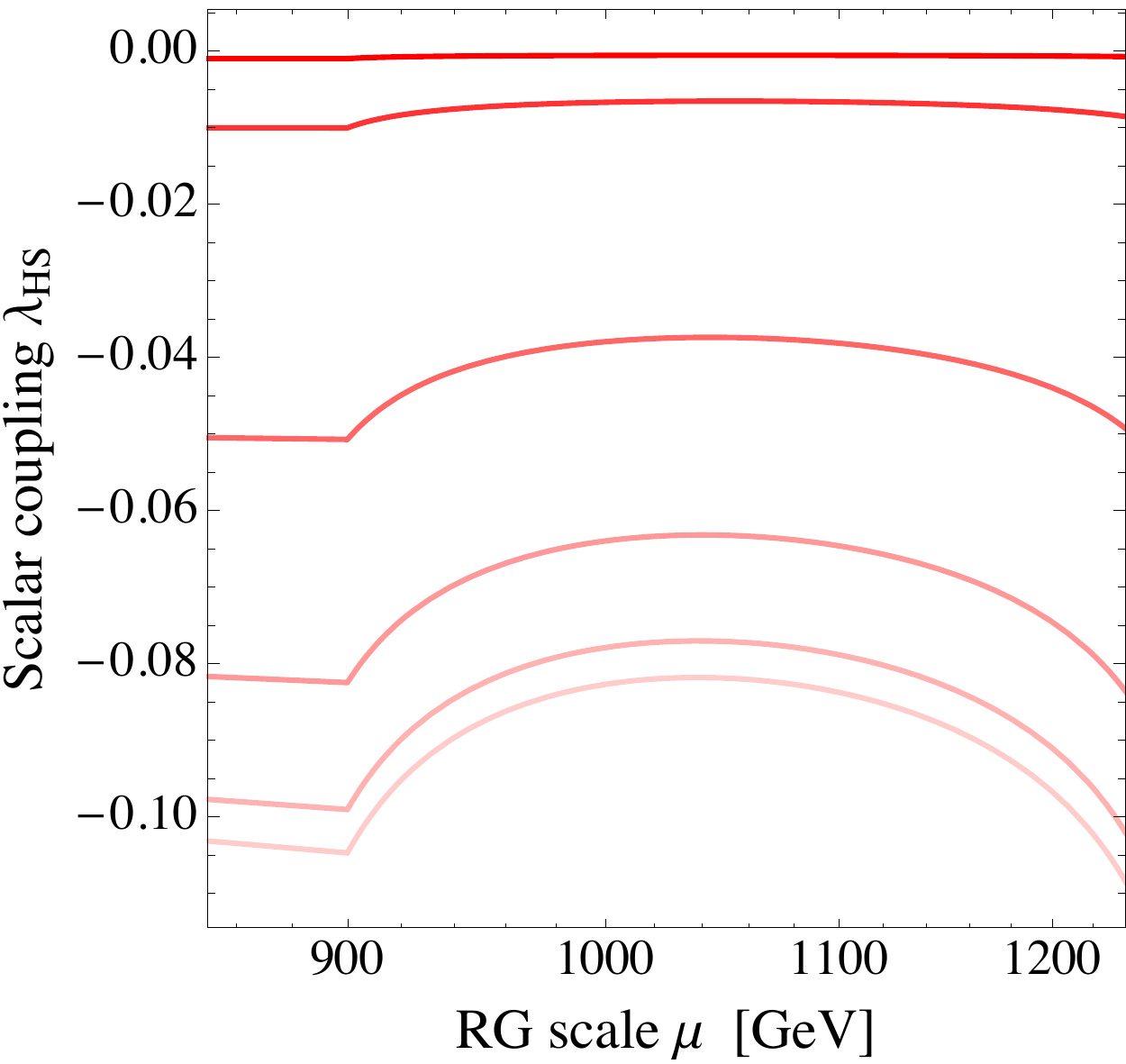}
\endminipage\hfill
\minipage{0.325\textwidth}
  \includegraphics[width=1.\linewidth]{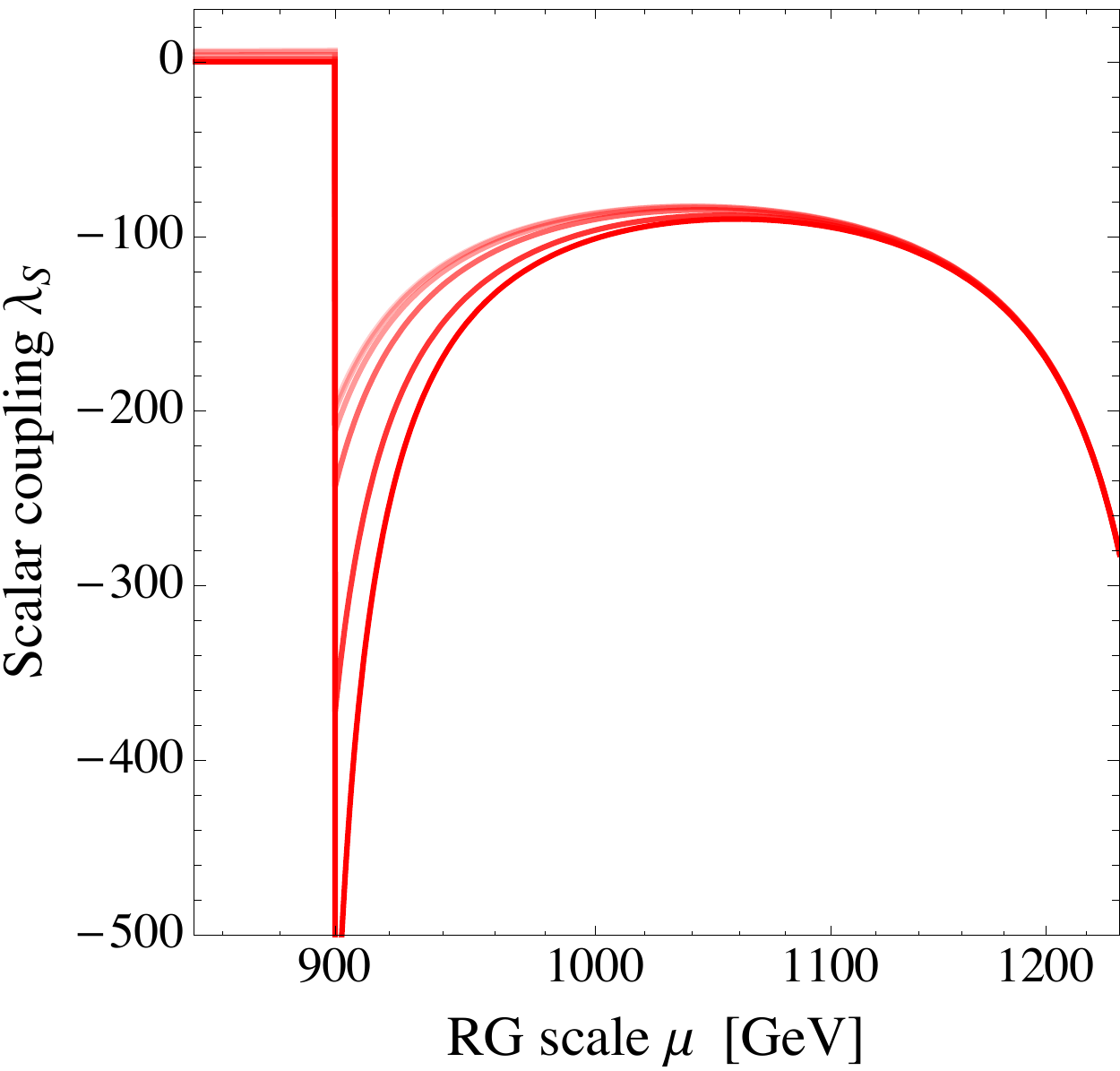}
\endminipage\hfill
\minipage{0.3\textwidth}
  \includegraphics[width=1.\linewidth]{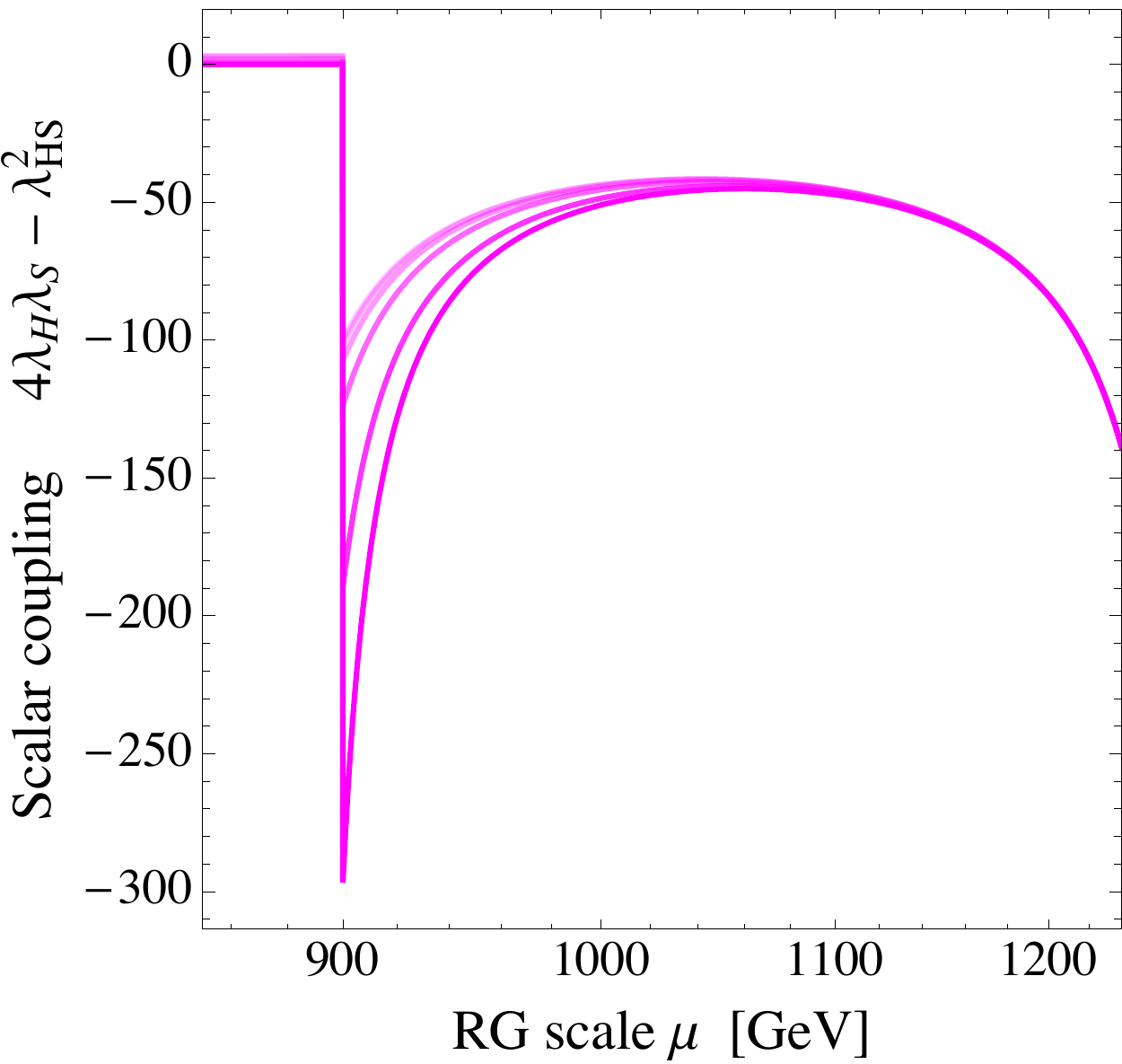}
\endminipage
 \caption{
The same as in Fig.~\ref{fig:Mixing} but for negative $\lambda_{HS}$. In the right panel we show, instead of the running of $y_X$ as in Fig.~\ref{fig:Mixing}, the running of the combination 
$4\lambda_H\lambda_ S - \lambda_{HS}$. We analyze the same values of $\lambda_{HS}(\mu_0)$ in absolute value but with opposite sign.
  }
 \label{fig:Mixing2}
\end{figure}
In this case we have the additional constraint  $4\lambda_H(\Lambda)\lambda_S(\Lambda) - \lambda_{HS}(\Lambda) >0$ (see Section~\ref{sec:Instability}).
We show our results in Fig.~\ref{fig:Mixing2}, in which we focus again on $N_X = 1$, $Q_X = 5/3$, $y_X = 5$, $m_X = 900$ GeV.
As expected, the pathological behavior of the theory already emphasized in the case $\lambda_{HS}(\mu_0) >0$ still persists.
In particular, in addition to $\lambda_S < 0$, also the  negative direction $4\lambda_H\lambda_S - \lambda_{HS} < 0$  is generated along the RG flow.

\bibliography{lit}



\end{document}